\documentclass[paper,toc,nohyper]{JHEP3}
\pdfoutput=1
\usepackage{epsfig}
\usepackage{amsbsy,ams} 
\usepackage{subfigure,epsfig}
\usepackage{graphicx}
\usepackage{pifont}
\usepackage{amsmath}
\usepackage{booktabs}
\usepackage{rotating}
\usepackage{multirow}
\usepackage{cite}

\def\dsigma{{\rm d} \hat\sigma}
\def\ph#1{\phantom{.}}
\def\JET{J}
\def\Poles{{\cal P}oles}

\def\d{\hbox{d}}

\def\doubletilde#1{\widetilde{\vphantom{\raise 1.5pt \hbox{#1}}\smash{\kern -2pt\widetilde{#1}}}}
\def\e{\epsilon}

\def\eps{\epsilon}

\def\spa#1.#2{\left\langle\mskip-1mu#1\,#2\mskip-1mu\right\rangle}
\def\spb#1.#2{\left[\mskip-1mu#1\,#2\mskip-1mu\right]}

\def\A{{\cal A}}

\def\NF{N_F}
\def\dbar#1{\bar{\bar{#1}}}
\def\qb{\bar{q}}

\def\hb#1{\hat{\bar{#1}}}
\def\hbb#1{\hat{\bar{\bar{#1}}}}
\def\nn{\nonumber}
\def\mc#1#2#3{\mathcal{#1}_3^0(s_{#2#3})}

\def\sterm#1#2{\mathcal{S}(s_{#1}, s_{#2}, x_{#1,#2})}
\def\stermi#1{\mathcal{S}(s_{#1}, s_{#1}, 1)}

\def\bs{\boldsymbol}

\def\wt{\widetilde}
\def\t{\tilde}

\def\b#1{\bar{#1}}

\def\hb#1{\hat{\bar{#1}}}
\def\hbb#1{\hat{\bar{\bar{#1}}}}

\newcommand{\beq}{\begin{equation}}
\newcommand{\eeq}{\end{equation}}
\newcommand{\bea}{\begin{eqnarray}}
\newcommand{\eea}{\end{eqnarray}}

\def\n2f{{n^{\,2}_{\! f}}}
\def\pgg(#1){p_{\rm{gg}}(#1)}
\def\H(#1){{\rm{H}}_{#1}}
\def\Hh(#1,#2){{\rm{H}}_{#1,#2}}

\def\ba{\begin{eqnarray}}
\def\ea{\end{eqnarray}}
\def\nn{\nonumber}

\def\NNNLORR{{\cal N}_{NNLO}^{RR}}
\def\NNNLORV{{\cal N}_{NNLO}^{RV}}
\def\NNNLOVV{{\cal N}_{NNLO}^{VV}}

\def\dPSxx#1#2{\int{\rm d}\Phi_{#1}(p_3,\ldots,p_{#2};x_1 p_1,x_2 p_2)\frac{1}{#1!}\,\frac{{\rm d}x_1}{x_1}\frac{{\rm d}x_2}{x_2}}

\def\mptwops {{\NNNLORR\ \sum_{n+2}\ \text{d}\Phi_{n+2}(p_{3},\ldots,p_{n+4};p_{1},p_{2})\ \frac{1}{S_{n+2}}}}
\def\mponeps {{\NNNLORV\ \sum_{n+1}\ \int\frac{\text{d}x_{1}}{x_{1}}\frac{\text{d}x_{2}}{x_{2}}\ \text{d}\Phi_{n+1}(p_{3},\ldots,p_{n+3};x_{1}{p}_{1},x_{2}{p}_{2})\ \frac{1}{S_{n+1}}}}
\def\mps {{\NNNLOVV\ \sum_{n}\ \int\frac{\text{d}z_{1}}{z_{1}}\frac{\text{d}z_{2}}{z_{2}}\ \text{d}\Phi_{n}(p_{3},\ldots,p_{n+2};z_{1}{p}_{1},z_{2}{p}_{2})\ \frac{1}{S_{n}}}}

\def\Mtreeqq(#1,#2){M_{2}^{0}(#1_{q},#2_{q})}
\def\Mtreeqgq(#1,#2,#3){M_{3}^{0}(#1_{q},#2_{g},#3_{q})}
\def\Mtreeqggq(#1,#2,#3,#4){M_{4}^{0}(#1_{q},#2_{g},#3_{g},#4_{q})}
\def\Mttreeqggq(#1,#2,#3,#4){\wt{M}_{4}^{0}(#1_{q},#2_{g},#3_{g},#4_{q})}

\def\Mtreeqqb(#1,#2){M_{2}^{0}(#1_{q},#2_{\bar{q}})}
\def\Mtreeqgqb(#1,#2,#3){M_{3}^{0}(#1_{q},#2_{g},#3_{\bar{q}})}
\def\Mtreeqggqb(#1,#2,#3,#4){B_{4}^{0}(#1_{q},#2_{g},#3_{g},#4_{\bar{q}})}
\def\Mttreeqggqb(#1,#2,#3,#4){\wt{B}_{4}^{0}(#1_{q},#2_{g},#3_{g},#4_{\bar{q}})}

\def\Mtreeqgggqb(#1,#2,#3,#4,#5){B_{5}^{0}(#1_{q},#2_{g},#3_{g},#4_{g},#5_{\bar{q}})}

\def\Mloopqq(#1,#2){M_{2}^{1}(#1_{q},#2_{q})}
\def\Mtloopqq(#1,#2){\wt{M}_{2}^{1}(#1_{q},#2_{q})}
\def\Mloopqgq(#1,#2,#3){M_{3}^{1}(#1_{q},#2_{g},#3_{q})}
\def\Mtloopqgq(#1,#2,#3){\wt{M}_{3}^{1}(#1_{q},#2_{g},#3_{q})}
\def\Mhatloopqgq(#1,#2,#3){\widehat{M}_{3}^{1}(#1_{q},#2_{g},#3_{q})}

\def\Mloopqqb(#1,#2){M_{2}^{1}(#1_{q},#2_{\bar{q}})}
\def\Mtloopqqb(#1,#2){\wt{M}_{2}^{1}(#1_{q},#2_{\bar{q}})}
\def\Mloopqgqb(#1,#2,#3){M_{3}^{1}(#1_{q},#2_{g},#3_{\bar{q}})}
\def\Mtloopqgqb(#1,#2,#3){\wt{M}_{3}^{1}(#1_{q},#2_{g},#3_{\bar{q}})}
\def\Mhatloopqgqb(#1,#2,#3){\widehat{M}_{3}^{1}(#1_{q},#2_{g},#3_{\bar{q}})}

\def\bs{\boldsymbol}
\def\bra[#1]{\langle#1|}
\def\ket[#1]{|#1\rangle}
\def\braket[#1,#2]{\langle#1|#2\rangle}

\allowdisplaybreaks

\preprint{
  IPPP/12/82, ZU-TH 26/12\\
  \\
  \today}

\title{Infrared Structure at NNLO Using Antenna Subtraction}

\author{James Currie$^a$, E.W.N. Glover$^{b}$, Steven Wells$^b$\\
$^a$Institut f\"ur Theoretische Physik, Universit\"at Z\"urich, Wintherturerstrasse 190,\\
CH-8057, Z\"urich, Switzerland\\
$^b$Institute for Particle Physics Phenomenology, University of Durham,
South Road,\\ Durham DH1 3LE, England}

\abstract{We consider the infrared structure of hadron-hadron collisions at next-to-next-to leading order using the antenna subtraction method.  The general form of the subtraction terms is presented for double real, real-virtual and double virtual contributions.  At NLO and NNLO it is shown that the virtual and double virtual subtraction terms can be written in terms of \emph{integrated dipoles}, formed by systematically combining the mass factorisation contributions and integrated antenna functions.  The integrated dipoles describing $\ell$ unresolved partons, denoted $\bs{J}_{2}^{(\ell)}$, are related to Catani's IR singularity operators, $\bs{I}_{ij}^{(\ell)}(\eps)$.  It is shown that the IR pole structure of the virtual and double virtual contributions can be written as a sum over integrated dipoles within the antenna subtraction formalism and the master expressions analogous to Catani's one- and two-loop factorisation formulae are derived.  To demonstrate the techniques described in this paper, we apply antenna subtraction to the production of two gluon jets via quark-antiquark scattering at NLO and NNLO.  Double real, real-virtual and double virtual subtraction terms are explicitly derived for the leading colour NNLO contribution.
\\
\today
}

\begin{document}
\maketitle
\tableofcontents
\newpage

\section{Introduction}
\label{sec:intro}

In QCD, the (renormalised and mass factorised) inclusive
cross section for a hard scattering process in proton-proton collisions that produces a particular final state, $X$, has the factorised form,
\begin{equation}
\label{eq:totsig}
{\rm d}\sigma =\sum_{i,j} \int   
\frac{d\xi_1}{\xi_1} \frac{d\xi_2}{\xi_2} f_i(\xi_1,\mu_F^2) f_j(\xi_2,\mu_F^2) \dsigma_{ij}(\alpha_s(\mu_R),\mu_R,\mu_F) \nonumber
\end{equation}
where the probability of finding a parton of type $i$ in the proton, carrying a momentum fraction $\xi$, is described by the parton distribution function $f_i(\xi,\mu_F^2)d\xi$ and the partonic cross section ${\rm d}\hat\sigma_{ij}$  for parton $i$ to scatter off parton $j$ and produce $X$, normalised to the hadron-hadron flux\footnote{The partonic cross section normalised to the parton-parton flux is obtained by absorbing the inverse factors of $\xi_1$ and $\xi_2$ into $\dsigma_{ij}$.} is summed over the possible parton types $i$ and $j$. As usual $\mu_R$ and $\mu_F$ are the renormalisation and factorisation scales which are frequently set to be equal for simplicity, $\mu_{R}=\mu_{F}=\mu$.

For suitably high centre of mass scattering energies, the infrared-finite partonic cross section has the perturbative expansion 
\begin{equation}
\label{eq:sigpert}
\dsigma_{ij} = {\rm d}\hat\sigma_{ij}^{LO}
+\left(\frac{\alpha_s(\mu_R)}{2\pi}\right)\dsigma_{ij}^{NLO}
+\left(\frac{\alpha_s(\mu_R)}{2\pi}\right)^2\dsigma_{ij}^{NNLO}
+{\cal O}(\alpha_s^3)
\end{equation}
where the next-to-leading order (NLO) and next-to-next-to-leading order (NNLO) strong corrections are identified. In general, the leading-order cross section will be proportional to an overall power of the strong coupling, $\alpha_{s}^{m}$, with the value of $m$ depending on the process under consideration.  NLO computations for multiparticle processes are becoming   standard for comparison with LHC data.  However, for processes with only one or two objects in the final state, the spectacular performance of the LHC and LHC experiments means that NNLO accuracy is mandatory.

Calculations of higher order QCD corrections to the underlying Born-level process encounter additional integrals that must be combined before physical predictions can be extracted: loop integrals in the case of virtual corrections and phase-space integrals for real corrections, reflecting the additional particles present in higher order calculations.  It is well known that for theories containing massless states, both the virtual and real radiative corrections are peppered with IR singularities which conspire to mutually cancel to form the finite physical cross section.  

In order to achieve this cancellation, the IR singularities must be extracted from the real emission phase space integrals in the form of a Laurent expansion in the dimensional regularisation parameter $\eps$, where $D=4-2\eps$.  
Subtraction schemes are a well established solution to this problem and several methods for systematically constructing general subtraction terms have been proposed in the literature at NLO \cite{Catani:1996vz,Frixione:1995ms,Nagy:1996bz,Frixione:1997np,Somogyi:2006cz} and at NNLO \cite{ Weinzierl:2003fx,Kilgore:2004ty,Frixione:2004is,GehrmannDeRidder:2005cm,Somogyi:2005xz,Somogyi:2006da,Somogyi:2006db,Somogyi:2008fc,Aglietti:2008fe,Somogyi:2009ri,Bolzoni:2009ye,Bolzoni:2010bt,Czakon:2010td,Anastasiou:2010pw,Czakon:2011ve,babishiggs3,Boughezal:2011jf,DelDuca:2013kw,Somogyi:2013yk}. 
So far,
successful applications of subtraction at NNLO to specific observables were
accomplished with antenna
subtraction~\cite{GehrmannDeRidder:2005cm}, $q_T$-subtraction~\cite{Catani:2007vq}, sector
decomposition~\cite{Binoth:2000ps,Heinrich:2002rc,Anastasiou:2003gr,
Binoth:2004jv}, non-linear transformations~\cite{Anastasiou:2010pw},    and  most recently with an approach
based on sector-improved  residue subtraction~\cite{Czakon:2010td,Czakon:2011ve}.
The $q_T$-subtraction method is  restricted
to processes with colourless final states at leading order and is  based on the
universal infrared structure of the real emissions, which can be inferred from
transverse momentum resummation. This method has been applied at NNLO to the
Higgs production~\cite{grazzinihiggs}, vector boson 
production~\cite{grazzinidy1,grazzinidy2}, associated
$VH$-production~\cite{grazziniwh}, photon pair production~\cite{grazzinigg} and
in modified form to top quark decay~\cite{scettop}.
The sector decomposition approach relies on an iterated decompostion of the final
state phase space and matrix  element, allowing an expansion in distributions,
followed by a numerical  evaluation of the sector integrals. It has been
successfully applied to Higgs
production~\cite{babishiggs1,babishiggs2,babishiggs3} and vector boson 
production~\cite{kirilldy} at NNLO. 
The sector-improved residue subtraction  extends NLO residue
subtraction~\cite{Frixione:1995ms}, combined with a numerical  evaluation of the
integrated subtraction terms. It has been applied to  compute the NNLO
corrections to top quark pair production~\cite{czakontop1,czakontop2}. 

The antenna subtraction formalism that is the focus of this paper has been fully
developed for all types of scattering processes at
NLO~\cite{GehrmannDeRidder:2005cm,GehrmannDeRidder:2005hi,GehrmannDeRidder:2005aw,Daleo:2006xa}
and further developed at NNLO~\cite{GehrmannDeRidder:2005cm,
Daleo:2009yj,Boughezal:2010mc,Gehrmann:2011wi,GehrmannDeRidder:2012ja}.
Within this method,  the subtraction terms are constructed from so-called
antenna functions which describe all unresolved partonic radiation (soft and
collinear) between a hard pair of radiator partons. The hard radiators may be in
the initial- or final-state, and in the most general case, final-final (FF),
initial-final  (IF) and initial-initial  (II) antennae need to be considered.
The subtraction terms and therefore the antennae also need to be integrated over
the unresolved phase space, which is different in the three configurations.
Recently, all of the integrals relevant for processes at NNLO with massless
quarks have been analytically
evaluated~\cite{Daleo:2009yj,Boughezal:2010mc,Gehrmann:2011wi,GehrmannDeRidder:2012ja}.  

The antenna subtraction approach has been successfully applied to the infrared
structure of three-jet events at NNLO~\cite{our3j1,weinzierl3j1,weinzierl3j2}
and the subsequent numerical calculation of the NNLO corrections to event shape
distributions~\cite{ourevent1,ourevent2,weinzierlevent1,ourevent3,weinzierlevent2},
the moments of event shapes~\cite{ourevent3,weinzierlevent2} and jet
rates~\cite{our3j2,weinzierlevent1}.   For hadron colliders, the antenna method
has been applied at NNLO to the all gluon contribution to dijet
production~\cite{Glover:2010im,GehrmannDeRidder:2011aa,GehrmannDeRidder:2012dg} 
and the production of heavy
particles~\cite{Abelof:2011jv,Abelof:2011ap,Abelof:2012rv,Abelof:2012he}.

There is however another aspect to the use of the antenna subtraction method.  The antenna approach successfully isolates the infrared singularities, which themselves have a very particular structure.   Therefore, one can ask the questions:
\begin{enumerate}
\item How does the infrared structure guide the construction of the real radiation subtraction terms?
\item How do the integrated subtraction terms relate to the known infrared structure of the loop amplitudes?
\end{enumerate}
The main goal of this paper is to develop the systematics of the antenna subtraction scheme at NLO and particularly at NNLO by focussing on the structure of the subtraction terms for the double real, real-virtual and double virtual channels.    

It is well known that the explicit poles of the virtual contributions are described by Catani's one- and
two-loop factorisation formulae~\cite{Catani:1998bh,Sterman:2002qn} and follow a colour-dipole structure, reflected by the IR
singularity operators $\bs{I}_{ij}^{(1)}(\eps)$ and $\bs{I}_{ij}^{(2)}(\eps)$ which describe the IR singularities associated with
two colour connected hard particles. It is similarly well established (and in fact the keystone of all
subtraction methods) that real radiation amplitudes factorise in IR divergent limits following an antenna
factorisation pattern whereby a pair of hard partons radiate unresolved partons.  The unintegrated antenna functions are used to mimic the implicit divergence of the real
contributions while the poles associated with the integrated antenna functions directly cancel the explicit
poles of the virtual contributions.   

The implicit divergence of the NLO and NNLO cross sections is captured by
distinct blocks of subtraction terms which follow a predictive structure based on the colour ordering of the
matrix elements under consideration.  After integration and combination with the relevant mass factorisation
contributions, a number of structures emerge which we call \emph{integrated dipoles}.  At one-loop,  
$\bs{J}_{2}^{(1)}$ is related to an integrated three-particle tree-level antenna and describes the unresolved radiation between two colour connected particles. Similarly, at two-loops, $\bs{J}_{2}^{(2)}$ fulfills the same role for double unresolved radiation and involves integrated four-particle tree-level antennae, three-particle one-loop antennae and products of three-parton tree-level antennae. 

The IR structure of any two-loop contribution is given by one- and two-loop \emph{integrated antenna strings} which are simply formed from  $\bs{J}_{2}^{(1)}$ and $\bs{J}_{2}^{(2)}$. For a
given colour ordering of a particular $n$-particle process, $\bs{J}_{n}^{(1)}$ and $\bs{J}_{n}^{(2)}$ can immediately be written down as a sum over integrated dipoles that involve colour connected particles,
\ba
\bs{J}_{n}^{(\ell)}(1,\cdots,n)&=&\sum_{(i,j)}\bs{J}_{2}^{(\ell)}(i,j),
\ea
where $\ell=1,2$ denotes the single and double unresolved integrated antenna strings respectively and the sum is over colour connected pairs of partons.  Equivalently, the pole structure is identified as a sum of dipole-like terms proportional to $\left(|s_{ij}|\right)^{-\ell\eps}$  that link the colour connected particles $i$ and $j$. 

Of course, the form of  $\bs{J}_{n}^{(1)}$ and $\bs{J}_{n}^{(2)}$ in terms of integrated antennae imposes a particular
structure on the unintegrated antennae that make up the real radiation subtraction terms. Understanding the explicit pole structure of virtual amplitudes in terms of integrated antenna
strings gives a direct connection between the block structure of the unintegrated subtraction terms and
the explicit pole structure of virtual contributions, thereby simplifying the construction of the double real, real-virtual and double virtual subtraction terms.

The paper is organised in the following way.  In Section 2 the general structure of the antenna subtraction terms relevant for NLO calculations is discussed, in both unintegrated and integrated forms.  This procedure leads to a natural definition of the NLO integrated dipole $\bs{J}_{2}^{(1)}$.  Section 3 then extends this analysis to NNLO where we construct groups of antenna subtraction terms that remove the double and single unresolved singularities from the double real and real-virtual contributions.  The NNLO integrated dipole $\bs{J}_{2}^{(2)}$  naturally emerges from the groups of integrated antenna functions and mass factorisation kernels and (along with combinations of $\bs{J}_{2}^{(1)}$) reproduces and properly subtracts the explicit poles of the double virtual contribution to the NNLO cross section.  

As an example of the power of the structures motivated in general terms in Sections 2 and 3, Section 4 considers the production of gluon jets from quark-antiquark scattering at NLO and NNLO. We derive the double real, real-virtual and double virtual subtraction terms relevant for the NNLO corrections to the $q\bar q \to gg$ channel, which is one of many channels that contribute to the dijet rate at the LHC. In Section~5 we revisit the double virtual subtraction term for $e^{+}e^{-}\to3$ jets at NNLO and show that it has the same structure in terms of integrated antenna strings.  Finally, our findings are summarised in Section~\ref{sec:conclusions}.  For completeness a number of appendices are enclosed detailing the collinear splitting kernels in the $N$, $\NF$ basis we employ and giving an explicit expression for the integrated soft factor with an initial-final mapping.


\section{Antenna subtraction at NLO}
\label{sec:nloant}

The structure of NLO calculations within the antenna subtraction formalism is particularly simple. Nevertheless it demonstrates some of the key features relevant to NNLO calculations which are fully explained in this section.  

We are interested in the higher-order corrections to the underlying leading-order (LO) cross section, which for the hadro-production of $n$ jets is given by,
\ba
\dsigma_{ij,LO}(\xi_{1}H_{1},\xi_{2}H_{2})&=&\int_{n}\dsigma_{ij,LO}^{B}(\xi_{1}H_{1},\xi_{2}H_{2}),\label{eq:loxsec}
\ea
where the Born-level cross section is obtained by evaluating the tree-level contributions for $(n+2)$-parton scattering with partons $i,j$ in the initial-state carrying a fraction $\xi_{1,2}$ of the parent hadron's momentum $H_{1,2}$.  This quantity is then integrated over the final-state phase space where we keep track of the number $n$ of final state particles using the shorthand notation,
\ba
\int_{n}.\label{eq:dphi}
\ea

Explicitly,  the Born-level cross section is given by,
\ba
\label{eq:sigB}
\dsigma_{ij,LO}^{B}&=&{\cal{N}}_{LO}\   \sum_{\sigma}\ \text{d}\Phi_{n}(p_{3},\cdots,p_{n+2};p_{1},p_{2})\ \frac{1}{S_{n}}\nn\\
&\times&\bigg[M_{n+2}^{0}(\sigma(1,\cdots,n+2))\ J_{n}^{(n)}(\{p\}_{n})+{\cal{O}}\bigg(\frac{1}{N^{2}}\bigg)\bigg],
\ea
where $S_{n}$ is a final-state symmetry factor, and $\{p\}_{n}$ denotes the set of $(n)$ final-state momenta.  $M_{n+2}^{0}$ denotes an $(n+2)$-parton squared partial amplitude for a given colour ordering denoted by $\sigma$ and $i$ labels the parton with momentum $p_{i}$. The factor ${\cal{N}}_{LO}$ will in general contain all non-QCD factors and some overall QCD factors such as the overall power of the coupling.  The $2\to n$ particle phase space is defined by,
\ba
&&{\rm{d}}\Phi_{n}(p_{3},\cdots,p_{n+2};p_{1},p_{2})=\nn\\
&&\ \frac{{\rm{d}}^{d-1}p_{3}}{2E_{3}(2\pi)^{d-1}}\cdots\frac{{\rm{d}}^{d-1}p_{n+2}}{2E_{n+2}(2\pi)^{d-1}}(2\pi)^{d}\delta^{d}(p_{1}+p_{2}-p_{3}-\cdots-p_{n+2}).
\ea
The jet algorithm $J_n^{(m)}$ builds $n$ jets from $m$ final-state partons with momenta labelled by the set $\{p\}_{m}$.  In the case of the leading-order cross section defined in Eq.~\eqref{eq:sigB} $m=n$ and there is an exact parton-jet correspondence.

The NLO correction to the $n$-jet cross section contains three contributions and is given by,
\ba
\dsigma_{ij,NLO}&=&\int_{n+1}\dsigma_{ij,NLO}^{R}\ +\ \int_{n}\biggl(\dsigma_{ij,NLO}^{V}+\dsigma_{ij,NLO}^{MF}\biggr),
\ea
where $\dsigma_{ij}^{R}$ and $\dsigma_{ij,NLO}^{V}$ are the real and virtual NLO corrections and $\dsigma_{ij,NLO}^{MF}$ is the NLO mass factorisation contribution,
\ba
\dsigma_{ij,NLO}^{MF}(\xi_{1}H_{1},\xi_{2}H_{2})&=&-
 \int\frac{{\rm{d}}z_{1}}{z_{1}}\frac{{\rm{d}}z_{2}}{z_{2}}\, 
 \left(\frac{\alpha_s N}{2\pi}\right)\, \bar{C}(\eps)\,\Gamma_{ij;kl}^{(1)}(z_{1},z_{2})\
 \dsigma_{kl,LO}(z_{1}\xi_{1}H_{1},z_{2}\xi_{2}H_{2}),\label{eq:nlomf}\nonumber \\
\ea
where $\bar{C}(\eps)=(4\pi)^{\eps}e^{-\eps\gamma}$.
The mass factorisation contribution serves to remove all initial-state collinear singularities from the cross section by absorbing them into the redefined physical Parton Distribution Function (PDF) and can be written in terms of the one-loop Altarelli-Parisi kernels,
\ba
\Gamma_{ij;kl}^{(1)}(z_{1},z_{2})&=&
\delta(1-z_{2})\delta_{lj}\Gamma_{ki}^{(1)}(z_{1})
+\delta(1-z_{1})\delta_{ki}\Gamma_{lj}^{(1)}(z_{2}).
\ea

The real cross section is known to contain soft and collinear IR divergences and so a subtraction term, $\dsigma_{ij,NLO}^{S}$, is constructed from antenna functions and reduced multiplicity matrix elements.  This subtraction term must remove all implicit singularities from the real emission cross section without introducing further spurious singularities of its own and be local in the sense that the subtraction is successful point-by-point in phase space.  A secondary requirement of a subtraction term is that it be analytically integrable in order to convert the implicit divergence of the subtraction term into the explicit singularities of the integrated subtraction term.\footnote{One could equally well numerically integrate the subtraction term as in Ref.~\cite{Czakon:2010td} or \cite{DelDuca:2013kw,Somogyi:2013yk}.} When combined with the mass factorisation contribution, these terms conspire to cancel the explicit poles of the virtual contribution and produce a finite NLO cross section.  The requirement of analytic integrability is automatically satisfied in the antenna subtraction method as all three- and four-parton antenna functions have been successfully integrated for FF, IF and II configurations.

Given these considerations, the NLO cross section can be reorganised in the following way such that each square bracket is free from both implicit divergence and explicit poles,
\ba
\dsigma_{ij,NLO}&=&\int_{n+1}\bigl[\dsigma_{ij,NLO}^{R}-\dsigma_{ij,NLO}^{S}\bigr]+\int_{n}\bigl[\dsigma_{ij,NLO}^{V}-\dsigma_{ij,NLO}^{T}\bigr],
\ea
where the virtual subtraction term is given by the real subtraction term integrated over the single unresolved phase space and the NLO mass factorisation contribution,
\ba
\dsigma_{ij,NLO}^{T}&=&-\int_{1}\dsigma_{ij,NLO}^{S}-\dsigma_{ij,NLO}^{MF}.
\ea

\subsection{Construction of the real emission subtraction term}
\label{sec:nloreal}

The single real emission cross section takes the form,
\ba
\dsigma_{ij,NLO}^{R}&=&{\cal{N}}_{NLO}^{R}\  \sum_{\sigma}\ \text{d}\Phi_{n+1}(p_{3},\cdots,p_{n+3};p_{1},p_{2})\ \frac{1}{S_{n+1}}\nn\\
&\times&\bigg[M_{n+3}^{0}(\sigma(1,\cdots,n+3))\ J_{n}^{(n+1)}(\{p\}_{n+1})+{\cal{O}}\bigg(\frac{1}{N^{2}}\bigg)\bigg],
\ea
where, as in Eq.~\eqref{eq:sigB}, $S_{n+1}$ is a final-state symmetry factor,  $\{p\}_{n+1}$ denotes the set of $(n+1)$ final-state momenta, $M_{n+3}^{0}$ denotes an $(n+3)$-parton squared partial amplitude for a given colour ordering denoted by $\sigma$ and $i$ labels the parton with momentum $p_{i}$. 
The overall coupling is given by
\ba
{\cal{N}}_{NLO}^{R} &=& {\cal{N}}_{LO} \left(\frac{\alpha_s N}{2\pi}\right) \ \frac{\bar{C}(\eps)}{C(\eps)}
\ea
where $C(\eps) = \bar{C}(\eps)/8\pi^2$.

\subsubsection{Colour ordered strings}

Before discussing the various subtraction terms, it is useful to review the possible structures available for colour ordered matrix elements.  There are two basic objects from which all colour structures can be assembled: gluon strings and quark strings.  A gluon string is characterised by a trace over $SU(N)$ generators, where $a_{k}$ denotes the generator associated with external gluon $k$, e.g., an $n$-gluon colour ordered matrix element is associated with the colour structure,
\ba
{\rm{Tr}}(T^{a_{1}}T^{a_{2}}\cdots T^{a_{n-1}}T^{a_{n}})\equiv(a_{1}a_{2}\cdots a_{n-1}a_{n}).
\ea
The colour ordered partial amplitude associated with this colour structure contains IR divergences between colour connected partons, i.e., those external legs whose associated generators are adjacent in the colour structure.  Due to the trace nature of the colour factors the partial amplitudes display cyclic symmetry, which means that the colour connection not only exists between adjacent partons in the notation defined above, but also between the endpoints $a_{1}$ and $a_{n}$.

The other basic colour structure is the quark string, given by a string of gluon generators bookended by fundamental quark indices, e.g., a quark string containing a quark-antiquark pair and $n$ gluons has the colour structure,
\ba
(T^{a_{1}}T^{a_{2}}\cdots T^{a_{n-1}}T^{a_{n}})_{ij}\equiv(a_{1}a_{2}\cdots a_{n-1}a_{n})_{ij},
\ea
where $i,j$ denote the fundamental (antifundamental) colour indices carried by the quark (antiquark).  The IR divergences of the partial amplitude associated with this colour structure also exist for unresolved configurations involving colour connected partons.  The gluons at each end of the quark string are not colour connected to each other but are colour connected to the quark endpoints instead.  

If the quark-antiquark pair become collinear to form a composite gluon,  the colour structure pinches down to a trace structure and the surviving reduced matrix element (multiplied by the universal splitting function) tends to a gluonic partial amplitude,
\ba
(a_{1}a_{2}\cdots a_{n-1}a_{n})_{ij}&\stackrel{q_{i}||\b{q}_{j}}{\longrightarrow}&(a_{1}a_{2}\cdots a_{n-1}a_{n}a_{(ij)}).
\ea

Colour structures involving multiple quark-antiquark pairs are built from multiple quark and gluon strings where the various colour ordered amplitudes are distinguished by the ordering of the gluons within a quark string. Sub-leading colour structures also arise at the amplitude level according to the distribution of fundamental indices among the endpoints.  For example, the four-quark two-gluon partial amplitudes have twelve colour structures.  Six of the structures are given according to the permutations of gluons within the structures,
\ba
(a_{1}a_{2})_{il}(\oslash)_{kj}\hspace{1cm}(a_{1})_{il}(a_{2})_{kj}\hspace{1cm}(\oslash)_{il}(a_{1}a_{2})_{kj},
\ea
where $i,k$ and $j,l$ denote the fundamental and anti-fundamental indices of the quarks and antiquarks respectively and $(\oslash)_{il}=\delta_{il}$ denotes an empty quark string whose colour factor is simply a Kronecker delta in fundamental indices.  The six sub-leading colour structures are obtained by the substitution $j\leftrightarrow l$.

As discussed above, in the case of a quark string, a collinear limit between the quarks causes the quark string to collapse onto a gluon string.  With multiple quark strings, a quark-antiquark collinear limit can cause quark strings to merge, with the composite gluon contributing to the gluon string by connecting the fundamental indices of the quark endpoints, e.g.,
\ba
(a_{i_{1}}\cdots a_{i_{n}})_{ij}(a_{j_{1}}\cdots a_{j_{m}})_{kl}&\stackrel{q_{k}||\b{q}_{j}}{\longrightarrow}&(a_{i_{1}}\cdots a_{i_{n}}a_{(kj)}a_{j_{1}}\cdots a_{j_{m}})_{il}.
\ea
The behaviour of the colour structures in various unresolved limits is mirrored by the resulting reduced partial amplitudes.  When multiple strings are present in a single colour ordered partial amplitude, the partons involved in each quark are separated from those involved in other strings by a semi-colon, e.g., 
\ba
M_{n}^{0}(\cdots;q_{1},g_{1},g_{2},\b{q}_{1};q_{3},g_{1},\b{q}_{2};\cdots).
\ea

\subsubsection{Sub-leading colour}

The sub-leading colour contribution cannot, in general, be written in the form of an incoherent sum of squared partial amplitudes. However for low multiplicity final-states the sub-leading colour contribution can often be rewritten as the square of a coherent sum of QCD partial amplitudes where one or more gluons behave in an Abelian fashion.  

Matrix elements containing ``Abelian gluons'' obey the usual QCD factorisation formulae in all the unresolved limits of the non-Abelian gluons.  The Abelian gluons do not couple to the non-Abelian gluons and only couple to quarks; therefore they can only be considered colour-connected to the quarks. In this way a sub-leading colour matrix element can be considered to contain two colour structures: a pure QCD colour structure formed from all quarks and non-Abelian gluons, the factorisation properties of which are identical to those of leading colour squared matrix element, and a QED-like colour structure containing quark pairs and colour disconnected Abelian gluons~\cite{our3j1}.  For example, a squared partial amplitude with two quarks, $n$ non-Abelian gluons and one Abelian gluon, $\tilde{i}$, is denoted $M_{n+3}^{0}(q,g_{1},\cdots,\tilde{g}_{i},\cdots,g_{n},\bar{q})$ and schematically has the colour structure,
\ba
(q,1,\cdots,\tilde{i},\cdots,n,\bar{q})_{ij}&\sim&(q,1,\cdots,n,\bar{q})_{ij}\otimes(q,\tilde{i},\bar{q})_{ij}.
\ea
The Abelian gluon can have collinear limits with the quarks and also become soft,
\ba
M_{n+3}^{0}(q,g_{1},\cdots,\tilde{g}_{i},\cdots,g_{n},\bar{q})&\stackrel{i||q}{\longrightarrow}&\frac{1}{s_{qi}}\ P_{qg\rightarrow Q}(z)\ M_{n+2}^{0}(Q,g_{1},\cdots,g_{n},\bar{q}),\\
M_{n+3}^{0}(q,g_{1},\cdots,\tilde{g}_{i},\cdots,g_{n},\bar{q})&\stackrel{i\rightarrow0}{\longrightarrow}&S_{qi\bar{q}}\ M_{n+2}^{0}(q,g_{1},\cdots,g_{n},\bar{q}),
\ea
where $P_{qg\rightarrow Q}(z)$ and $S_{qi\bar{q}}$ are the time-like quark-gluon splitting function and soft function respectively.  The collinear limit with the antiquark is given by exchanging $q\leftrightarrow\bar{q}$.  When the sub-leading colour matrix elements  can be re-written in terms of squared matrix elements involving Abelian gluons then the IR divergent limits can be subtracted using antenna functions just as at leading colour.  For six or more coloured particles, the sub-leading colour contribution cannot be written purely in terms of squared matrix elements with Abelian gluons. However, because the IR divergent limits of the remaining contribution are known, antenna subtraction can be readily applied. Notwithstanding these complications, the sub-leading colour contributions will in general contain fewer and less intricate divergent limits than their leading colour counterparts.  Any finite contribution to the cross section requires no subtraction and is integrated numerically.  The remainder of this section will focus on the leading colour contribution for clarity but it is clear that a similar treatment can be reproduced for sub-leading colour contributions.

\subsubsection{Subtraction terms}

The NLO subtraction term is based on the single unresolved limits of the relevant matrix elements.  In our case, the subtraction term reflects the antenna factorisation of the squared partial amplitude.  

The full subtraction term is a sum of contributions of the type,
\ba
\dsigma_{NLO}^{S}&=&{\cal{N}}_{NLO}^{R}\   \sum_{\text{perms}}\sum_{j}\ \text{d}\Phi_{n+1}(p_{3},\cdots,p_{n+3};p_{1},p_{2})\ \frac{1}{S_{n+1}}\nn\\
&\times&X_{3}^{0}(\cdot,j,\cdot)\ M_{n+2}^{0}(\cdots,j,\cdots)\ J_{n}^{(n)}(\{p\}_{n}).\label{eq:subterm}
\ea
where $X_3^0$ is the NLO antenna function, $M_{n+2}^{0}$ the reduced matrix element with a particular colour ordering and the sum runs over all the possible unresolved particles $j$.   There are three separate cases that correspond to the hard radiators being in the final state (FF), both in the initial state (II) or one in the initial-state and one in the final-state (IF).
For DIS processes $\dsigma_{NLO}^{S,II}=0$ while for $e^{+}e^{-}$ annihilation $\dsigma_{NLO}^{S,IF}=\dsigma_{NLO}^{S,II}=0$. At NLO it is possible to write the subtraction terms explicitly for all configurations at leading colour and for an arbitrary number of partons.

When the unresolved parton $j$ is colour connected to the final-state hard radiators $i$ and $k$, the subtraction term for the partial amplitude  $M_{n+3}^{0}(\cdots,i,j,k,\cdots)$ takes the form,
\ba
M_{n+3}^{0}(\cdots,i,j,k,\cdots) \longrightarrow X_{3}^{0}(i,j,k)\ M_{n+2}^{0}(\cdots,I,K,\cdots).\label{eq:ffsubterm}
\ea
The species of antenna function is determined by the identities of the partons contained within the antenna, as shown in Tab.~\ref{tab:ffsubterms}.  The sum over $j$ takes into account all the possible unresolved partons in the colour-ordered matrix element fitting the final-final configuration.  The final-final phase space map~\cite{Kosower:1997zr}, $(i,j,k)\to (I,K)$, ensures that the momenta involved in the antenna function are mapped onto two hard composite momenta.  The IR divergence associated with the configuration where $j$ becomes unresolved is described by the appropriate antenna function and, in the singular limit, the subtraction term tends to the value of the real emission cross section.  The various antennae and reduced matrix elements appropriate for a particular colour ordered matrix element in the real emission are listed in Tab.~\ref{tab:ffsubterms}. There are five possibilities reflecting the different particle assignments and possible colour structures.  
\begin{table}[h]
\centering
\begin{tabular}{|c|c|c|}
\hline
\multicolumn{3}{|l|}{{\bf Final-Final Unintegrated Antennae}}\\\hline
\multirow{2}{*}{Matrix element, $M_{n+3}^{0}$}&\multirow{2}{*}{Antenna, $X_{3}^{0}$}&\multicolumn{1}{|l|}{Reduced matrix}\\
& &\multicolumn{1}{|l|}{element, $M_{n+2}^{0}$}\\
\hline
$(\cdots;i_{q},j_{g},k_{\b{q}};\cdots)$&$A_{3}^{0}(i,j,k)$&$(\cdots;I_{q},K_{\b{q}};\cdots)$\\
\hline
$(\cdots;i_{q},j_{g},k_{g},\cdots)$&$d_{3}^{0}(i,j,k)$&$(\cdots;I_{q},K_{g},\cdots)$\\
\hline
$(\cdots;i_{q'},j_{\b{q}};k_{q},\cdots)$&$E_{3}^{0}(i,j,k)$&$(\cdots;I_{q'},K_{g},\cdots)$\\
\hline
$(\cdots,i_{g},j_{g},k_{g},\cdots)$&$f_{3}^{0}(i,j,k)$&$(\cdots,I_{g},K_{g},\cdots)$\\
\hline
$(\cdots,i_{g},j_{\b{q}};k_{q},\cdots)$&$G_{3}^{0}(i,j,k)$&$(\cdots,I_{g},K_{g},\cdots)$\\
\hline
\end{tabular}
\caption{The NLO antennae, $X_3^0$, and reduced matrix elements, $M_{n+2}^0$, appropriate for the various particle assignments and colour structures in the real radiation partial amplitudes, $M_{n+3}^0$ for the final-final configuration.}
\label{tab:ffsubterms}
\end{table}

In configurations with initial-state partons, an initial-final antenna is necessary and the initial-final phase space map~\cite{Daleo:2006xa} $(\hat{1},i,j)\to (\hb{1},J)$ is employed to generate the composite momenta for the reduced matrix element so that
\ba
M_{n+3}^{0}(\cdots,\hat{1}_a,i,j,\cdots) \longrightarrow X_{3,a\to b}^{0}(\hat{1},i,j)\ M_{n+2}^{0}(\cdots,\hb{1}_b,J,\cdots)\label{eq:ifsubterm}
\ea
where the initial-state parton labelled $\hat{1}_{a}$ where $a$ is the species of the initial-state parton, either $q$ or $g$ for a quark or gluon respectively.  The various antennae and reduced matrix elements appropriate for a particular colour ordered matrix element in the real emission are listed in Tab.~\ref{tab:ifsubterms}.  In hadron-hadron collisions, an analogous subtraction term exists for the second initial-state parton, i.e., $\hat{1}_{a}\to\hat{2}_{b}$ where $b$ is the species of the second initial-state parton.  For processes where $a = b$, we drop the subscript $a \to b$ for simplicity and only retain the label in the species changing cases.

\begin{table}[h]
\centering
\begin{tabular}{|c|c|c|}
\hline
\multicolumn{3}{|l|}{{\bf Initial-Final Unintegrated Antennae}}\\\hline
\multirow{2}{*}{Matrix element, $M_{n+3}^{0}$}&\multirow{2}{*}{Antenna, $X_{3}^{0}$}&\multicolumn{1}{|l|}{Reduced matrix}\\
& &\multicolumn{1}{|l|}{element, $M_{n+2}^{0}$}\\
\hline
$(\cdots;\hat{1}_{q},i_{g},j_{\b{q}};\cdots)$&$A_{3}^{0}(\hat{1},i,j)$&$(\cdots;\hb{1}_{q},J_{\b{q}};\cdots)$\\
\hline
$(\cdots;\hat{1}_{q},i_{g},j_{g},\cdots)$&$d_{3}^{0}(\hat{1},i,j)$&$(\cdots;\hb{1}_{q},J_{g},\cdots)$\\
\hline
$(\cdots;i_{q},j_{g},\hat{1}_{g},\cdots)$&$d_{3}^{0}(i,j,\hat{1})$&$(\cdots;J_{q},\hb{1}_{g},\cdots)$\\
\hline
$(\cdots;\hat{1}_{q'},i_{\b{q}};j_{q},\cdots)$&$E_{3}^{0}(\hat{1},i,j)$&$(\cdots;\hb{1}_{q'},J_{g},\cdots)$\\
\hline
$(\cdots,\hat{1}_{g},i_{g},j_{g},\cdots)$&$f_{3}^{0}(\hat{1},i,j)$&$(\cdots,\hb{1}_{g},J_{g},\cdots)$\\
\hline
$(\cdots,\hat{1}_{g},i_{\b{q}};j_{q},\cdots)$&$G_{3}^{0}(\hat{1},i,j)$&$(\cdots,\hb{1}_{g},J_{g},\cdots)$\\
\hline & &\\ [-1.2em]\hline
$(\cdots;i_{q},\hat{1}_{g},j_{\b{q}},\cdots)$&$-a_{3,g\to q}^{0}(i,\hat{1};j)$&$(\cdots;\hb{1}_{q},J_{g},\cdots)$\\
\hline
$(\cdots;i_{q},\hat{1}_{g},j_{g},\cdots)$&$-d_{3,g\to q}^{0}(i,\hat{1};j)$&$(\cdots;\hb{1}_{q},J_{g},\cdots)$\\
\hline
$(\cdots;i_{q'},\hat{1}_{q};j_{q},\cdots)$&$-E_{3,q\to g}^{0}(i,\hat{1},j)$&$(\cdots;J_{q'},\hb{1}_{g},\cdots)$\\
\hline
$(\cdots,i_{g},\hat{1}_{q};j_{q},\cdots)$&$-G_{3,q\to g}^{0}(i,\hat{1},j)$&$(\cdots,J_{g},\hb{1}_{g},\cdots)$\\
\hline
\end{tabular}
\caption{The NLO antennae, $X_3^0$, and reduced matrix elements, $M_{n+2}^0$, appropriate for the various particle assignments and colour structures in the real radiation partial amplitudes, $M_{n+3}^0$ for the initial-final configuration.  The identity changing antennae are collected at the bottom of this table.}
\label{tab:ifsubterms}
\end{table}

The last configuration that is allowed for hadron-hadron collisions is the initial-initial configuration where an unresolved parton is emitted between two initial-state partons of species $a$ and $b$.  The subtraction term uses an initial-initial antenna function and the appropriate initial-initial phase space map~\cite{Daleo:2006xa} $(\hat{1},i,\hat{2})\to (\hb{1},\hb{2})$,
\ba
M_{n+3}^{0}(\cdots,\hat{1}_a,i,\hat{2}_b,\cdots) \longrightarrow X_{3,a\to c, b\to d}^{0}(\hat{1},i,\hat{2})\ M_{n+2}^{0}(\cdots,\hb{1}_c,\hb{2}_d,\cdots).\label{eq:iisubterm}
\ea
The various antennae and reduced matrix elements appropriate for a particular colour ordered matrix element in the real emission are listed in Tab.~\ref{tab:iisubterms}. As in the initial-final case, the subscripts are only retained in the species changing cases, $ a \neq c$ or $b \neq d$.

\begin{table}[h]
\centering
\begin{tabular}{|c|c|c|}
\hline
\multicolumn{3}{|l|}{{\bf Initial-Initial Unintegrated Antennae}}\\\hline
\multirow{2}{*}{Matrix element, $M_{n+3}^{0}$}&\multirow{2}{*}{Antenna, $X_{3}^{0}$}&\multicolumn{1}{|l|}{Reduced matrix}\\
& &\multicolumn{1}{|l|}{element, $M_{n+2}^{0}$}\\
\hline
$(\cdots;\hat{1}_{q},i_{g},\hat{2}_{\b{q}},\cdots)$&$A_{3}^{0}(\hat{1},i,\hat{2})$&$(\cdots;\hb{1}_{q},\hb{2}_{\b{q}};\cdots)$\\
\hline
$(\cdots;\hat{1}_{q},i_{g},\hat{2}_{g},\cdots)$&$D_{3}^{0}(\hat{1},i,\hat{2})$&$(\cdots;\hb{1}_{q},\hb{2}_{g},\cdots)$\\
\hline
$(\cdots,\hat{1}_{g},i_{g},\hat{2}_{g},\cdots)$&$F_{3}^{0}(\hat{1},i,\hat{2})$&$(\cdots,\hb{1}_{g},\hb{2}_{g},\cdots)$\\
\hline & &\\ [-1.2em]\hline
$(\cdots;\hat{1}_{q},\hat{2}_{g},i_{\b{q}};\cdots)$&$-A_{3,g\to q}^{0}(\hat{1},\hat{2},i)$&$(\cdots;\hb{1}_{q},\hb{2}_{\b{q}};\cdots)$\\
\hline
$(\cdots;i_{q},\hat{1}_{g},\hat{2}_{g},\cdots)$&$-d_{3,g\to q}^{0}(i,\hat{1};\hat{2})$&$(\cdots;\hb{1}_{q},\hb{2}_{g},\cdots)$\\
\hline
$(\cdots;\hat{1}_{q'},\hat{2}_{\b{q}};i_{q},\cdots)$&$-E_{3,q\to g}^{0}(\hat{1},\hat{2},i)$&$(\cdots;\hb{1}_{q'},\hb{2}_{g},\cdots)$\\
\hline
$(\cdots,\hat{1}_{g},\hat{2}_{\b{q}};i_{q},\cdots)$&$-G_{3,q\to g}^{0}(\hat{1},\hat{2},i)$&$(\cdots,\hb{1}_{g},\hb{2}_{g},\cdots)$\\
\hline
\end{tabular}
\caption{The NLO antennae, $X_3^0$, and reduced matrix elements, $M_{n+2}^0$, appropriate for the various particle assignments and colour structures in the real radiation partial amplitudes, $M_{n+3}^0$ for the initial-initial configuration.}
\label{tab:iisubterms}
\end{table}


\subsection{NLO mass factorisation term}

The notation for the NLO mass factorisation contribution was presented at the beginning of this section in Eq. \eqref{eq:nlomf}. Writing the Born cross section out in terms of the matrix elements being integrated over the final-state phase space the mass factorisation contribution at NLO is given by,
\ba
\dsigma_{ij,NLO}^{MF}&=&-{\cal{N}}_{NLO}^{V}\ \sum_{\text{perms}}\ \text{d}\Phi_{n}(p_{3},\cdots,p_{n+1};x_{1}{p}_{1},x_{2}{p}_{2})\ \int\frac{\text{d}z_{1}}{z_{1}} \frac{\text{d}z_{2}}{z_{2}}\ \frac{1}{S_{n}}\nn\\
&\times& \Gamma_{ij;kl}^{(1)}(z_{1},z_{2})\ M_{n}^{0}(\cdots,\hat{k},\cdots,\hat{l},\cdots)\ J_{n}^{(n)}(p_{3},\cdots,p_{n+1}).
\label{eq:mfnlo}
\ea
In this formula there is an implicit sum over $k$ and $l$, denoting the particle types of the initial-state partons.  The explicit forms for the various mass factorisation kernels and their associated colour decompositions are listed in Appendix \ref{sec:splitcoltree}. The overall factor is given by,
\ba
{\cal{N}}_{NLO}^{V}\ = {\cal{N}}_{LO} \left(\frac{\alpha_s N}{2\pi}\right) \bar{C}(\eps) = {\cal{N}}_{NLO}^{R}\ C(\eps).
\ea

\subsection{Construction of the virtual subtraction term}

\begin{figure}[t]
\centering
\includegraphics[width=0.5cm]{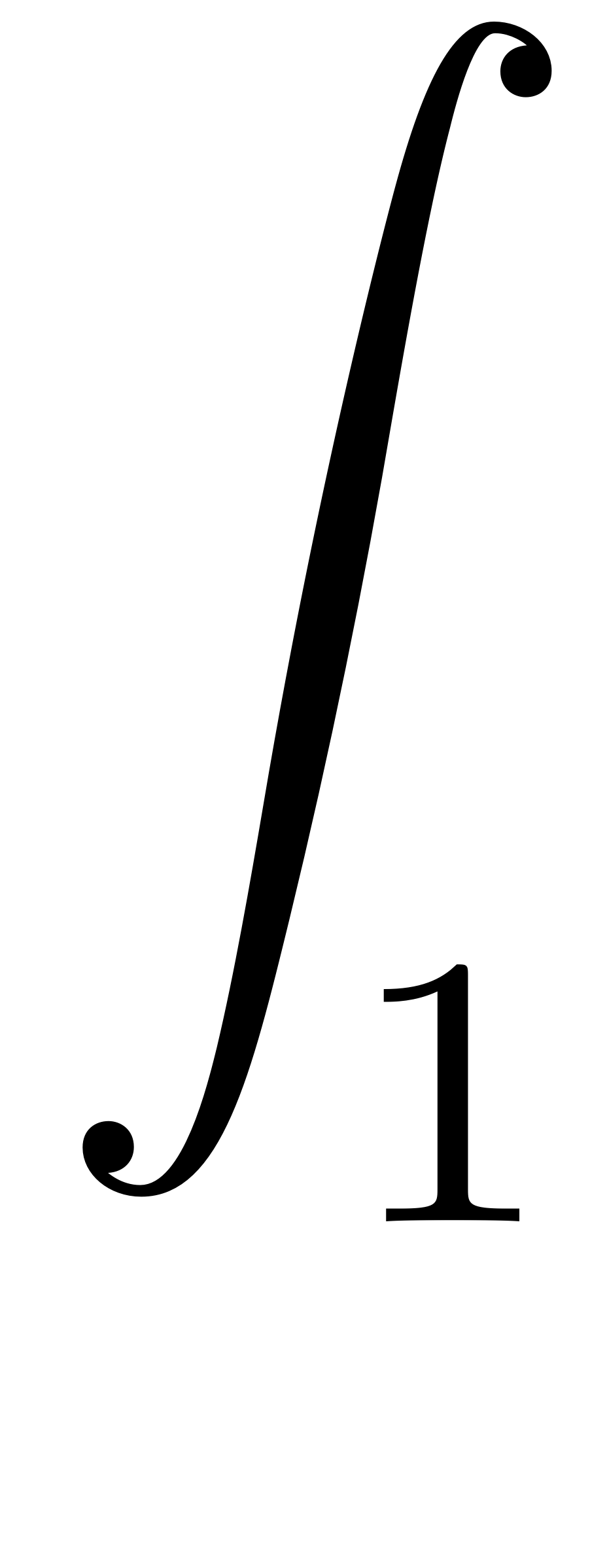}
\includegraphics[width=5.5cm]{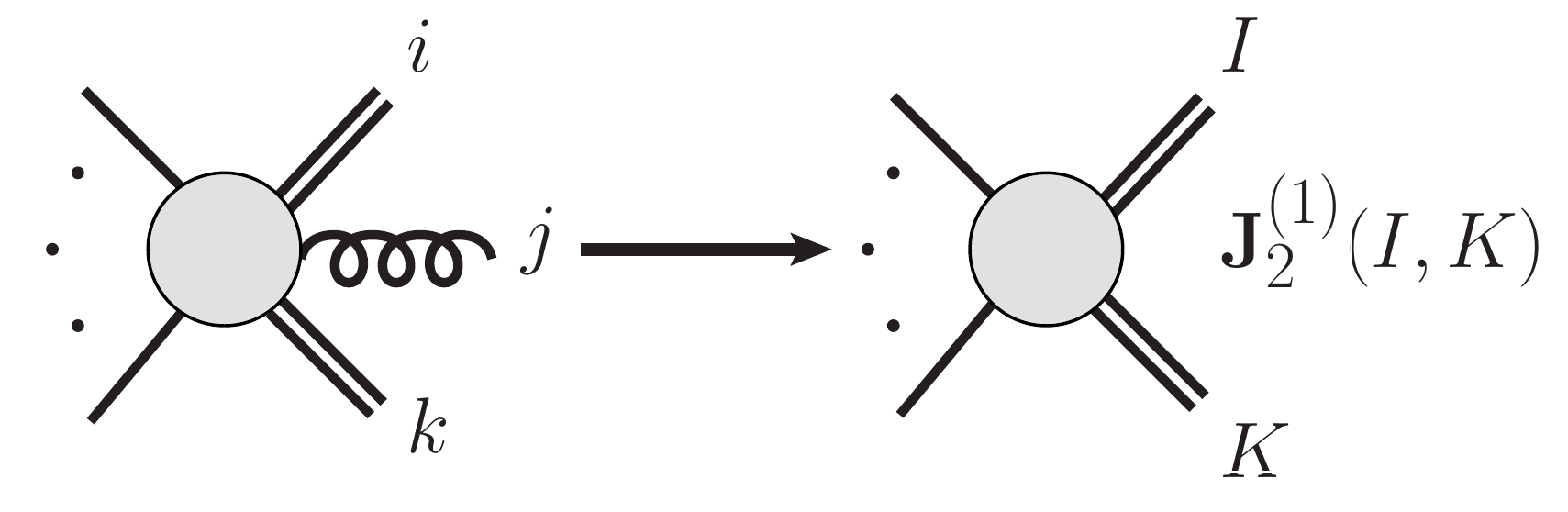}
\caption{NLO virtual structure.  Integration of the tree-level real emission over the single unresolved phase space for particle $j$ generates an integrated dipole $\bs{J}_{2}^{(1)}(I,K)$ for each dipole pair $I,K$, containing the explicit IR poles of the virtual contribution.}
\label{fig:j1pic}
\end{figure}

The virtual subtraction term is constructed from the integrated real subtraction term and the mass factorisation contribution.  For a single colour connected pair in the colour ordering of the virtual matrix element, a virtual subtraction term can be constructed of the type,
\ba
\dsigma_{NLO}^{T}&=&-{\cal{N}}_{NLO}^{V}\  \int\frac{\text{d}x_{1}}{x_{1}} \frac{\text{d}x_{2}}{x_{2}}\ \frac{1}{S_{n}}\ \sum_{\text{perms}}\ \text{d}\Phi_{n}(p_{3},\cdots,p_{n};x_{1}{p}_{1},x_{2}{p}_{2})\nn\\
&\times&\sum_{I,K}\,\bs{J}_{2}^{(1)}(I,K)\ M_{n+2}^{0}(\cdots,I,K,\cdots)\ J_{n}^{(n)}(\{p\}_{n}).\label{eq:virtsubterm}
\ea
As illustrated in Fig.~\ref{fig:j1pic}, the singular integrated dipole factor, $\bs{J}_{2}^{(1)}$ is directly related to the integration of the $X_{3}^{0}$ antenna function over the single unresolved phase space, and is defined to all orders in $\e$.  The precise form of $\bs{J}_{2}^{(1)}$ depends on the species of the resolved hard radiator partons in the real radiation matrix elements  that produced this integrated dipole contribution and the kinematic configuration, i.e., final-final, initial-final or initial-initial.  The full virtual subtraction term is formed by joining together the various integrated dipoles to form a single integrated antenna string for each reduced matrix element. 

In the final-final configuration, the generic virtual subtraction term for a pair of colour connected partons $I,K$ is produced by integrating the subtraction term associated with the real radiation matrix element $M_{n+3}^{0}(\cdots,i,j,k,\cdots)$ with the mapping $(i,j,k)\to (I,K)$ and has the form, 
\ba
M_{n+3}^{0}(\cdots,i,j,k,\cdots) \longrightarrow \bs{J}_{2}^{(1)}(I,K)\ M_{n+2}^{0}(\cdots,I,K,\cdots).\label{eq:virtff}
\ea
This equation is the analogue of Eq.~\eqref{eq:ffsubterm}.  The details of the integrated dipoles depends on the species of parton under consideration and all possibilities are listed in Tab.~\ref{tab:ffintsubterms}.
\begin{table}[h]
\centering
\begin{tabular}{|c|l|c|}
\hline
\multicolumn{3}{|l|}{{\bf Final-Final Integrated Antennae}}\\\hline
\multirow{2}{*}{Matrix element, $M_{n+3}^{0}$}&\multirow{2}{*}{Integrated dipole, $\bs{J}_{2}^{(1)}$}&\multicolumn{1}{|l|}{Reduced matrix}\\
& &\multicolumn{1}{|l|}{element, $M_{n+2}^{0}$}\\
\hline
$(\cdots;i_{q},j_{g},k_{\b{q}};\cdots)$&$\bs{J}_{2}^{(1)}(I_{q},K_{\b{q}})={\cal{A}}_{3}^{0}(s_{IK})$&$(\cdots;I_{q},K_{\b{q}};\cdots)$\\
\hline
$(\cdots;i_{q},j_{g},k_{g},\cdots)$&$\bs{J}_{2}^{(1)}(I_{q},K_{g})=\frac{1}{2}{\cal{D}}_{3}^{0}(s_{IK})$&$(\cdots;I_{q},K_{g},\cdots)$\\
\hline
$(\cdots;i_{q'},j_{\b{q}};k_{q},\cdots)$&$\bs{J}_{2,\NF}^{(1)}(I_{q},K_{g})=\frac{1}{2}{\cal{E}}_{3}^{0}(s_{IK})$&$(\cdots;I_{q},K_{g},\cdots)$\\
\hline
$(\cdots,i_{g},j_{g},k_{g},\cdots)$&$\bs{J}_{2}^{(1)}(I_{g},K_{g})=\frac{1}{3}{\cal{F}}_{3}^{0}(s_{IK})$&$(\cdots,I_{g},K_{g},\cdots)$\\
\hline
$(\cdots,i_{g},j_{\b{q}};k_{q},\cdots)$&$\bs{J}_{2,\NF}^{(1)}(I_{g},K_{g})={\cal{G}}_{3}^{0}(s_{IK})$&$(\cdots,I_{g},K_{g},\cdots)$\\
\hline
\end{tabular}
\caption{The correspondence between the real radiation matrix elements, $M_{n+3}^0$ and the 
integrated NLO dipoles $\bs{J}_{2}^{(1)}$ and reduced matrix elements, $M_{n+2}^0$ for various particle assignments and colour structures for the final-final configuration.}
\label{tab:ffintsubterms}
\end{table}

When one of the partons making up the hard dipole is in the initial-state then the analogous formula to Eq.~\eqref{eq:virtff} now contains the appropriate mass factorisation kernels from Eq.~\eqref{eq:mfnlo},
\ba
M_{n+3}^{0}(\cdots,\hat{1},i,j,\cdots) \longrightarrow \bs{J}_{2}^{(1)}(\hb{1},J)\ M_{n+2}^{0}(\cdots,\hb{1},J,\cdots),\label{eq:virtif}
\ea
where the only modification is the form of the integrated dipoles, listed in Tab.~\ref{tab:ifintsubterms}.  In this equation the subtraction term involving parton $\hat{1}$ is presented; in hadron-hadron collisions the initial-final virtual subtraction term involving the other initial-state parton is obtained by the substitution $\hat{1}\to\hat{2}$
\begin{table}[h]
\centering
\begin{tabular}{|c|l|c|}
\hline
\multicolumn{3}{|l|}{{\bf Initial-Final Integrated Antennae}}\\\hline
\multirow{2}{*}{Matrix element, $M_{n+3}^{0}$}&\multirow{2}{*}{Integrated dipole, $\bs{J}_{2}^{(1)}$}&\multicolumn{1}{|l|}{Reduced matrix}\\
& &\multicolumn{1}{|l|}{element, $M_{n+2}^{0}$}\\
\hline
$(\cdots;\hat{1}_{q},i_{g},j_{\b{q}};\cdots)$&$\bs{J}_{2}^{(1)}(\hb{1}_{q},J_{\b{q}})={\cal{A}}_{3,q}^{0}(s_{\b{1}J})-\Gamma_{qq}^{(1)}(x_{1})\delta_{2}$&$(\cdots;\hb{1}_{q},J_{\b{q}};\cdots)$\\
\hline
$(\cdots;\hat{1}_{q},i_{g},j_{g},\cdots)$&$\bs{J}_{2}^{(1)}(\hb{1}_{q},J_{g})=\frac{1}{2}{\cal{D}}_{3,q}^{0}(s_{\b{1}J})-\Gamma_{qq}^{(1)}(x_{1})\delta_{2}$&$(\cdots;\hb{1}_{q},J_{g},\cdots)$\\
\hline
$(\cdots;i_{q},j_{g},\hat{1}_{g},\cdots)$&$\bs{J}_{2}^{(1)}(J_{q},\hb{1}_{g})={\cal{D}}_{3,g,gq}^{0}(s_{\b{1}J})-\frac{1}{2}\Gamma_{gg}^{(1)}(x_{1})\delta_{2}$&$(\cdots;J_{q},\hb{1}_{g},\cdots)$\\
\hline
$(\cdots;\hat{1}_{q'},i_{\b{q}};j_{q},\cdots)$&$\bs{J}_{2,\NF}^{(1)}(\hb{1}_{q},J_{g})=\frac{1}{2}{\cal{E}}_{3,q',q\b{q}}^{0}(s_{\b{1}J})$&$(\cdots;\hb{1}_{q},J_{g},\cdots)$\\
\hline
$(\cdots,\hat{1}_{g},i_{g},j_{g},\cdots)$&$\bs{J}_{2}^{(1)}(\hb{1}_{g},J_{g})=\frac{1}{2}{\cal{F}}_{3,g}^{0}(s_{\b{1}J})-\frac{1}{2}\Gamma_{gg}^{(1)}(x_{1})\delta_{2}$&$(\cdots,\hb{1}_{g},J_{g},\cdots)$\\
\hline
$(\cdots,\hat{1}_{g},i_{\b{q}};j_{q},\cdots)$&$\bs{J}_{2,\NF}^{(1)}(\hb{1}_{g},J_{g})={\cal{G}}_{3,g}^{0}(s_{\b{1}J})$&$(\cdots,\hb{1}_{g},J_{g},\cdots)$\\
\hline & &\\[-1.2em]\hline
$(\cdots;i_{q},\hat{1}_{g},j_{\b{q}};\cdots)$&$\bs{J}_{2,g\to q}^{(1)}(\hb{1}_{q},J_{\b{q}})=-\frac{1}{2}{\cal{A}}_{3,g,q\b{q}}^{0}(s_{\b{1}J})-\Gamma_{qg}^{(1)}(x_{1})\delta_{2}$&$(\cdots;\hb{1}_{q},J_{\b{q}};\cdots)$\\
\hline
$(\cdots;i_{q},\hat{1}_{g},j_{g},\cdots)$&$\bs{J}_{2,g\to q}^{(1)}(\hb{1}_{q},J_{g})=-{\cal{D}}_{3,g,qg}^{0}(s_{\b{1}J})-\Gamma_{qg}^{(1)}(x_{1})\delta_{2}$&$(\cdots;\hb{1}_{q},J_{g},\cdots)$\\
\hline
$(\cdots;i_{q'},\hat{1}_{q};j_{q},\cdots)$&$\bs{J}_{2,q\to g}^{(1)}(J_{q},\hb{1}_{g})=-{\cal{E}}_{3,q,qq'}^{0}(s_{\b{1}J})-\Gamma_{gq}^{(1)}(x_{1})\delta_{2}$&$(\cdots;J_{q},\hb{1}_{g},\cdots)$\\
\hline
$(\cdots,i_{g};\hat{1}_{q},j_{q},\cdots)$&$\bs{J}_{2,q\to g}^{(1)}(J_{g},\hb{1}_{g})=-{\cal{G}}_{3,q}^{0}(s_{\b{1}J})-\Gamma_{gq}^{(1)}(x_{1})\delta_{2}$&$(\cdots,J_{g},\hb{1}_{g},\cdots)$\\
\hline
\end{tabular}
\caption{The correspondence between the real radiation matrix elements, $M_{n+3}^0$ and the 
integrated NLO dipoles $\bs{J}_{2}^{(1)}$ and reduced matrix elements, $M_{n+2}^0$ for various particle assignments and colour structures for the initial-final configuration.  For brevity $\delta(1-x_{i})=\delta_{i}$ for $i=1,2$.}
\label{tab:ifintsubterms}
\end{table}

The final kinematic configuration is when both hard radiators in the dipole are in the initial-state.  In this case the relevant subtraction term is,
\ba
M_{n+3}^{0}(\cdots,\hat{1},i,\hat{2},\cdots) \longrightarrow \bs{J}_{2}^{(1)}(\hb{1},\hb{2})\ M_{n+2}^{0}(\cdots,\hb{1},\hb{2},\cdots),\label{eq:virtii}
\ea
where once again, only the form of the integrated dipole is different from Eqs.~\eqref{eq:virtff} and~\eqref{eq:virtif} and the initial-initial integrated dipoles are listed in Tab.~\ref{tab:iiintsubterms}.
\begin{table}[h]
\centering
\begin{tabular}{|l|l|c|}
\hline
\multicolumn{3}{|l|}{{\bf Initial-Initial Integrated Antennae}}\\\hline
Matrix element,&\multirow{2}{*}{Integrated dipole, $\bs{J}_{2}^{(1)}$}&\multicolumn{1}{|l|}{Reduced matrix}\\
$M_{n+3}^{0}$& &\multicolumn{1}{|l|}{element, $M_{n+2}^{0}$}\\
\hline
$(\cdots;\hat{1}_{q},i_{g},\hat{2}_{\b{q}},\cdots)$&$\bs{J}_{2}^{(1)}(\hb{1}_{q},\hb{2}_{\b{q}})={\cal{A}}_{3,q\b{q}}^{0}(s_{\b{1}\b{2}})-\Gamma_{qq}^{(1)}(x_{1})\delta_{2}-\Gamma_{qq}^{(1)}(x_{2})\delta_{1}$&$(\cdots;\hb{1}_{q},\hb{2}_{\b{q}};\cdots)$\\
\hline
$(\cdots;\hat{1}_{q},i_{g},\hat{2}_{g},\cdots)$&$\bs{J}_{2}^{(1)}(\hb{1}_{q},\hb{2}_{g})={\cal{D}}_{3,qg}^{0}(s_{\b{1}\b{2}})-\Gamma_{qq}^{(1)}(x_{1})\delta_{2}-\frac{1}{2}\Gamma_{gg}^{(1)}(x_{2})\delta_{1}$&$(\cdots;\hb{1}_{q},\hb{2}_{g},\cdots)$\\
\hline
$(\cdots,\hat{1}_{g},i_{g},\hat{2}_{g},\cdots)$&$\bs{J}_{2}^{(1)}(\hb{1}_{g},\hb{2}_{g})={\cal{F}}_{3,gg}^{0}(s_{\b{1}\b{2}})-\frac{1}{2}\Gamma_{gg}^{(1)}(x_{1})\delta_{2}-\frac{1}{2}\Gamma_{gg}^{(1)}(x_{2})\delta_{1}$&$(\cdots,\hb{1}_{g},\hb{2}_{g},\cdots)$\\
\hline & &\\[-1.2em]\hline
$(\cdots;\hat{1}_{q},\hat{2}_{g},i_{\b{q}};\cdots)$&$\bs{J}_{2,g\to q}^{(1)}(\hb{1}_{q},\hb{2}_{\b{q}})=-{\cal{A}}_{3,qg}^{0}(s_{\b{1}\b{2}})-\Gamma_{qg}^{(1)}(x_{2})\delta_{1}$&$(\cdots;\hb{1}_{q},\hb{2}_{\b{q}};\cdots)$\\
\hline
$(\cdots;i_{q},\hat{1}_{g},\hat{2}_{g},\cdots)$&$\bs{J}_{2,g\to q}^{(1)}(\hb{1}_{q},\hb{2}_{g})=-{\cal{D}}_{3,gg}^{0}(s_{\b{1}\b{2}})-\Gamma_{qg}^{(1)}(x_{1})\delta_{2}$&$(\cdots;\hb{1}_{q},\hb{2}_{g},\cdots)$\\
\hline
$(\cdots;\hat{1}_{q'},\hat{2}_{\b{q}};i_{q},\cdots)$&$\bs{J}_{2,q\to g}^{(1)}(\hb{1}_{q},\hb{2}_{g})=-{\cal{E}}_{3,q'q,q}^{0}(s_{\b{1}\b{2}})-\Gamma_{gq}^{(1)}(x_{2})\delta_{1}$&$(\cdots;\hb{1}_{q'},\hb{2}_{g},\cdots)$\\
\hline
$(\cdots,\hat{1}_{g},\hat{2}_{\b{q}};i_{q},\cdots)$&$\bs{J}_{2,q\to g}^{(1)}(\hb{1}_{g},\hb{2}_{g})=-{\cal{G}}_{3,gq}^{0}(s_{\b{1}\b{2}})-\Gamma_{gq}^{(1)}(x_{2})\delta_{1}$&$(\cdots,\hb{1}_{g},\hb{2}_{g},\cdots)$\\
\hline
\end{tabular}
\caption{The correspondence between the real radiation matrix elements, $M_{n+3}^0$ and the 
integrated NLO dipoles $\bs{J}_{2}^{(1)}$ and reduced matrix elements, $M_{n+2}^0$ for various particle assignments and colour structures for the initial-initial configuration.  For brevity $\delta(1-x_{1})=\delta_{1}$, $\delta(1-x_{2})=\delta_{2}$.}
\label{tab:iiintsubterms}
\end{table}

\subsection{Summary}

The relationship betweem the real radiation subtraction term $\dsigma^S$ and the virtual subtraction term $\dsigma^T$ is shown in Fig.~2.
The full virtual subtraction term is generated by summing over the colour connected pairs in the virtual matrix element's colour ordering.  For example, consider an $(n+3)$-parton squared partial amplitude, containing a quark-antiquark pair and $(n+1)$ gluons, that maps onto the $(n+2)$-parton squared partial amplitude $M_{n+2}^{0}(\hat{\bar{1}}_{q},i_{g},\cdots,j_{g},\hat{\bar{2}}_{\bar{q}})$.  Once the leading colour single unresolved subtraction term for this matrix element has been integrated and combined with the mass factorisation terms, the resulting virtual subtraction term is simply given by,
\ba
\dsigma_{NLO}^{T}&=&-{\cal{N}}_{NLO}^{V}\   \int\frac{\text{d}x_{1}}{x_{1}} \frac{\text{d}x_{2}}{x_{2}}\ \frac{1}{S_{n}}\ \sum_{\text{perms}}\ \text{d}\Phi_{n}(p_{3},\cdots,p_{n};x_{1}{p}_{1},x_{2}{p}_{2})\nn\\
&\times&\bs{J}_{n+2}^{(1)}(\hat{\bar{1}}_{q},i_{g},\cdots,j_{g},\hat{\bar{2}}_{\bar{q}})\ M_{n+2}^{0}(\hat{\bar{1}}_{q},i_{g},\cdots,j_{g},\hat{\bar{2}}_{\bar{q}})\ J_{n}^{(n)}({p}_{3},\cdots,{p}_{n}),
\label{eq:dstnlo}
\ea
where the full integrated antenna string is formed from the sum of integrated dipoles,
\ba
\bs{J}_{n+2}^{(1)}(\hat{\bar{1}}_{q},i_{g},j_{g},\cdots,k_{g},l_{g},\hat{\bar{2}}_{\bar{q}})&=&\bs{J}_{2}^{(1)}(\hb{1}_{q},i_{g})+\bs{J}_{2}^{(1)}(i_{g},j_{g})+\cdots\nn\\
&+&\bs{J}_{2}^{(1)}(k_{g},l_{g})+\bs{J}_{2}^{(1)}(\hb{2}_{q},l_{g}).
\ea
Each term in $\bs{J}_{n+2}^{(1)}$ is proportional to a particular kinematic factor, $(|s_{ij}|)^{-\eps}$.  We therefore see that the singularities present in each term correspond to a particular singular contribution $\bs{I}_{ij}^{(1)}$.  The full singularity structure is simply obtained by summing the integrated dipoles to form $\bs{J}_{n+2}^{(1)}$, or equivalently by summing $\bs{I}_{ij}^{(1)}$.  Using Catani's one-loop factorisation formula~\cite{Catani:1998bh},
\ba
\Poles\big(M_{n+2}^{1}(1,\cdots,n+2)\big)&=&2\bs{I}_{n+2}^{(1)}(\eps;1,\cdots,n+2)M_{n+2}^{0}(1,\cdots,n+2),
\ea
where $\bs{I}_{n+2}^{(1)}(\eps;1,\cdots,n+2)$ is the $n$-parton IR singularity operator, which can be written as a sum of dipole contributions,
\ba
\bs{I}_{n+2}^{(1)}(\eps;1,\cdots,n)&=&\sum_{(i,j)}\bs{I}_{ij}^{(1)}(\eps;s_{ij}),\label{eq:catani1}
\ea
and the sum runs over colour connected pairs of partons. It is clear that the virtual subtraction term, formulated in terms of integrated antenna strings, reproduces the explicit poles of the virtual cross section to complete the IR finite calculation at NLO.  It is noted that now the correspondence between the poles of $\bs{J}_{n+2}^{(1)}$ and the one-loop matrix elements has been established, the explicit pole cancellation can be carried out without recourse to $\bs{I}_{n+2}^{(1)}$.  This correspondence is particularly clear because of the way the integrated antenna string has been constructed as a sum of dipole-like antenna contributions and by recognising that the mass factorisation contributions naturally fit into the structure of the integrated antenna strings so that only the genuine final-state IR poles remain in the virtual subtraction term.  This puts the virtual subtraction term in a particularly convenient form for carrying out the explicit pole cancellation against the virtual contribution.
\begin{figure}[t]
\centering
\includegraphics[width=10cm]{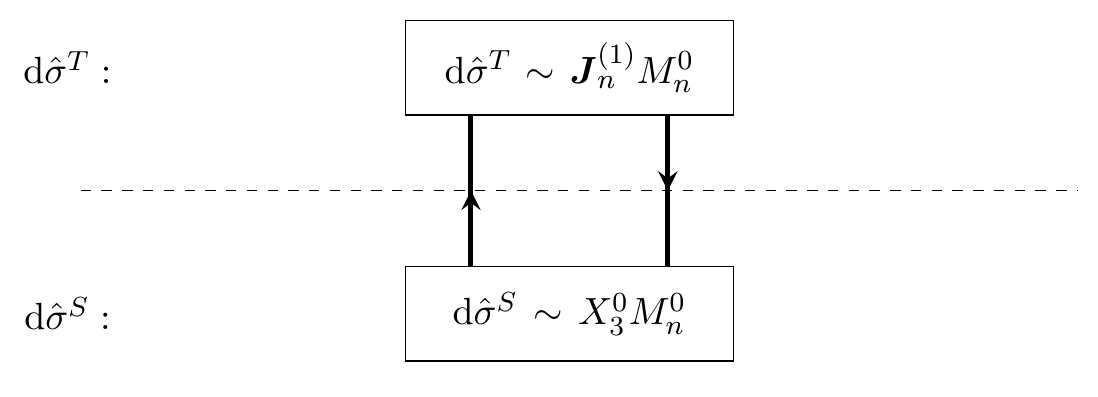}
\caption{The NLO correction to $n$-parton scattering requires the construction of the subtraction term $\dsigma_{NLO}^{S}$ to remove all IR divergent behaviour of the $(n+1)$-parton real emission.  Integrating this subtraction term and combining with the NLO mass factorisation terms generates the virtual subtraction term, $\dsigma_{NLO}^{T}$.  Antenna subtraction allows $\dsigma_{NLO}^{T}$ to be generated by first constructing $\dsigma_{NLO}^{S}$ or vice versa.}
\label{fig:nloflow}
\end{figure}


\section{Antenna subtraction at NNLO}
\label{sec:nnloant}

The NNLO cross section involves two additional particles in the scattering process, appearing as either real or virtual particles.   These additional particles lead to three distinct contributions: double real $\dsigma_{ij,NNLO}^{RR}$, real-virtual $\dsigma_{ij,NNLO}^{RV}$ or double virtual $\dsigma_{ij,NNLO}^{VV}$, containing $(n+4)$-, $(n+3)$-, $(n+2)$-partons and zero-, one-, two-loops respectively, where $n$ is the number of jets in the final-state,
\begin{eqnarray}
\dsigma_{ij,NNLO}&=&\int_{n+2}\dsigma_{ij,NNLO}^{RR}+\int_{n+1}\left(\dsigma_{ij,NNLO}^{RV}+\dsigma_{ij,NNLO}^{MF,1}\right)\nn\\
&+&\int_{n}\left(\dsigma_{ij,NNLO}^{VV}+\dsigma_{ij,NNLO}^{MF,2}\right),
\end{eqnarray}
where $\int_{{\rm{d}}\Phi_{n}}$ is defined in \eqref{eq:dphi}.  The NNLO mass factorisation terms are naturally partitioned into terms containing $(n+1)$- or $n$-parton final-states, which are given by,
\ba
&&\dsigma_{ij,NNLO}^{MF,1}(\xi_{1}H_{1},\xi_{2}H_{2})=-
\int\frac{{\rm{d}}z_{1}}{z_{1}}\frac{{\rm{d}}z_{2}}{z_{2}}\nn\\
&&\ \times\left(\frac{\alpha_sN}{2\pi}\right)\, \bar{C}(\eps) \,
\Gamma_{ij;kl}^{(1)}(z_{1},z_{2})\biggl(\dsigma_{kl,NLO}^{R}-\dsigma_{kl,NLO}^{S}\biggr)(z_{1}\xi_{1}H_{1},z_{2}\xi_{2}H_{2}),\label{eq:RVMF}\\
&&\dsigma_{ij,NNLO}^{MF,2}(\xi_{1}H_{1},\xi_{2}H_{2})=-\int\frac{{\rm{d}}z_{1}}{z_{1}}\frac{{\rm{d}}z_{2}}{z_{2}} \biggl\{\nonumber \\
&&\phantom{+}\left(\frac{\alpha_sN}{2\pi}\right)^2\ \bar{C}(\eps)^2\, \Gamma_{ij;kl}^{(2)}(z_{1},z_{2})\dsigma_{kl,LO}(z_{1}\xi_{1}H_{1},z_{2}\xi_{2}H_{2})\nn\\
&&+\left(\frac{\alpha_sN}{2\pi}\right)\, \bar{C}(\eps) \,
\Gamma_{ij;kl}^{(1)}(z_{1},z_{2})\biggl(\dsigma_{kl,NLO}^{V}-\dsigma_{kl,NLO}^{T}\biggr)(z_{1}\xi_{1}H_{1},z_{2}\xi_{2}H_{2})\biggr\},\label{eq:VVMF}
\ea
where the new ingredient at NNLO, $\Gamma_{ij;kl}^{(2)}$, can be written in terms of one- and two-loop Altarelli-Parisi kernels,
\ba
\Gamma_{ij;kl}^{(2)}(z_{1},z_{2})&=&
\delta(1-z_{2})\delta_{lj}\Gamma_{ki}^{(2)}(z_{1})+\delta(1-z_{1})\delta_{ki}\Gamma_{lj}^{(2)}(z_{2})+\Gamma_{ki}^{(1)}(z_{1})\Gamma_{lj}^{(1)}(z_{2}).
\ea

Using the antenna subtraction method, subtraction terms can be constructed from antenna functions (now including four-parton antennae to describe double unresolved radiation and one-loop three-parton antennae to describe the single unresolved radiation of one-loop matrix elements) to remove all implicit divergence of the double real and real-virtual contributions.  Introducing a double real subtraction term $\dsigma_{ij,NNLO}^{S}$, a real-virtual subtraction term $\dsigma_{ij,NNLO}^{VS}$, their corresponding integrated forms and the NNLO mass factorisation contributions, the NNLO cross section can be reorganised such that each term in square brackets is rendered free from implicit divergence and explicit poles and thus suitable for a numerical implementation.
\begin{eqnarray}
\dsigma_{ij,NNLO}&=&\int_{n+2}\left[\dsigma_{ij,NNLO}^{RR}-\dsigma_{ij,NNLO}^S\right]
\nonumber \\
&+& \int_{n+1}
\left[
\dsigma_{ij,NNLO}^{RV}-\dsigma_{ij,NNLO}^{T}
\right] \nonumber \\
&+&\int_{n\phantom{+1}}\left[
\dsigma_{ij,NNLO}^{VV}-\dsigma_{ij,NNLO}^{U}\right],
\label{eq:stu}
\end{eqnarray}
where the real-virtual and double virtual subtraction terms are given by,
\begin{eqnarray}
\label{eq:Tdef1}
\dsigma_{ij,NNLO}^{T} &=& \phantom{ -\int_1 }\dsigma_{ij,NNLO}^{VS}
- \int_1 \dsigma_{ij,NNLO}^{S,1} - \dsigma_{ij,NNLO}^{MF,1},  \\
\label{eq:Udef}
\dsigma_{ij,NNLO}^{U} &=& -\int_1 \dsigma_{ij,NNLO}^{VS}
-\int_2 \dsigma_{ij,NNLO}^{S,2}
-\dsigma_{ij,NNLO}^{MF,2}.
\end{eqnarray}
Note that because integration of the subtraction term $\dsigma_{NNLO}^S$ gives contributions to both the $(n+1)$- and $n$-parton final states, we have explicitly decomposed the integrated double real subtraction term into a piece that is integrated over the single unresolved particle phase space and a piece that is integrated over the phase space of two unresolved particles,
\begin{equation}
\int_{n+2}\dsigma_{ij,NNLO}^S
= \int_{n+1} \int_1 \dsigma_{ij,NNLO}^{S,1}
+\int_{n} \int_2 \dsigma_{ij,NNLO}^{S,2}.
\end{equation}

The construction of subtraction terms at the double real, real-virtual and double virtual levels requires an intimate understanding of the explicit IR singularity structure and implicit IR divergent behaviour of both the physical matrix elements and the subtraction terms for different final-state multiplicities.  The structures of these subtraction terms are dictated by two main considerations: the colour connection of partons and explicit IR pole cancellation.  The former completely informs the construction of the double real subtraction term as there are no explicit IR poles to consider; the latter is the sole concern of the double virtual subtraction term which is free from implicit IR divergences.  The real-virtual subtraction term contains both explicit poles and implicit divergent behaviour and so its structure must take into account both parton colour connection and explicit pole cancellation.

\begin{figure}[h!]
\centering
 \begin{sideways}
 \begin{minipage}{21cm}
 \includegraphics[width=22cm]{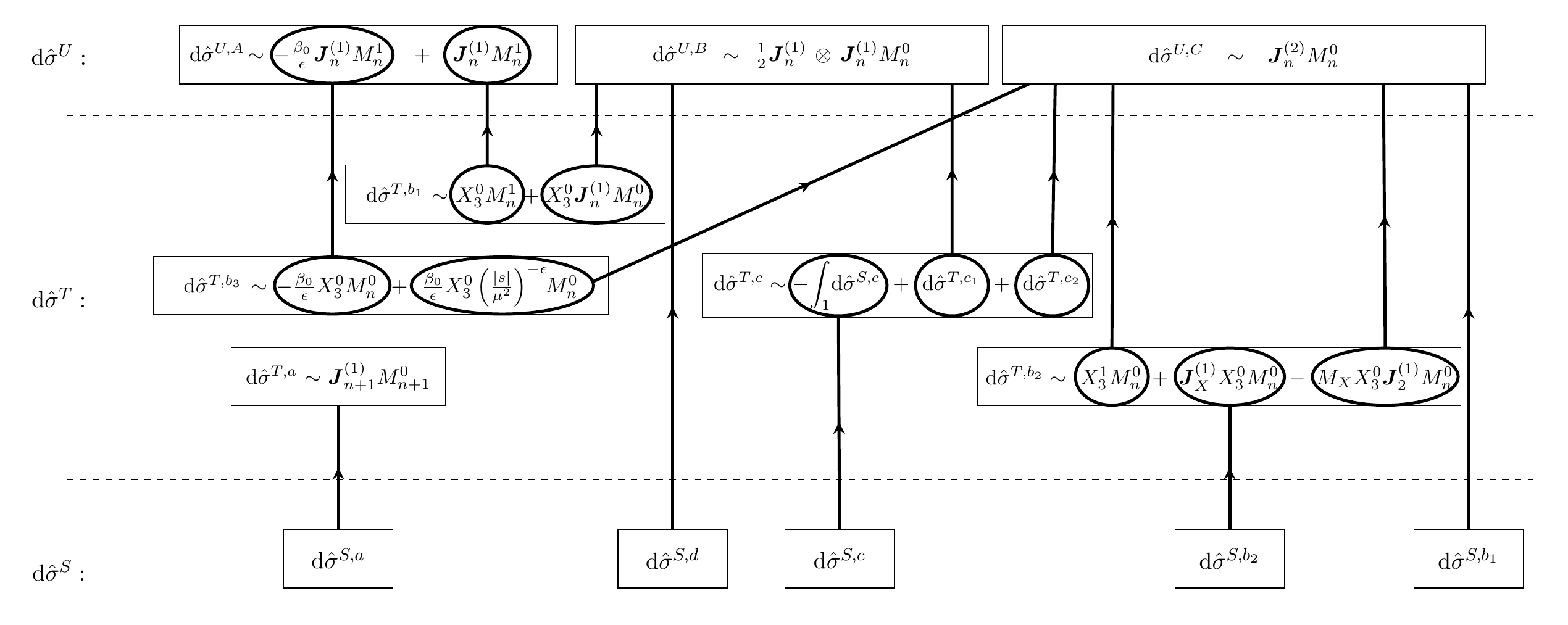}
  \caption{The NNLO correction to $n$-parton scattering contains three contributions: double real, real-virtual and double virtual, with the associated subtraction terms denoted $\dsigma_{NNLO}^{S}$, $\dsigma_{NNLO}^{T}$ and $\dsigma_{NNLO}^{U}$ discussed in Secs.~\ref{sec:RR},~\ref{sec:rv} and~\ref{sec:doublevirt} respectively. Each distinct contribution to the subtraction terms is shown in a rectangular box.  Individual terms within a box flow between boxes after integration, indicated by the arrows.}
 \label{fig:roadmap}
 \end{minipage}
 \end{sideways}
 \end{figure}

As we will discuss in Secs.~\ref{sec:RR},~\ref{sec:rv} and~\ref{sec:doublevirt} below, the $\dsigma_{NNLO}^{S}$, $\dsigma_{NNLO}^{T}$ and $\dsigma_{NNLO}^{U}$ subtraction terms each contain several components that each fulfill a particular function in constructing the correct infrared structure for NNLO computations.  It is the nature of any subtraction scheme, that whatever is subtracted from a final state with a particular multiplicity, must be added back in either integrated or unintegrated form to a final state with a different multiplicity.   To illustrate the connections between the various terms, Fig.~\ref{fig:roadmap} shows how the subtraction terms $\dsigma_{NNLO}^{S}$, $\dsigma_{NNLO}^{T}$ and $\dsigma_{NNLO}^{U}$ are linked.
The ``roadmap" shows how the various terms are related.  The five terms introduced in the double real subtraction term are directly linked with specific terms in the real-virtual and double virtual subtraction terms. Similarly, terms in the real-virtual subtraction term are composed partly of integrated terms from the double real subtraction term $\int_1 \dsigma_{ij,NNLO}^{S}$ and 
new virtual subtraction terms, $\dsigma_{ij,NNLO}^{VS}$.  The remaining terms, $\int_2 \dsigma_{ij,NNLO}^{S}$ and $\int_1 \dsigma_{ij,NNLO}^{VS}$, combine to produce the structures present in $\dsigma_{ij,NNLO}^{U}$.  As the discussion unfolds, we anticipate that it will help the reader to refer back to Fig.~\ref{fig:roadmap}.

\subsection{Construction of the double real subtraction term}

\label{sec:RR}

At NNLO the double real radiative $(n+4)$-parton correction to the $pp\rightarrow n$ jets process, must be included integrated over the full  $(n+2)$-parton phase space, subject to the constraint that $n$ jets are observed. In order for this to be achieved, the IR divergent behaviour of the partonic cross section must be isolated.  In the antenna subtraction formalism this is achieved by constructing a subtraction term from three- and four-parton antenna functions and reduced multiplicity matrix elements.

By construction the double real subtraction term mimics the IR divergence and factorisation of the real radiation matrix elements in all relevant single and double unresolved limits.  At NNLO there are four colour configurations to consider when constructing the subtraction term for $n$-jet production~\cite{GehrmannDeRidder:2005cm,our3j1,weinzierl3j2,Glover:2010im}:
\begin{enumerate}
\item $\dsigma^{S,a}$\\A single unresolved parton but the remaining $(n+1)$ final-state partons form $n$ jets.
\item $\dsigma^{S,b}$\\Two colour-connected unresolved partons, i.e., two unresolved partons radiated between a single pair of hard radiators.
\item $\dsigma^{S,c}$\\Two almost colour-connected unresolved partons, i.e., two colour disconnected unresolved partons sharing a common radiator, and including large angle soft radiation.
\item $\dsigma^{S,d}$\\Two colour disconnected unresolved partons, i.e., two colour disconnected unresolved partons with no radiators in common.
\end{enumerate}
Here we have suppressed the dependence of the subtraction terms on the initial-state partons $i$ and $j$.  The four possible colour configurations for the single and double unresolved limits of the $(n+4)$-parton matrix elements provide a natural way to divide the subtraction term into pieces, as has been emphasised in previous works~\cite{GehrmannDeRidder:2005cm,our3j1}.  It should be noted that although the factorisation of the matrix elements is strictly classified according to the colour connection of unresolved partons, the associated subtraction terms readily communicate with one another in most unresolved limits due to the existence and cross cancellation of spurious singularities.  These systematic cross cancellations suggest that structures exist in the subtraction term other than those dictated by the colour connection of unresolved partons.

In the following sections we make a further reorganisation of the double real subtraction term into five contributions,
\ba
\dsigma_{NNLO}^{S}&=&\dsigma_{NNLO}^{S,a}+\dsigma_{NNLO}^{S,b_{1}}+\dsigma_{NNLO}^{S,b_{2}}+\dsigma_{NNLO}^{S,c}+\dsigma_{NNLO}^{S,d}.
\ea
This manifestation of the subtraction term is broadly organised along the lines of colour connection but is also informed by considering how the double real subtraction terms appear in real-virtual and double virtual subtraction terms after integration.  Such an organisation of the double real subtraction term allows a transparent understanding of how various terms from the double real level cascade down the calculation upon integration.  Understanding this structure also permits a systematic construction of the double real subtraction term in terms of predictable blocks of terms.

\subsubsection{Single unresolved subtraction term, $\bs{\dsigma_{NNLO}^{S,a}}$}

The removal of single unresolved limits from the $(n+4)$-parton contribution to the $n$-jet cross section at NNLO strongly resembles the subtraction term constructed to isolate single unresolved limits from the $(n+4)$-parton contribution to the $(n+1)$-jet cross section at NLO.  The two subtraction terms differ only in the number of jets allowed by the jet function,
\ba
\dsigma_{NNLO}^{S,a}(\{p\}_{n+4})=\dsigma_{NLO}^{S}(\{p\}_{n+4})\biggr|_{J_{n+1}^{(n+1)}\rightarrow J_{n}^{(n+1)}}.
\ea
The detailed discussion about constructing the NLO subtraction terms given in Sec.~\ref{sec:nloant} is then sufficient to construct the single unresolved NNLO subtraction term $\dsigma_{NNLO}^{S,a}$.  Although this part of the subtraction term removes single unresolved divergences of the physical matrix elements for $n$-jet selecting observables, it also generates additional spurious singularities in the almost colour connected and colour disconnected limits, over-subtracting the divergence in each case.  This is because the $(n+3)$-parton reduced matrix elements allow an additional parton to become unresolved and yet still form $n$-jets.  This contribution is reintroduced at the real-virtual level upon analytic integration where they cancel the explicit poles in the real-virtual contribution. They form a distinct set of terms proportional to the $(n+3)$-parton matrix elements.

\subsubsection{Four-parton antenna subtraction term, $\bs{\dsigma_{NNLO}^{S,b_{1}}}$}
\label{sec:dssb1}

At NNLO there are essential new ingredients in the form of four-parton tree level antenna functions that are required to faithfully reproduce the colour connected double unresolved divergences of the physical matrix elements.  The momentum map associated with a four-parton antenna function is the $(n+4)\rightarrow(n+2)$ map that maps the four antenna momenta $i,j,k,l$ down to two composite momenta $I,L$, which is different for the final-final~\cite{Kosower:1997zr}, initial-final and initial-initial~\cite{Daleo:2006xa} configurations. This group of terms has the form,
\ba
\dsigma_{NNLO}^{S,b_{1}}&=&\mptwops\ \nn\\
&\times&\sum_{j,k}\ \ X_{4}^{0}(i,j,k,l)\ M_{n+2}^{0}(\cdots,I,L,\cdots)\ J_{n}^{(n)}(\{p\}_{n}),\label{eq:dssb1}
\ea
where ${\cal{N}}_{NNLO}^{RR}$ contains the overall QCD coupling, colour and non-QCD factors appropriate for the double real radiation contribution,
\ba
{\cal{N}}_{NNLO}^{RR} = {\cal{N}}_{LO} \left(\frac{\alpha_s N}{2\pi}\right)^2 \frac{\bar{C}(\eps)^2}{C(\eps)^2}.
\ea  
This term is generic to all three kinematic configurations, (FF, IF and II), with the specific antenna functions and momentum maps depending on which configuration the term belongs to.  The four-parton antenna functions display many different types of divergent behaviour including single unresolved and almost colour connected singularities all of which must be properly removed elsewhere in the subtraction term.  The analytic integration of the four-parton antenna functions is carried out over the double unresolved antenna phase space and so the terms in $\dsigma_{NNLO}^{S,b_{1}}$ are reintroduced as part of the double virtual subtraction term. 

\subsubsection{Four-parton single unresolved subtraction term, $\bs{\dsigma_{NNLO}^{S,b_{2}}}$}
\label{sec:dssb2}

As mentioned in the previous section, four-parton antenna functions contain single unresolved spurious singularities which must be removed to ensure a proper subtraction of the IR divergence in the physical matrix elements.  For each four-parton antenna the single unresolved limits are removed by constructing a subtraction term along the lines of $\dsigma_{NNLO}^{S,a}$ but now applying this method to the four-parton antenna function rather than the physical matrix elements.  As such, the terms are built from three-parton antennae (used to remove the single unresolved limits) multiplied by another three-parton antenna (the remnant of the four-parton antenna after the single unresolved limit is taken) and a reduced matrix element which in the single unresolved limit maps on to the matrix element associated with the four-parton antenna.  For a four-parton antenna function this block has the generic form
\ba
\dsigma_{NNLO}^{S,b_{2}}&=&-\mptwops\ \nn\\
&\times& \sum_{j}\ X_{3}^{0}(i,j,k)\ X_{3}^{0}(I,K,l)\ M_{n+2}^{0}(\cdots,I',L,\cdots)\ J_{n}^{(n)}(\{p\}_{n}),
\label{eq:dsb2}
\ea
where the sum is over all partons in the four-parton antenna which admit a single unresolved singularity.  As with other parts of the subtraction term, this block of terms may contain spurious almost colour connected singularities of its own which can arise when the secondary antenna in \eqref{eq:dsb2} also contains single unresolved limits.  The block of terms associated with each four-parton antenna encapsulates all of the single unresolved singularities of the four-parton antenna.  As discussed in Sec.~\ref{sec:rv}, a collection of terms which correctly mimic the single unresolved limits of the four-parton antenna functions performs a specific role when reintroduced as part of the real-virtual subtraction term.

\subsubsection{Almost colour connected subtraction term, $\bs{\dsigma_{NNLO}^{S,c}}$}
\label{sec:dssc}

The almost colour connected contribution only exists when there are at least five partons in the scattering process.  It is intimately linked to the four-parton antenna functions that contain almost colour connected unresolved partons, denoted by $\tilde{X}_{4}^{0}$; $X_{4}^{0}$ antennae do not give a contribution to $\dsigma_{NNLO}^{S,c}$.  The classification of the four-parton antenna functions into $X_{4}^{0}$ and $\tilde{X}_{4}^{0}$ types is displayed in Tab. \ref{tab:x4tildes}.  
\begin{table}[t]
\centering
\begin{tabular}{|l|c|c|}
\hline
Configuration & $X_{4}^{0}$ & $\tilde{X}_{4}^{0}$\\
\hline
Final-final  & $A_{4}^{0},B_{4}^{0},C_{4}^{0},D_{4,a}^{0},E_{4,a}^{0},\tilde{E}_{4}^{0},F_{4,a}^{0},G_{4}^{0},\tilde{G}_{4}^{0},H_{4}^{0}$ & $\tilde{A}_{4}^{0},D_{4,c}^{0},E_{4,b}^{0},F_{4,b}^{0},$\\
\hline
Initial-final &  $A_{4}^{0},B_{4}^{0},C_{4}^{0},G_{4}^{0},\tilde{G}_{4}^{0},H_{4}^{0}$ & $\tilde{A}_{4}^{0},D_{4}^{0},E_{4}^{0},F_{4}^{0}$\\
\hline
Initial-initial & $A_{4}^{0},B_{4}^{0},C_{4}^{0},D_{4,\text{adj}}^{0}F_{4,\text{adj}}^{0},G_{4}^{0},\tilde{G}_{4}^{0},H_{4}^{0}$ & $\tilde{A}_{4}^{0},D_{4,\text{n.adj}}^{0},E_{4}^{0},F_{4,\text{n.adj}}^{0}$ \\
\hline
\end{tabular}
\caption{The classification of the four-parton antenna functions into those containing almost colour connected limits or not.  The final state  $D_{4}^{0}$, $E_{4}^{0}$ and $F_{4}^{0}$ antennae are decomposed into sub-antennae~\cite{our3j1,Glover:2010im} for numerical implementation.}
\label{tab:x4tildes}
\end{table}

In almost colour connected limits, the terms in $\dsigma_{NNLO}^{S,b_{2}}$ tend to over-subtract the divergences of the associated $\tilde{X}_{4}^{0}$.  In the same limits $\dsigma_{NNLO}^{S,a}$ contributes twice the subtraction required by the matrix elements.  Both of these over-subtractions have to be accounted for by $\dsigma_{NNLO}^{S,c}$, which also includes the wide angle soft subtraction term~\cite{our3j1,Pires:2010jv}.  

\begin{figure}[t]
\centering
\includegraphics[width=5cm]{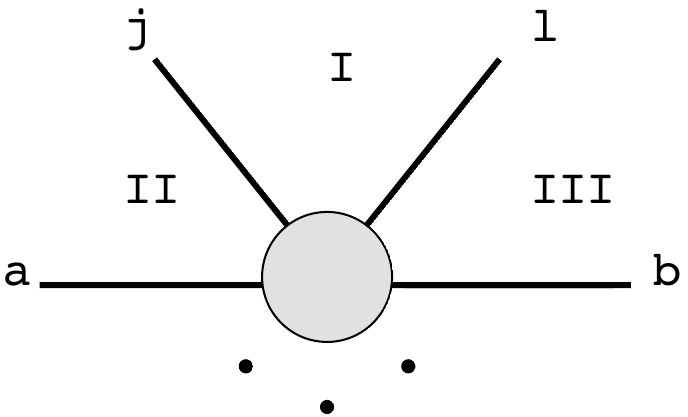}
\caption{The three regions associated with the radiation of the primary unresolved parton. Region I: between the hard radiators of the four-parton antenna, $j,l$.  Region II: one place to the left in the colour ordering between partons $a,j$.  Region III: one place to the right in the colour ordering between partons $l,b$.}
\label{fig:ordering}
\end{figure}

The terms appropriate for $\dsigma_{NNLO}^{S,c}$ can be generated in the following fashion. We consider a final-final double unresolved configuration for clarity but the strategy also applies to initial-final and initial-initial configurations.  Each $\tilde{X}_{4}^{0}$, in $\dsigma_{NNLO}^{S,b_{1}}$ generates a block of terms in $\dsigma_{NNLO}^{S,c}$.  For example, consider the antenna $\tilde{X}_{4}^{0}(j,i,k,l)$ with partons $i,k$ unresolved and $j,l$ the hard radiators, such that the reduced matrix element is given by $M_{n}^{0}(\ldots,a,\widetilde{(jik)},\widetilde{(ikl)},b,\ldots)$. To construct $\dsigma_{NNLO}^{S,c}$ we first consider the colour ordering with $i$ and $k$ removed, i.e., $(\cdots,a,j,l,b,\cdots)$ as depicted in Fig. \ref{fig:ordering}. We then allow partons $i$ and $k$ to be radiated from this underlying ordering; $i$ is radiated first, followed by $k$ (plus the reverse ordering).  The second radiation ($k$) always takes place between the antenna's hard radiators (region I) whilst the first radiation ($i$) may be inserted directly between (region I) and one place either side (regions II and III) of the antenna's radiators in the colour ordering, with a relative minus sign for these two contributions.  Written in terms of antennae this contribution has the form,
\ba
&+&\ \frac{1}{2}\ X_{3}^{0}(j,i,l)\ X_{3}^{0}(\widetilde{(ji)},k,\widetilde{(il)})\ M_{n}^{0}(\ldots,a,\widetilde{((ji)k)},\widetilde{(k(il))},b,\ldots)\nn\\
&-&\ \frac{1}{2}\ X_{3}^{0}(a,i,j)\ X_{3}^{0}(\widetilde{(ji)},k,l)\ M_{n}^{0}(\ldots,\widetilde{(ai)},\widetilde{((ji)k)},\widetilde{(kl)},b,\ldots)\nn\\
&-&\ \frac{1}{2}\ X_{3}^{0}(l,i,b)\ X_{3}^{0}(j,k,\widetilde{(li)})\ M_{n}^{0}(\ldots,a,\widetilde{(jk)},\widetilde{(k(li))},\widetilde{(ib)},\ldots) +(i\leftrightarrow k).
\label{eq:radpatt}
\ea

The large angle soft terms can also be included in this structure.  Two almost colour-connected unresolved partons are removed from the underlying colour ordering and we impose the same radiation pattern of these partons from the underlying colour ordering as in \eqref{eq:radpatt}.  The mapping applied to the primary antenna does not matter because this term only contributes in the soft limit where all mappings become identical.  For six or more parton processes, the large angle soft terms can be generated by allowing the first radiation to be between a pair of final-state partons~\cite{GehrmannDeRidder:2011aa}.  In the special case of hadron-hadron initiated six parton scattering there will only ever be two hard final-state partons so the choice of final-state partons, between which the first unresolved parton is radiated, is unambiguous. However for five-parton processes there simply are not enough final-state partons and the first mapping should be of initial-final type. Four parton processes would require an initial-initial mapping, however in such a simple case there is no almost colour connected contribution.  

In terms of the radiation pattern described in \eqref{eq:radpatt} and Fig.~\ref{fig:ordering}, the first radiation takes place either between two final-state particles, or between one final-state and one initial-state particle, in the underlying ordering and the second radiation between the hard radiators in the antenna, $j$ and $l$.    If the hard partons being considered are both in the final-state then both unresolved partons are radiated between those hard partons.

With the first mapping fixed to be of a particular type, the secondary antenna (which describes the radiation of the second unresolved parton) has common arguments across all terms, unlike in \eqref{eq:radpatt} where the arguments depend on the specific mapping inherited from the primary antenna.  As a consequence, the secondary antenna can be factored out with the reduced matrix element so that, for example, in the final-final mapping case, the large angle soft can have the form,
\ba
Y\cdot X_{3}^{0}(\widetilde{(ij)},k,\widetilde{(il)})\ M_{n}^{0}(\ldots,a,\widetilde{((ij)k)},\widetilde{(k(il))},b,\ldots),
\ea
where $Y$ is a sum of large angle soft antennae.  A similar structure emerges in the initial-final mapping case. The structure of this term can be understood with reference to Fig. \ref{fig:ordering}.  For each region into which an unresolved parton can be radiated there is a term in $Y$ given by,
\ba
-\frac{1}{2}\ \biggl[S_{\alpha i \beta}-S_{A I B}\ \biggr],
\ea
where $I$ denotes the momentum of the unresolved parton $i$ after the second mapping and for the example configuration considered in \eqref{eq:radpatt},
\ba
(\alpha,\beta)&=&\bigl\{\bigl(a,\widetilde{(ij)}\bigr),\bigl(\widetilde{(ij)},\widetilde{(il)}\bigr),\bigl(\widetilde{(il)},b\bigr)\bigr\},\nn\\
(A,B)&=&\bigl\{\bigl(a,\widetilde{((ij)k)}\bigr),\bigl(\widetilde{((ij)k)},\widetilde{(k(il))}\bigr),\bigl(\widetilde{(k(il))},b\bigr)\bigr\},
\ea
The pair of soft functions for each region comes with an overall sign depending on which region the primary antenna belongs to, i.e., a relative minus sign for the regions II and III.  Collecting the terms in \eqref{eq:radpatt} and the large angle soft terms produces a block of terms for a given $\tilde{X}_{4}^{0}$; the sum of such blocks constitutes $\dsigma_{NNLO}^{S,c}$,
\ba
\dsigma_{NNLO}^{S,c}&=&\mptwops\ \sum_{i,k}\biggl\{\nn\\
&+&\ \frac{1}{2}\ X_{3}^{0}(j,i,l)\ X_{3}^{0}(\widetilde{(ji)},k,\widetilde{(il)})\ M_{n}^{0}(\ldots,a,\widetilde{((ji)k)},\widetilde{(k(il))},b,\ldots)\nn\\
&-&\ \frac{1}{2}\ X_{3}^{0}(a,i,j)\ X_{3}^{0}(\widetilde{(ji)},k,l)\ M_{n}^{0}(\ldots,\widetilde{(ai)},\widetilde{((ji)k)},\widetilde{(kl)},b,\ldots)\nn\\
&-&\ \frac{1}{2}\ X_{3}^{0}(l,i,b)\ X_{3}^{0}(j,k,\widetilde{(li)})\ M_{n}^{0}(\ldots,a,\widetilde{(jk)},\widetilde{(k(li))},\widetilde{(ib)},\ldots)\nn\\
&-&\frac{1}{2}\biggl[\bigl(S_{\wt{(ij)},i,\wt{(il)}}-S_{\wt{((ij)k)}i\wt{(k(il))}}\bigr)\nn\\
&&-\ \bigl(S_{ai\wt{(ij)}}-S_{ai\wt{((ij)k)}}\bigr)-\bigl(S_{bi\wt{(il)}}-S_{bi\wt{((il)k)}}\bigr)\biggr]\nn\\
&&X_{3}^{0}(\widetilde{(ij)},k,\widetilde{(il)})\ M_{n}^{0}(\ldots,a,\widetilde{((ij)k)},\widetilde{(k(il))},b,\ldots)\biggr\}\ J_{n}^{(n)}(\{p\}_{n}),\label{eq:dssc}
\ea
where the sum over $i,k$ denotes the set of almost colour-connected pairs contained within the $\tilde{X}_{4}^{0}$ antennae in  $\dsigma_{NNLO}^{S,b_{1}}$.  The blocks in $\dsigma_{NNLO}^{S,c}$ have a common secondary antenna and so when integrated over the single unresolved phase space of the primary antenna, will stay together as a group of integrated antenna functions factoring onto a common unintegrated antenna.  
  
\subsubsection{Colour disconnected subtraction term, $\bs{\dsigma_{NNLO}^{S,d}}$}

For processes involving six or more coloured particles, colour disconnected double unresolved configurations can arise where unresolved partons are separated by more than one hard radiator in the colour ordering. The single unresolved subtraction term $\dsigma_{NNLO}^{S,a}$ admits colour disconnected unresolved limits when the unresolved parton in the antenna is colour disconnected from any other parton in the reduced matrix element that can become unresolved. 

The subtraction terms in $\dsigma_{NNLO}^{S,a}$ take into account all possible single unresolved partons and so for each colour disconnected pair of unresolved partons, $j$ and $m$, where parton $j$ lies in the antenna function and $m$ in the reduced matrix element, there is the corresponding subtraction term where $m$ lies in the antenna and $j$ in the reduced matrix element.  In this double unresolved limit both of these subtraction terms tend to the same value and $\dsigma_{NNLO}^{S,a}$ exactly double-counts the divergence of the matrix elements.

To correct for this over-subtraction, a block of terms is introduced with the form,
\ba
\dsigma_{NNLO}^{S,d}&=&-\mptwops\ \nn\\
&\times&\sum_{j,m}\ X_{3}^{0}(i,j,k)\ X_{3}^{0}(l,m,n)\ M_{n+2}^{0}(\cdots,I,K,\cdots,L,N,\cdots)\ J_{n}^{(n)}(\{p\}_{n}).\nn\\\label{eq:dssd}
\ea
where the sum runs over pairs of partons separated by more than one hard parton in the ordering.  

The disconnected nature of the terms in $\dsigma_{NNLO}^{S,d}$ allows this block of terms to be analytically integrated over the two disconnected regions of single unresolved antenna phase space.  The fact that both unresolved partons can be integrated out directly from the double real level dictates that this contribution should be reintroduced as part of the double virtual subtraction term. 

%
%

\subsection{Construction of the real-virtual subtraction term}
\label{sec:rv}

The real-virtual subtraction term must successfully remove the implicit IR divergent behaviour and explicit poles of the one loop $(n+3)$-parton matrix elements.  In addition to imitating the physical one-loop matrix elements, the real-virtual subtraction term inherits terms from the double real subtraction term after analytic integration over a single unresolved parton.  $\dsigma_{NNLO}^{T}$ contains three types of contribution:
\begin{enumerate}
\item $\dsigma^{T,a}$\\
Terms of the type ${\cal X}_3^0$ that cancel the explicit poles in the virtual one-loop $(m+3)$-parton matrix elements. 
\item $\dsigma^{T,b}$\\
Terms that describe the single unresolved limits of the virtual one-loop $(m+3)$-parton matrix elements.
\item $\dsigma^{T,c}$\\
Almost colour-connected contributions of the type ${\cal X}_3^0\,X_3^0$. 
\end{enumerate}
The subtraction term for the real-virtual contribution can be further sub-divided into five contributions,
\ba
\dsigma_{NNLO}^{T}&=&\dsigma_{NNLO}^{T,a}+\dsigma_{NNLO}^{T,b_{1}}+\dsigma_{NNLO}^{T,b_{2}}+\dsigma_{NNLO}^{T,b_{3}}+\dsigma_{NNLO}^{T,c}.
\ea
The structure and behaviour of each of these terms will be explained in the rest of this section, including whether the terms originate and terminate in the double real, real-virtual and double virtual contribution.

\subsubsection{Real-virtual mass factorisation term, $\bs{\dsigma_{ij,NNLO}^{MF,1}}$}

The NNLO real-virtual mass factorisation contribution of Eq.~\eqref{eq:RVMF} can be further divided,
\ba
&&\dsigma_{ij,NNLO}^{MF,1}=\dsigma_{ij,NNLO}^{MF,1,a}+\dsigma_{ij,NNLO}^{MF,1,b},
\ea
such that $\dsigma_{ij,NNLO}^{MF,1,a}$ is proportional to the $(n+3)$-parton matrix elements,
\ba
\dsigma_{ij,NNLO}^{MF,1,a}(\xi_{1}H_{1},\xi_{2}H_{2})&=&-\int\ \frac{\text{d}x_{1}}{x_{1}}\frac{\text{d}x_{2}}{x_{2}}\nn\\
&\times& \left(\frac{\alpha_s N}{2\pi}\right)\, \bar{C}(\eps)\, \Gamma_{ij;kl}^{(1)}(x_{1},x_{2})\ \dsigma_{kl,NLO}^{R}(x_{1}\xi_{1}H_{1},x_{2}\xi_{2}H_{2}),\nonumber \\
\ea
which will contribute to $\dsigma_{NNLO}^{T,a}$ and $\dsigma_{ij,NNLO}^{MF,1,b}$, which contributes to $\dsigma_{NNLO}^{T,b}$. 
\ba
\dsigma_{ij,NNLO}^{MF,1,b}&=&\dsigma_{ij,NNLO}^{MF,1,b_{2}},\label{eq:mf1bundivide}
\ea
where,
\ba
\dsigma_{ij,NNLO}^{MF,1,b_{2}}(\xi_{1}H_{1},\xi_{2}H_{2})&=&\int\ \frac{\text{d}x_{1}}{x_{1}}\frac{\text{d}x_{2}}{x_{2}}\nn\\
&\times&\left(\frac{\alpha_s N}{2\pi}\right)\, \bar{C}(\eps)\,\Gamma_{ij;kl}^{(1)}(x_{1},x_{2})\ \dsigma_{kl,NLO}^{S}(x_{1}\xi_{1}H_{1},x_{2}\xi_{2}H_{2}).
\label{eq:mf1b2}\nonumber \\
\ea
However, to simplify the book-keeping in the construction of the real-virtual subtraction terms it proves useful to trivially rewrite $\dsigma_{ij,NNLO}^{MF,1,b}$ in the following way,
\ba
\dsigma_{ij,NNLO}^{MF,1,b}&=&\dsigma_{ij,NNLO}^{MF,1,b_{1}}+\dsigma_{ij,NNLO}^{MF,1,b_{2}}+\dsigma_{ij,NNLO}^{MF,1,b_{3}},\label{eq:mf1bdivide}
\ea
where we have added and subtracted,
\ba
\dsigma_{ij,NNLO}^{MF,1,b_{1}}(\xi_{1}H_{1},\xi_{2}H_{2})&=&\int\ \frac{\text{d}x_{1}}{x_{1}}\frac{\text{d}x_{2}}{x_{2}}\nn\\
&\times&\left(\frac{\alpha_s N}{2\pi}\right)\, \bar{C}(\eps)\,\Gamma_{ab;ab}^{(1)}(x_{1},x_{2})\ \dsigma_{ij,NLO}^{S}(x_{1}\xi_{1}H_{1},x_{2}\xi_{2}H_{2}),
\label{eq:mf1b1}\nonumber \\ \\
\dsigma_{ij,NNLO}^{MF,1,b_{3}}(\xi_{1}H_{1},\xi_{2}H_{2})&=&-\dsigma_{ij,NNLO}^{MF,1,b_{1}}(\xi_{1}H_{1},\xi_{2}H_{2}).\label{eq:mf1b3}
\ea
The parton distributions that multiply these contributions to the partonic cross section are always fixed by the identity of partons $i$ and $j$, which, in the initial-final (and initial-initial) configurations, are always of the same type as one (or both) partons involved in the antenna.  However, in certain collinear limits, the type of parton can change, e.g. a gluon may transform into a quark or vice versa.  Eqs.~\eqref{eq:mf1b1} and \eqref{eq:mf1b3} describe these situations. In either case, the antenna is designed to correctly describe the singularity, but the parton emerging from the antenna (and therefore participating in the reduced matrix element) will be of a different type.
In Eq.~\eqref{eq:mf1b1} $a$ and $b$ represent the species of the initial-state partons in the {\em reduced matrix element} in $\dsigma_{ij,NLO}^{S}$ rather than the initial-state partons $i$ and $j$ that are involved in the {\em antenna}.  In unresolved limits that do not change the identities of initial-state partons then $a=i$ and $b=j$. 
However, in an identity changing initial-state collinear limit there will be a mismatch between the parton species in the antenna, $i,j$ and those in the reduced matrix element, $a,b$.  For example, consider a subtraction term $\dsigma_{qg,NLO}^{S}$ that contains a term,
\ba
\dsigma_{qg,NLO}^{S}&\sim&\cdots+A_{3}^{0}(\hat{1}_{q},\hat{2}_{g},3_{\b{q}})\ M_{n}^{0}(\cdots;\hb{1}_{q},\hb{2}_{\b{q}};\cdots)\ J_{n}^{(n)}(\{p\}_{n}),
\ea
where in the $\hat{2}_{g}||3_{\b{q}}$ limit, the initial state gluon transforms into an antiquark. 
In this case the initial-state partons involved in the antenna
$A_{3}^{0}(\hat{1}_{q},\hat{2}_{g},3_{\b{q}})$ are
$(i,j)\equiv (q,g)$ while the initial-state parton species in the reduced matrix element $M_{n}^{0}(\cdots;\hb{1}_{q},\hb{2}_{\b{q}};\cdots)$ are $(a,b)\equiv(q,\b{q})$. In other words, $i = a = q$, while the labels associated with parton $\hat{2}$ differ, $j=g$ and $b=\b{q}$. In this example the mass factorisation kernel used in Eq.~\eqref{eq:mf1b1} is given by $\Gamma_{q\b{q};q\b{q}}^{(1)}(x_{1},x_{2})$.

\subsubsection{One-loop explicit pole subtraction term, $\bs{\dsigma_{NNLO}^{T,a}}$}

It was emphasised in section \ref{sec:RR} that the construction of $\dsigma_{NNLO}^{S,a}$ follows the same lines as constructing $\dsigma_{NLO}^{S}$ with a modified jet function and an additional parton.  This matches the interpretation of the mass factorisation contribution $\dsigma_{NNLO}^{MF,1,a}$ and so integrating the antennae in $\dsigma_{NNLO}^{S,a}$ and combining with the mass factorisation kernels in $\dsigma_{NNLO}^{MF,1,a}$ will generate precisely the same type of integrated antenna strings seen at NLO but with one additional parton,
\ba
\dsigma_{NNLO}^{T,a}&=&-\int_{1}\dsigma_{NNLO}^{S,a}-\dsigma_{NNLO}^{MF,1,a}\nn\\
&=&-\mponeps\ \nn\\
&\times&\bs{J}_{n+3}^{(1)}(1,\ldots,n+3)\ M_{n+3}^{0}(1,\ldots,n+3)\ J_{n}^{(n+1)}(\{p\}_{n+1}),
\ea
where $\NNNLORV=C(\eps)\NNNLORR$.  It was shown in Sec.~\ref{sec:nloant} that the poles of $\bs{J}_{n+2}^{(1)}$ are simply related to the poles of the $(n+2)$-parton one-loop matrix elements.  Using this fact, it is clear that the term $\dsigma_{NNLO}^{T,a}$ correctly subtracts the explicit poles of the real-virtual matrix elements.  The only divergence remaining in the one-loop matrix elements is the implicit divergence arising from single unresolved configurations.

\subsubsection{Tree $\times$ loop subtraction term, $\bs{\dsigma_{NNLO}^{T,b_{1}}}$}
\label{sec:dstb1}

One-loop matrix elements factorise in implicit IR singular limits into two terms~\cite{Bern:1994zx} which can be schematically understood as,
\ba
\text{1-loop}\ \longrightarrow\ (\text{tree}\times\text{loop})\ +\ (\text{loop}\times\text{tree}).\nn
\ea
The first term is the product of  a tree-level singular function factoring onto a one-loop reduced matrix element and is subtracted using a tree level antenna function and a one-loop reduced matrix element.  This term by itself removes part of the implicit IR divergence of the one-loop matrix elements but introduces explicit poles associated with the one-loop $(n+2)$-parton reduced matrix element.  The explicit poles of this matrix element can be removed by introducing the appropriate integrated antenna string so that, after including the mass factorisation contribution $\dsigma_{NNLO}^{MF,1,b_{1}}$, defined in Eq.~\eqref{eq:mf1b1},
\ba
&&\dsigma_{NNLO}^{T,b_{1}}=\mponeps\ \nn\\
&&\times\ \sum_{j}\ X_{3}^{0}(i,j,k)\biggl\{\delta(1-x_{1})\delta(1-x_{2})\ M_{n+2}^{1}(1,\ldots,n+2)\nn\\
&&\hspace{3.1cm}+\ c_J\bs{J}_{n+2}^{(1)}(1,\ldots,n+2)\ M_{n+2}^{0}(1,\ldots,n+2)\biggr\}\ J_{n}^{(n)}(\{p\}_{n}),
\label{eq:dstb1}
\ea
is free from explicit IR poles and where the sum is over the final-state unresolved partons.  $c_J$ is a constant equal to unity unless $n=0$ and all particles are gluons in which case $c_J = 2$ to account for both colour orderings of the gluons.  This formula can be applied to final-final, initial-final and initial-initial configurations.

\subsubsection{Loop $\times$ tree subtraction term, $\bs{\dsigma_{NNLO}^{T,b_{2}}}$}
\label{sec:dstb2}

The factorisation of one-loop matrix elements in IR limits also requires a term given by a one-loop universal singular function factoring onto a tree-level matrix element.  To properly account for this contribution, a subtraction term is constructed from a one-loop antenna function and a tree-level reduced matrix element.  The one-loop antenna function contains explicit IR poles which must be removed to ensure a finite total contribution.  This goal is achieved by a difference of integrated antenna strings which contains the IR explicit poles of the one-loop antenna, and reflects the construction of $X_{3}^{1}$~\cite{GehrmannDeRidder:2005cm}, 
\ba
X_{3}^{1}(i,j,k)&=&{\cal{S}}_{ijk/IK}\frac{M_{3}^{1}(i,j,k)}{M_{2}^{0}(I,K)}-X_{3}^{0}(i,j,k)\frac{M_{2}^{1}(I,K)}{M_{2}^{0}(I,K)}.
\label{eq:1loopant}
\ea
A block of terms can then be constructed to remove the remaining implicit IR divergence from the one-loop matrix elements, which incorporates the mass factorisation terms $\dsigma_{NNLO}^{MF,1,b_{2}}$ and $\dsigma_{NNLO}^{MF,1,b_{3}}$, defined in Eqs.~\eqref{eq:mf1b2} and~\eqref{eq:mf1b3} respectively, and is free from explicit poles,
\ba
\dsigma_{NNLO}^{T,b_{2}}&=&\mponeps\ \nn\\
&\times&\ \sum_{j}\ \biggl[X_{3}^{1}(i,j,k)\delta(1-x_{1})\delta(1-x_{2})+\bs{J}_{X}^{(1)}(i,j,k)X_{3}^{0}(i,j,k)\nn\\
&-&M_{X} X_{3}^{0}(i,j,k)\bs{J}_{2}^{(1)}(I,K)\biggr]\ M_{n+2}^{0}(\cdots,I,K,\cdots)\ J_{n}^{(n)}(\{p\}_{n}),
\label{eq:dstb2}
\ea
where $M_{X}$ is a constant which depends on the species of one-loop antenna, $X_{3}^{1}$.  The string $\bs{J}_{X}^{(1)}$ is constructed as a sum of $N_{X}$ two-particle integrated antenna strings,
\ba
\bs{J}_{X}^{(1)}&=&\sum_{(i,j)=1}^{N_{X}}\bs{J}_{2}^{(1)}(i,j),
\ea
where $N_{X}$ counts the number of colour-connected pairs of partons in the one-loop antenna.  The values of $M_{X}$ and $N_{X}$ are displayed in Tab.~\ref{tab:mxnx}.  This subtraction term applies to all kinematic configurations and particle combinations. The mass factorisation kernels used to form the $\bs{J}_{X}^{(1)}$ come from $\dsigma_{NNLO}^{MF,1,b_{2}}$ and those used to generate $\bs{J}_{2}^{(1)}$ come from $\dsigma_{NNLO}^{MF,1,b_{3}}$. 
\begin{table}[t]
\begin{center}
\begin{tabular}{|l||c|c|c|c|c|c|c|c|c|c|c|c|c|}
\hline
$X_{3}^1$ & $A_3^1$ & $\tilde{A}_3^1$ & $\hat{A}_3^1$ & $D_3^1$ & $\hat{D}_3^1$ & $E_3^1$ & $\tilde{E}_3^1$ & $\hat{E}_3^1$  & $F_3^1$ & $\hat{F}_3^1$ & $G_3^1$ & $\tilde{G}_3^1$ & $\hat{G}_3^1$ \\\hline
$N_{X}$ & 2 & 1 & 2 & 3 & 3 & 2 & 1 & 0 & 3 & 3 & 2 & 1 & 0 \\
$M_{X}$ & 1 & 1 & 0 & 2 & 2 & 2 & 0 & 2 & 2 & 2 & 2 & 0 & 2 \\\hline
\end{tabular}
\caption{Number of colour-connected pairs $N_X$ for the one-loop antenna $X_3^1$, and the associated constant $M_X$.}
\label{tab:mxnx}
\end{center}
\end{table}
The term proportional to $\bs{J}_{X}^{(1)}$ is derived from the integral of the double real subtraction term.  Following the discussion in Sec.~\ref{sec:nloant}, the poles of a four-parton matrix element integrated over the single unresolved phase space may be encapsulated by an integrated antenna string.  Schematically,
\ba
\Poles\bigg[\int_{1}M_{4}^{0}\bigg]&=&\Poles\bigg[\bs{J}_{3}^{(1)}M_{3}^{0}\bigg]=-\Poles\bigg[M_{3}^{1}\bigg],
\ea
and therefore the single unresolved poles of a four-parton antenna are also described by,
\ba
\Poles\bigg[\int_{1}X_{4}^{0}\bigg]&=&\Poles\bigg[\bs{J}_{X}^{(1)}X_{3}^{0}\bigg]=-\Poles\bigg[X_{3}^{1}\bigg].
\ea
In Sec.~\ref{sec:RR}, $\dsigma_{NNLO}^{S,b}$ was constructed to have no single unresolved limits such that,
\ba
\Poles\bigg[\int_{1}\dsigma_{NNLO}^{S,b}\bigg]&=&0.
\ea
Equivalently,
\ba
\Poles\bigg[\int_{1}\dsigma_{NNLO}^{S,b_{1}}\bigg]&=&-\Poles\bigg[\int_{1}\dsigma_{NNLO}^{S,b_{2}}\bigg].
\ea
For each $X_{4}^{0}$ or $\tilde{X}_{4}^{0}$ contributing to $\dsigma_{NNLO}^{S,b_{1}}$, the associated block of iterated antennae in $\dsigma_{NNLO}^{S,b_{2}}$ integrated over the single unresolved phase space therefore systematically generates a singular contribution of the form, $\bs{J}_{X}^{(1)}(i,j,k)X_{3}^{0}(i,j,k)$, as required by Eq.~\eqref{eq:dstb2}.

The one-loop antenna term, $X_{3}^{1}$, and the two-parton integrated antenna string, $\bs{J}_{2}^{(1)}$, in Eq.~\eqref{eq:dstb2} do not come from the double real subtraction term and so must be compensated for in the double virtual subtraction term.  

\subsubsection{One-loop renormalisation subtraction term, $\bs{\dsigma_{NNLO}^{T,b_{3}}}$}
\label{sec:dstb3}

When introducing one-loop quantities it is important to ensure they are properly renormalised to guarantee a complete cancellation of explicit poles.  In the real-virtual subtraction term two one-loop quantities are introduced: the one-loop reduced matrix elements in \eqref{eq:dstb1} and the one-loop antenna functions in \eqref{eq:dstb2}.  The one-loop matrix elements are renormalised at the renormalisation scale $\mu^{2}$ whereas the one-loop antenna is renormalised at the mass scale of the antenna $s_{ijk}$.  To ensure both quantities are renormalised at the same scale, the replacement is made,
\ba
X_{3}^{1}(i,j,k)&\rightarrow&X_{3}^{1}(i,j,k) + \frac{\beta_{0}}{\epsilon}\ X_{3}^{0}(i,j,k)\ \biggl(\biggl(\frac{|s_{ijk}|}{\mu^{2}}\biggr)^{-\epsilon}-1\biggr).
\label{eq:x31shift}
\ea
The systematic inclusion of these terms requires that each one loop antenna (whose pole structure is proportional to a tree-level antenna) used in the real-virtual subtraction term is compensated by a term proportional to $\beta_{0}$, such that a block of subtraction terms is generated of the form,
\ba
\dsigma_{NNLO}^{T,b_{3}}&=&\mponeps\ \nn\\
&\times&\ \sum_{j}\ \beta_{0}\log\biggl(\frac{\mu^{2}}{|s_{ijk}|}\biggr) X_{3}^{0}(i,j,k)\delta(1-x_{1})\delta(1-x_{2})\nn\\
&\times&\ M_{n+2}^{0}(\cdots,I,K,\cdots)\ J_{n}^{(n)}(\{p\}_{n}),
\label{eq:dstb3}
\ea
where the sum is the same as in \eqref{eq:dstb2}.  The colour decomposition of $\beta_{0}$ into $b_{0}$ and $b_{0,\NF}$ dictates to which orders in the colour decomposition these terms contribute,
\ba
\beta_{0}&=&b_{0}N+b_{0,F}\NF,
\ea
where $b_{0}=11/6$ and $b_{0,F}=-1/3$.  The terms in $\dsigma_{NNLO}^{T,b_{3}}$ originate in the real-virtual subtraction term and so, by integrating the three-parton antenna over the single-unresolved phase space, are reintroduced as part of the double virtual subtraction term.
\subsubsection{Integrated almost colour connected subtraction term, $\bs{\dsigma_{NNLO}^{T,c}}$}
\label{sec:dstc}

\par The final term that contributes to the real-virtual subtraction term is mostly derived from the analytic integration of $\dsigma_{NNLO}^{S,c}$ over the single unresolved antenna phase space with additional predictable terms to ensure an IR finite contribution.
Integrating Eq.~\eqref{eq:dssc} over the single unresolved phase space and introducing three additional terms to ensure all explicit IR poles are cancelled yields,
\ba
\dsigma_{NNLO}^{T,c}&=&-\mponeps\ \biggl\{\nn\\
 &&\frac{1}{2}\ \sum_{j}\ \biggl[\ \biggl(\ \bigl({\cal{X}}_{3}^{0}(s_{ik})-{\cal{X}}_{3}^{0}(s_{(ij)(jk)})\bigr)\nn\\
 &&-\ \bigl({\cal{X}}_{3}^{0}(s_{ai})-{\cal{X}}_{3}^{0}(s_{a(ij)})\bigr)-\ \bigl({\cal{X}}_{3}^{0}(s_{kb})-{\cal{X}}_{3}^{0}(s_{(kj)b})\bigr)\biggr)\nn\\
&-&\biggl(\bigl({\cal{S}}(s_{ik},s_{ik},1)-{\cal{S}}(s_{(ij)(jk)},s_{ik},x_{(ij)(jk),ik})\bigr)\nn\\
&-&\bigl({\cal{S}}(s_{ai},s_{ik},x_{ai,ik})-{\cal{S}}(s_{a(ij)},s_{ik},x_{a(ij),ik})\bigr)\nn\\
&-&\bigl({\cal{S}}(s_{kb},s_{ik},x_{kb,ik})-{\cal{S}}(s_{(jk)b},s_{ik},x_{(jk)b,ik})\bigr)\biggr)\delta(1-x_{1})\delta(1-x_{2})\biggr]\biggr\} \nn\\
&\times&X_{3}^{0}(i,j,k)\ M_{n+2}^{0}(\cdots,I,K,\cdots)\ J_{n}^{(n)}(\{p\}_{n}).
\label{eq:dstc}
\ea
In this equation, the terms introduced to ensure IR finiteness are those involving the integrated antennae with mapped momenta, ${\cal{X}}_{3}^{0}(s_{(ij)(jk)})$, ${\cal{X}}_{3}^{0}(s_{a(ij)})$ and ${\cal{X}}_{3}^{0}(s_{(kj)b})$. These terms must therefore appear in the double virtual subtraction term.  For future reference we label these as,
\ba
\dsigma_{NNLO}^{T,c_{1}}&=&-\mponeps\ \nn\\
 &\times&\frac{1}{2}\ \sum_{j}\ \big[{\cal{X}}_{3}^{0}(s_{(ij)(jk)})+{\cal{X}}_{3}^{0}(s_{a(ij)})+{\cal{X}}_{3}^{0}(s_{(kj)b})\big]\nn\\
&\times& X_{3}^{0}(i,j,k)\ M_{n+2}^{0}(\cdots,I,K,\cdots)\ J_{n}^{(n)}(\{p\}_{n}),\\
\label{eq:dstc1}
\dsigma_{NNLO}^{T,c_{2}}&=&+\mponeps\ \nn\\
 &\times&\ \sum_{j}\ {\cal{X}}_{3}^{0}(s_{(ij)(jk)})\ X_{3}^{0}(i,j,k)\ M_{n+2}^{0}(\cdots,I,K,\cdots)\ J_{n}^{(n)}(\{p\}_{n}).
\label{eq:dstc2}
\ea
The integrated soft function for the final-final mapping is denoted by ${\cal{S}}(s_{ac},s_{ik},x_{ac,ik})$. An explicit expression can be found in Ref.~\cite{GehrmannDeRidder:2011aa}. As discussed in Sec.~\ref{sec:dssc}, for five-parton processes it is necessary to use a soft function with an initial-final map.  In this case one finds an analogous equation to Eq.~\eqref{eq:dstc} involving the integrated initial-final soft function, ${\cal S}^{IF}(s_{ac},s_{IK},y_{ac,iK})$, which is given explicitly in App.~\ref{sec:soft}.


\subsection{Double virtual subtraction term structure}
\label{sec:doublevirt}

The double virtual contribution to the $pp\rightarrow n$-jet cross section involves the two-loop $(n+2)$-parton matrix elements which have no implicit IR divergence in any regions of the appropriate $n$-parton phase space.  All that remains is to reintroduce the integrated forms of the appropriate subtraction terms, and combine them with the double virtual mass factorisation terms, so that the explicit IR poles of the two-loop contribution are cancelled. The main result of this section is that the double virtual subtraction term is constructed from three terms,
\ba
\dsigma_{ij,NNLO}^{U}&=&\dsigma_{ij,NNLO}^{U,A}+\dsigma_{ij,NNLO}^{U,B}+\dsigma_{ij,NNLO}^{U,C}.
\ea
The following sections will demonstrate the origin of each of these terms.  This construction reflects the dipole-like singularity structure for two-loop amplitudes made apparent in Catani's two-loop factorisation formula~\cite{Catani:1998bh},
\ba
\Poles\big(M_{n}^{2}(1,\cdots,n)\big)&=&2\bs{I}_{n}^{(1)}(\eps;1,\cdots,n)\bigg(M_{n}^{1}(1,\cdots,n)-\frac{\beta_{0}}{\eps}M_{n}^{0}(1\cdots,n)\bigg)\nn\\
&-&2\bs{I}_{n}^{(1)}(\eps;1,\cdots,n)^{2}M_{n}^{0}(1,\cdots,n)\nn\\
&+&2e^{-\eps\gamma}\frac{\Gamma(1-2\eps)}{\Gamma(1-\eps)}\bigg(\frac{\beta_{0}}{\eps}+K\bigg)\bs{I}_{n}^{(1)}(2\eps;1\cdots,n)M_{n}^{0}(1\cdots,n)\nn\\
&+&2\bs{H}^{(2)}(\eps)M_{n}^{0}(1,\cdots,n),\label{eq:catani2}
\ea
with $\bs{I}_{n}^{(1)}(\eps)$ given by Eq.~\eqref{eq:catani1} and where the form of the hard function, $\bs{H}^{(2)}(\eps)$ and the constant $K$ depends on the particle content and order in $N$ under consideration.

\subsubsection{Double virtual mass factorisation terms, $\bs{\dsigma_{NNLO}^{MF,2}}$}

The general form of the double virtual mass factorisation contribution was given in Eq.~\eqref{eq:VVMF}.  To simplify the construction of the double virtual subtraction term, $\Gamma_{ij;kl}^{(2)}(z_{1},z_{2})$ may be decomposed in the following way,
\ba
\Gamma_{ij;kl}^{(2)}(z_{1},z_{2})&=&\overline{\Gamma}_{ij;kl}^{(2)}(z_{1},z_{2})-\frac{\beta_{0}}{\eps}\Gamma_{ij;kl}^{(1)}(z_{1},z_{2})+\frac{1}{2}\big[\Gamma_{ij;ab}^{(1)}\otimes\Gamma_{ab;kl}^{(1)}\big](z_{1},z_{2}),
\ea
such that,
\ba
\overline{\Gamma}_{ij;kl}^{(2)}(z_{1},z_{2})&=&\overline{\Gamma}_{ik}^{(2)}(z_{1})\delta_{jl}\delta(1-z_{2})+\overline{\Gamma}_{jl}^{(2)}(z_{2})\delta_{ik}\delta(1-z_{1}),
\ea
where,
\ba
\overline{\Gamma}_{ij}^{(2)}(z)&=&-\frac{1}{2\eps}\bigg(p_{ij}^{1}(z)+\frac{\beta_{0}}{\eps}p_{ij}^{0}(z)\bigg).\label{eq:bargam2}
\ea
The double virtual mass factorisation contribution can be recast into three terms,
\ba
\dsigma_{ij,NNLO}^{MF,2}&=&\dsigma_{ij,NNLO}^{MF,2,A}+\dsigma_{ij,NNLO}^{MF,2,B}+\dsigma_{ij,NNLO}^{MF,C},
\ea
where the individual contributions are given by:
\ba
\dsigma_{ij,NNLO}^{MF,2,A}&=&
-\int\frac{{\rm{d}}z_{1}}{z_{1}}\frac{{\rm{d}}z_{2}}{z_{2}}\ \left(\frac{\alpha_s N}{2\pi}\right)\, \bar{C}(\eps)\,
\Gamma_{ij;kl}^{(1)}\bigg(\dsigma_{kl,NLO}^{V}-\frac{\beta_{0}}{\eps}\dsigma_{kl,LO}\bigg),\label{eq:dsmf2A}\\
\dsigma_{ij,NNLO}^{MF,2,B}&=&
+\int\frac{{\rm{d}}z_{1}}{z_{1}}\frac{{\rm{d}}z_{2}}{z_{2}}\
\left(\frac{\alpha_s N}{2\pi}\right)\, \bar{C}(\eps)\,
\Gamma_{ij;kl}^{(1)}\dsigma_{kl,NLO}^{T}\nonumber \\
&&-\int\frac{{\rm{d}}z_{1}}{z_{1}}\frac{{\rm{d}}z_{2}}{z_{2}}\
\left(\frac{\alpha_s N}{2\pi}\right)^2\, \bar{C}(\eps)^2\,
\frac{1}{2}\big[\Gamma_{ij;ab}^{(1)}\otimes\Gamma_{ab;kl}^{(1)}\big] \dsigma_{kl,LO},\label{eq:dsmf2B}\\
\dsigma_{ij,NNLO}^{MF,2,C}&=&
-\int\frac{{\rm{d}}z_{1}}{z_{1}}\frac{{\rm{d}}z_{2}}{z_{2}}\ \left(\frac{\alpha_s N}{2\pi}\right)^2\, \bar{C}(\eps)^2\, \overline{\Gamma}_{ij;kl}^{(2)}\ \dsigma_{kl,LO},\label{eq:dsmf2C}
\ea
and the convolution of two functions $f(x_{1},x_{2})$ and $g(y_{1},y_{2})$ is defined in the following way,
\ba
\big[f\otimes g\big](z_{1},z_{2})\equiv\int{\rm{d}}x_{1}{\rm{d}}x_{2}{\rm{d}}y_{1}{\rm{d}}y_{2}f(x_{1},x_{2})g(y_{1},y_{2})\delta(z_{1}-x_{1}y_{1})\delta(z_{2}-x_{2}y_{2}).
\ea
Each of these terms naturally fits into the corresponding piece of the double virtual subtraction term $\dsigma_{ij,NNLO}^{U,A}$, $\dsigma_{ij,NNLO}^{U,B}$, and $\dsigma_{ij,NNLO}^{U,C}$, to render each contribution free from initial-state collinear poles.

\subsubsection{$\bs{\dsigma_{NNLO}^{U,A}}$ subtraction term}
\label{sec:jm1}

The first double virtual subtraction term is generated by integrating two terms from the real-virtual subtraction term and combining these with the appropriate mass factorisation term.  The first term to be integrated comes from $\dsigma_{NNLO}^{T,b_{1}}$, defined in Eq.~\eqref{eq:dstb1} as the collection of terms proportional to the one-loop matrix element.  After integration, this contribution yields a sum of integrated antenna functions factoring onto a one-loop matrix element. The second term comes from the integrated form of $\dsigma_{NNLO}^{T,b_{3}}$,
\ba
\int_{1}\dsigma_{ij,NNLO}^{T,b_{3}}&=&-\mps\nn\\
&\times&\sum_{\{i,j\}}\frac{\beta_{0}}{\eps}\biggl(\biggl(\frac{|s_{ij}|}{\mu^{2}}\biggr)^{-\eps}-1\biggr)\ {\cal{X}}_{3}^{0}(s_{ij})M_{n+2}^{0}(1,\cdots,n+2)J_{n}^{(n)}(\{p\}_{n}),\nn\\
\ea
where $\NNNLOVV=C(\eps)\NNNLORV=C(\eps)^2\NNNLORR$.
The bracket may be expanded to produce two terms. The first term (proportional to $(|s_{ij}|^{-\eps}$)) will become part of $\dsigma_{NNLO}^{U,C}$.  The second, proportional to $-1$, is combined with the one-loop matrix element term from the integral of $\dsigma_{NNLO}^{T,b_{1}}$ and the mass factorisation term $\dsigma_{NNLO}^{MF,2,A}$, to produce the first double virtual subtraction term,
\ba
\lefteqn{\dsigma_{ij,NNLO}^{U,A}=-\mps}\nn\\
&\times&\bs{J}_{n+2}^{(1)}(1,\cdots,n+2)\biggl(M_{n+2}^{1}(1,\cdots,n+2)-\frac{\beta_{0}}{\eps}\ M_{n+2}^{0}(1,\cdots,n+2)\biggr)J_{n}^{(n)}(\{p\}_{n}).\nn\\ \label{eq:dsuA}
\ea
The poles of the integrated antenna string are directly related to the poles of Catani's one-loop insertion operator, therefore this contribution to the double virtual cross section contains the $1/\eps^{4}$ and $1/\eps^{3}$ singular contributions given in the first line of Eq.~\eqref{eq:catani2}.  

It is important to note that $\dsigma_{NNLO}^{U,A}$ does not contain precisely the same singularities at $1/\eps^{2}$ and above as the first line of Eq.~\eqref{eq:catani2} because the finite difference between $\bs{J}_{n}^{(1)}$ and $\bs{I}_{n}^{(1)}$ is formally of order $\eps^{0}$.  These differences ultimately cancel against similar terms in $\dsigma_{NNLO}^{U,B}$ and $\dsigma_{NNLO}^{U,C}$.

\subsubsection{$\bs{\dsigma_{NNLO}^{U,B}}$ subtraction term}
\label{sec:jxj}

The second double virtual subtraction term is formed from the integral of the remaining term from $\dsigma_{ij,NNLO}^{T,b_{1}}$ \eqref{eq:dstb1} proportional to $M_{n}^{0}$, the integral of $\dsigma_{NNLO}^{T,c_{1}}$ \eqref{eq:dstc1} and $\dsigma_{NNLO}^{S,d}$ \eqref{eq:dssd}, and the mass factorisation contribution $\dsigma_{NNLO}^{MF,2,B}$.  The resultant subtraction term is given by,
\ba
\dsigma_{ij,NNLO}^{U,B}&=&-\mps\nn\\
&\times&\frac{1}{2}\big[\bs{J}_{n+2}^{(1)}(1,\cdots,n+2)\otimes\bs{J}_{n+2}^{(1)}(1,\cdots,n+2)\big](z_{1},z_{2})\nn\\
&\times& M_{n+2}^{0}(1,\cdots,n+2)\ J_{n}^{(n)}(\{p\}_{n}).\label{eq:dsuB}
\ea
It is easily seen that this expression contains the pole structure of the second line of Eq.~\eqref{eq:catani2} with additional singular contributions arising from the finite differences between $\bs{J}_{n}^{(1)}$ and $\bs{I}_{n}^{(1)}$ at ${\cal{O}}(\eps^{0})$ and higher orders.

Note that the combination of Eqs.~\eqref{eq:dsuA} and \eqref{eq:dsuB} reproduce the pole structure of the first two lines of Eq.~\eqref{eq:catani2} up to ${\cal O}(1/\e)$.   

\subsubsection{$\bs{\dsigma_{NNLO}^{U,C}}$ subtraction term}
\label{sec:j2}

The third double virtual subtraction term is generated by integrating appropriate terms in the remaining double real and real-virtual subtraction terms, $\dsigma_{NNLO}^{S,b_{1}}$ \eqref{eq:dssb1}, $\dsigma_{NNLO}^{T,b_{2}}$ \eqref{eq:dstb2}, $\dsigma_{NNLO}^{T,c_{2}}$ \eqref{eq:dstc2}, the term proportional to $|s_{ij}|^{-\eps}$ generated by expanding the bracket in Sec.~\ref{sec:jm1} and the mass factorisation contribution $\dsigma_{NNLO}^{MF,2,C}$ \eqref{eq:dsmf2C}.  The final subtraction term is given by,
\ba
\dsigma_{NNLO}^{U,C}&=&-\mps\nn\\
&&\bs{J}_{n+2}^{(2)}(1,\cdots,n+2)\ M_{n+2}^{0}(1,\cdots,n+2)\ J_{n}^{(n)}(\{p\}_{n}),\label{eq:dsuC}
\ea
where we have introduced the double unresolved integrated antenna string, $\bs{J}_{n+2}^{(2)}$ as a sum over double unresolved integrated dipoles, $\bs{J}_{2}^{(2)}$,
\ba
\bs{J}_{n+2}^{(2)}(1,\cdots,n+2)&=&\sum_{(i,j)}\bs{J}_{2}^{(2)}(i,j),
\ea
and the sum runs over colour connected pairs of partons in the $(n+2)$-parton ordering.  The double unresolved integrated dipoles depend on the type of particle in the scattering process, the order in the colour decomposition under consideration and the kinematic configuration (FF, IF or II) of the dipole.  Schematically, the final-final double unresolved integrated dipole has the form,
\ba
\bs{J}_{2}^{(2)}(I,J)&=&c_{1}^{FF}{\cal{X}}_{4}^{0}(s_{IJ})+c_{2}^{FF}\tilde{\cal{X}}_{4}^{0}(s_{IJ})+c_{3}^{FF}{\cal{X}}_{3}^{1}(s_{IJ})\nn\\
&+&c_{4}^{FF}\frac{\beta_{0}}{\eps}\biggl(\frac{|s_{IJ}|}{\mu^{2}}\biggr)^{-\eps}{\cal{X}}_{3}^{0}(s_{IJ})+ c_{5}^{FF}{\cal{X}}_{3}^{0}(s_{IJ})\otimes{\cal{X}}_{3}^{0}(s_{ij}),
\ea
where $c_{n}^{FF}$, are constants associated with each integrated dipole.  For an initial-final integrated dipole, the form of the integrated dipole is modified by the use of initial-final integrated antennae and the inclusion of the necessary mass factorisation contribution to remove all initial-state collinear poles,
\ba
\bs{J}_{2}^{(2)}(\hb{1},I)&=&c_{1}^{IF}{\cal{X}}_{4}^{0}(s_{\b{1}I})+c_{2}^{IF}\tilde{\cal{X}}_{4}^{0}(s_{\b{1}I})+c_{3}^{IF}{\cal{X}}_{3}^{1}(s_{\b{1}I})+c_{4}^{IF}\frac{\beta_{0}}{\eps}\biggl(\frac{|s_{\b{1}I}|}{\mu^{2}}\biggr)^{-\eps}{\cal{X}}_{3}^{0}(s_{\b{1}I})\nn\\
&+&c_{5}^{IF}{\cal{X}}_{3}^{0}(s_{\b{1}I})\otimes{\cal{X}}_{3}^{0}(s_{\b{1}I})-\overline{\Gamma}_{ik}^{(2)}(z_{1})\delta(1-z_{2}),
\ea
where $i$ labels the species of parton in the initial-state and $k$ is the parton involved in the matrix element in Eq.~\eqref{eq:dsuC}.   The initial-initial integrated dipole is formed from initial-initial antennae and includes a mass factorisation contribution with non-trivial $z_{1}$ and $z_{2}$ dependence,
\ba
\bs{J}_{2}^{(2)}(\hb{1},\hb{2})&=&c_{1}^{II}{\cal{X}}_{4}^{0}(s_{\b{1}\b{2}})+c_{2}^{II}\tilde{\cal{X}}_{4}^{0}(s_{\b{1}\b{2}})+c_{3}^{II}{\cal{X}}_{3}^{1}(s_{\b{1}\b{2}})+\frac{\beta_{0}}{\eps}\biggl(\frac{|s_{\b{1}\b{2}}|}{\mu^{2}}\biggr)^{-\eps}c_{4}^{II}{\cal{X}}_{3}^{0}(s_{\b{1}\b{2}})\nn\\
&+&c_{5}^{II}{\cal{X}}_{3}^{0}(s_{\b{1}\b{2}})\otimes{\cal{X}}_{3}^{0}(s_{\b{1}\b{2}})-\overline{\Gamma}_{ij;kl}^{(2)}(z_{1})\delta(1-z_{2}),
\ea
where the $i,j$ labels carried by $\overline{\Gamma}^{(2)}_{ij;kl}$ are the species of initial-state parton carried by the cross section and $k,l$ denote those carried by the matrix element in Eq.~\eqref{eq:dsuC}.   

\subsubsection{The full double virtual subtraction term, $\bs{\dsigma_{NNLO}^{U}}$}

\begin{figure}[t]
\centering
\includegraphics[width=0.5cm]{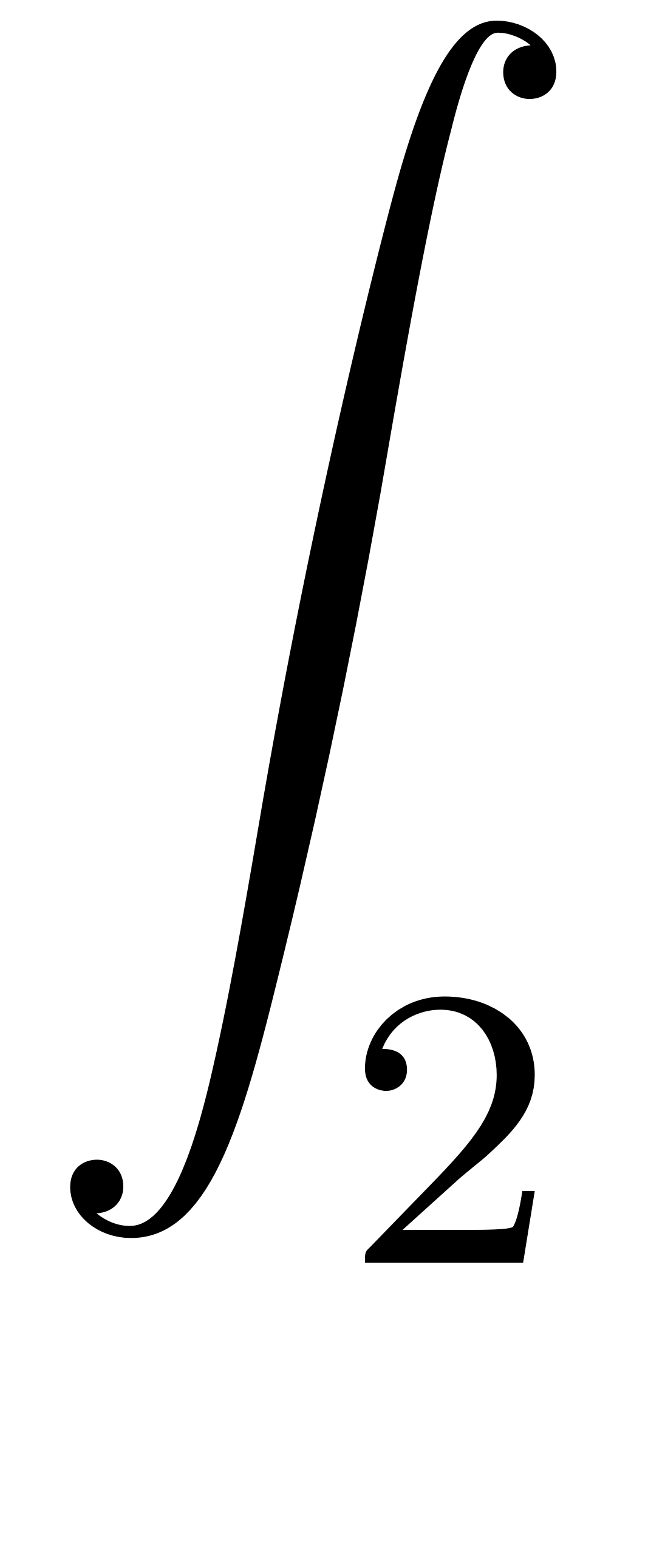}
\includegraphics[width=2.6cm]{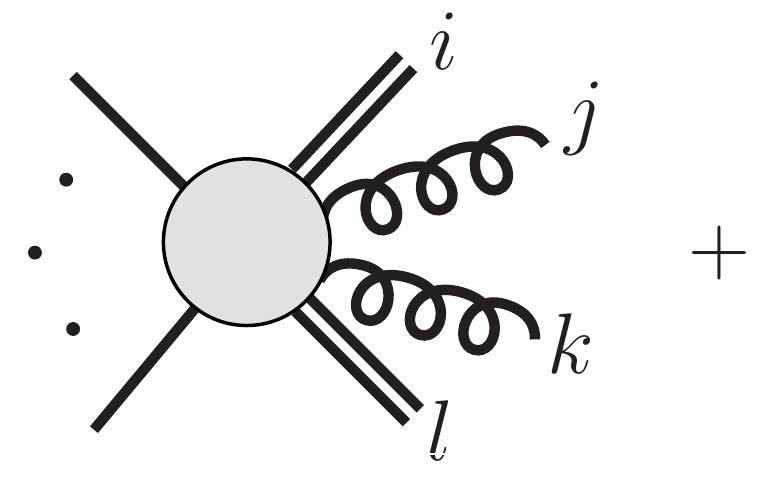}
\includegraphics[width=0.5cm]{int1}
\includegraphics[width=3.2cm]{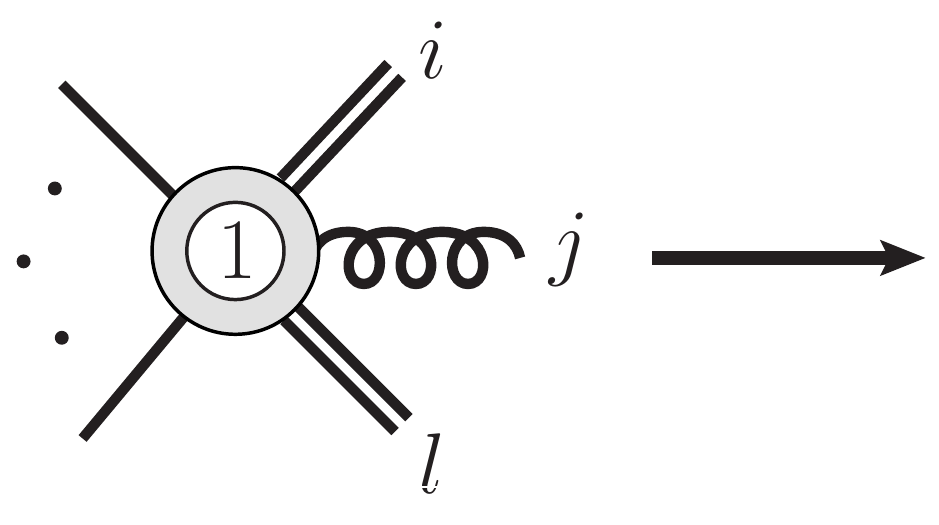}
\includegraphics[width=2.8cm]{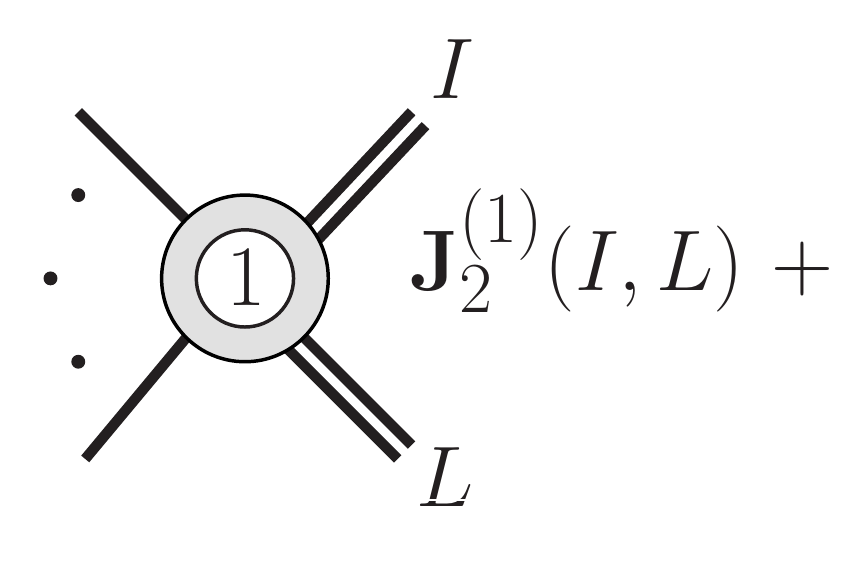}
\includegraphics[width=1.8cm]{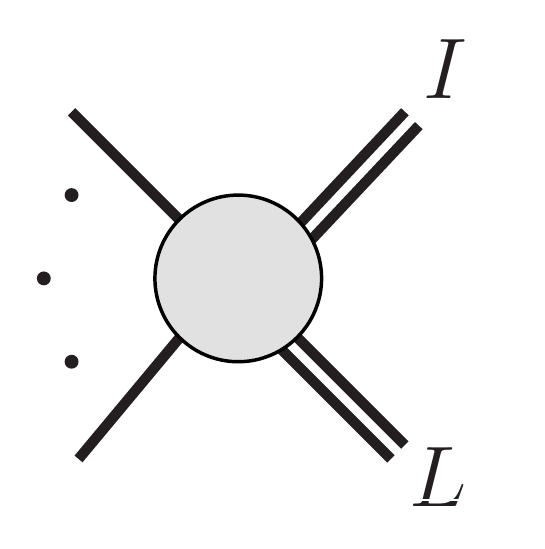}
\includegraphics[width=2.7cm]{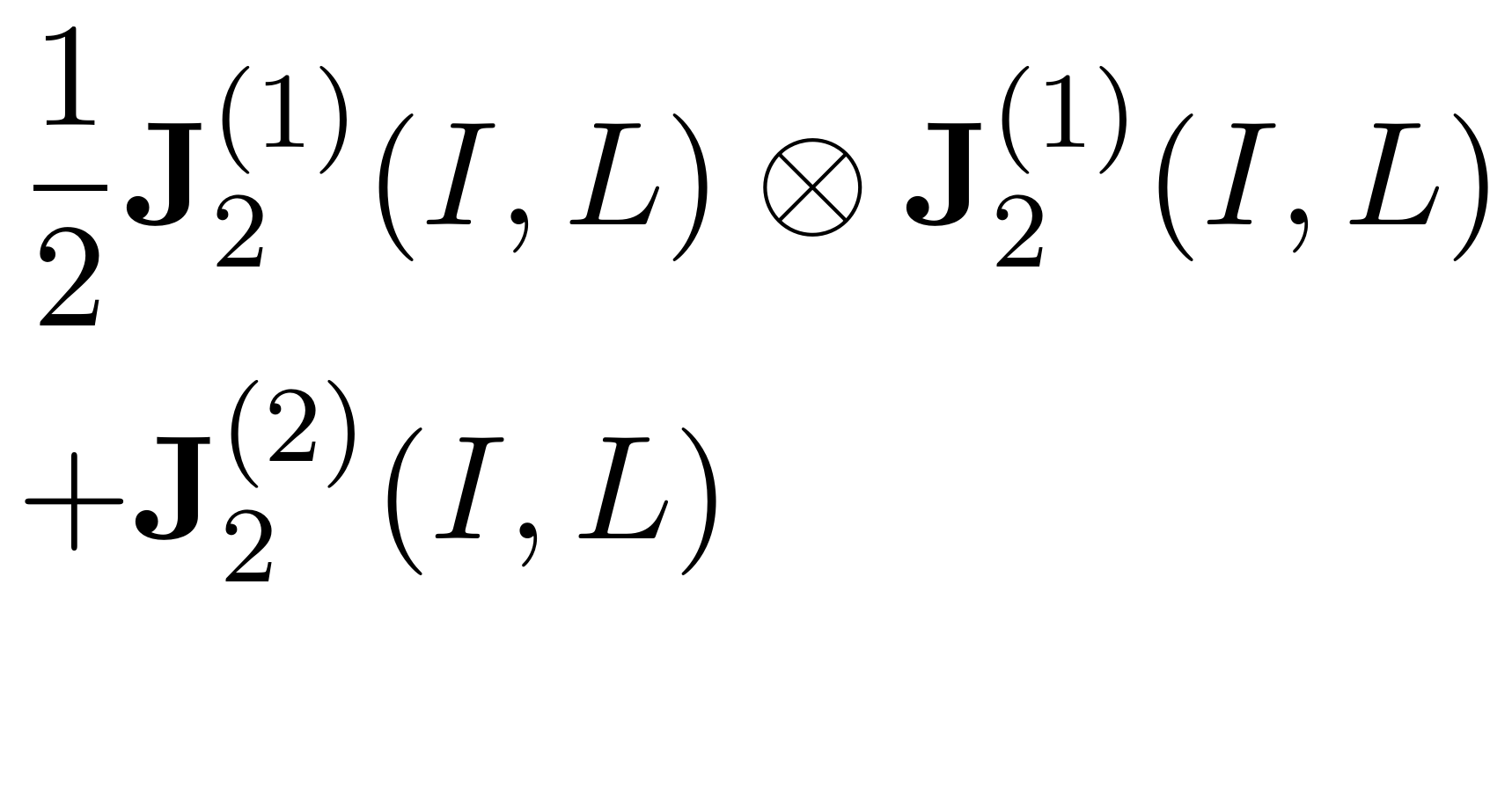}
\caption{NNLO double virtual structure.  Integration of the tree-level double real emission over the double unresolved phase space, plus the integration of the one-loop real-virtual contribution over the single unresolved phase space, generates the double virtual singularity structure described in Eq.~\eqref{eq:dsu}, for each colour connected dipole pair $I,L$.  The one-loop diagram on the right of this figure implicitly includes the $\beta_{0}$ term in Eq.~\eqref{eq:dsu}.}\label{fig:j2pic}
\end{figure}

The full double virtual subtraction term is given by the sum of $\dsigma_{NNLO}^{U,A}$ \eqref{eq:dsuA}, $\dsigma_{NNLO}^{U,B}$ \eqref{eq:dsuB} and $\dsigma_{NNLO}^{U,C}$ \eqref{eq:dsuC} so that,
\ba
&&\dsigma_{NNLO}^{U}=-\mps\biggl\{\nn\\
&&\ \ \bs{J}_{n+2}^{(1)}(1,\cdots,n+2)\biggl(M_{n+2}^{1}(1,\cdots,n+2)-\frac{\beta_{0}}{\eps}\ M_{n+2}^{0}(1,\cdots,n+2)\biggr)\nn\\
&&+ \frac{1}{2}\big[\bs{J}_{n+2}^{(1)}(1,\cdots,n+2)\otimes\bs{J}_{n+2}^{(1)}(1,\cdots,n+2)\big](z_{1},z_{2})\ M_{n+2}^{0}(1,\cdots,n+2)\nn\\
&&+\bs{J}_{n+2}^{(2)}(1,\cdots,n+2)\ M_{n+2}^{0}(1,\cdots,n+2)\biggr\} J_{n}^{(n)}(\{p\}_{n}).\label{eq:dsu}
\ea
This master equation encapsulates the IR poles generated by the integration of the tree-level double real emission over the double unresolved phase space, plus the integration of the one-loop real-virtual contribution over the single unresolved phase space and is summarised in Fig.~\ref{fig:j2pic}.  

As previously discussed, the first term in Eq.~\eqref{eq:dsu} reflects the pole structure of the first line in Eq.~\eqref{eq:catani2}.  The second term similarly reflects the structure of the second line in Eq.~\eqref{eq:catani2}.  The third term introduces the double unresolved integrated antenna string, $\bs{J}_{n+2}^{(2)}$, which is summed over colour-connected dipoles.  This term corresponds to the terms proportional to $\bs{I}_{n+2}^{(1)}(2\eps)$ and $\bs{H}^{(2)}(\eps)$ in the last two lines of Eq.~\eqref{eq:catani2}.  

Of course, it is known that for six or more partons, there are potentially multiple particle correlators in the $1/\e$ contribution.  Nevertheless, once the scale dependent functions proportional to $|s_{ij}|^{-\e}$ are expanded, the difference between 
Eq.~\eqref{eq:dsu} and the two-loop matrix elements is finite.

Note also that when adding together all contributions in the double-virtual channel, we observe that contributions of order $\epsilon$ or higher to the one-loop amplitude cancel, even though they are multiplied with divergent factors in individual terms (one-loop self-interference and integrated
real-virtual subtraction terms). This cancellation can be understood in detail from the NNLO infrared singularity structure~\cite{Catani:1998bh,Weinzierl:2011uz}. 

\section{Quark-antiquark scattering to gluonic dijets at NLO}
\label{sec:dijetnlo}

In this section we review the IR structure of gluonic dijet production via quark-antiquark scattering at NLO (to all orders in $N$ and setting $\NF=0$) using the antenna subtraction method. 

Here we first review our notation for amplitudes involving quarks.  In Secs.~\ref{sec:nloant} and \ref{sec:nnloant} the symbol $M_{n}^{l}$ was used to denote a generic $l$-loop, $n$-parton, squared partial amplitude.  In previous works~\cite{Pires:2010jv,GehrmannDeRidder:2011aa,GehrmannDeRidder:2012dg} the symbol $A_{n}^{l}$ was used to denote the $l$-loop, $n$-parton all-gluon squared partial amplitudes. Following this example, squared partial amplitudes involving one quark-antiquark pair and $(n-2)$ gluons will be represented by the symbol $B_{n}^{l}$ and we will reserve $C_{n}^{l}$ ($D_{n}^{l}$) for processes involving two non-identical (identical) quark-antiquark pairs and $(n-4)$ gluons respectively.

In our notation, the leading colour $l$-loop amplitude for $(m+2)$-parton scattering with one quark-antiquark pair in the initial-state is given by,
\ba
\lefteqn{\bs{\cal{B}}_{m+2}^{l}(\{p_{i},\lambda_{i},a_{i},i,j\})=2^{m/2}g^{m}\bigg(\frac{g^{2}NC(\eps)}{2}\bigg)^{l}\sum_{\sigma\in S_{m}}}\nn\\
&\times&\big(T^{a_{\sigma(1)}}\cdots T^{a_{\sigma(m)}}\big)_{ij}\ {\cal{B}}_{m+2}^{l}(\hat{1}_{q},\sigma(3)_{g},\cdots,\sigma(m+2)_{g},\hat{2}_{\b{q}}),
\ea
where the permutation sum $S_{m}$ is the group of permutations of $m$ symbols.  The $SU(N)$ algebra is normalised according to the convention, ${\rm{Tr}}(T^{a}T^{b})=\delta^{ab}/2$.  Squaring and summing over helicities and colours gives the leading colour  $l$-loop $q\b{q}\rightarrow m$ gluon contribution to the $n$-jet cross section,
\begin{eqnarray}
\label{eq:Nmaster}
\dsigma&=&{\cal N}_{m+2}^l 
{\rm d}\Phi_{m}(p_3,\dots,p_{m+2};p_1,p_2)\frac{1}{m!}\nonumber \\
&\times&
\sum_{\sigma\in S_{m}}B_{m+2}^l(\hat{1}_q,\sigma(3)_g,\dots,\sigma(m+2)_g,\hat{2}_{\bar q})\JET_n^{(m)}(\{p\}_{m}),
\end{eqnarray}
where $B_{m+2}^{l}$ denotes the $l$-loop, $(m+2)$-parton, colour-ordered helicity summed, squared partial amplitude.  The tree-level, one-loop and two-loop squared partial amplitudes are defined according to:
\ba
B_{m+2}^{0}(\sigma)&=&|{\cal{B}}_{m+2}^{0}(\sigma)|^{2},\nn\\
B_{m+2}^{1}(\sigma)&=&2{\rm{Re}}\big[{\cal{B}}_{m+2}^{0,\dagger}(\sigma){\cal{B}}_{m+2}^{1}(\sigma)\big],\nn\\
B_{m+2}^{2}(\sigma)&=&2{\rm{Re}}\big[{\cal{B}}_{m+2}^{0,\dagger}(\sigma){\cal{B}}_{m+2}^{2}(\sigma)\big]+|{\cal{B}}_{m+2}^{1}(\sigma)|^{2},\label{eq:medefs}
\ea
where $\sigma$ denotes a given colour ordering.  The normalisation factor ${\cal N}_{m+2}^l$ includes the average over initial spins and colours and is given by,
\begin{eqnarray}
{\cal N}_{m+2}^l &=& {\cal N}_{LO} \times \left(\frac{\alpha_s N}{2\pi}\right)^{m+l-2} \frac{\bar{C}(\epsilon)^{m+l-2}}{C(\epsilon)^{m-2}},\label{eq:nfac}
\label{eq:NNLORR}
\end{eqnarray}
where for the $q\b{q}\rightarrow gg$ process,
\begin{eqnarray}
{\cal N}_{LO} &=& \frac{1}{2s} \times \frac{1}{4N^2}\times \left(g^2 N\right)^{2} \frac{(N^2-1)}{N}.
\end{eqnarray}
The coupling $g^2$ has been converted into $\alpha_s$ using the factors $C(\epsilon)$ and $\bar{C}(\epsilon)$,
\ba
g^2 N C(\epsilon) = \left(\frac{\alpha_s N}{2\pi}\right) \bar{C}(\epsilon).
\ea
For low multiplicity matrix elements ($m\leq 3$), the sub-leading colour contributions can be written as an incoherent sum of squared partial amplitudes.

\subsection{Leading order cross section, $\bs{\dsigma_{LO}}$}

The leading order contribution to $q\b{q}\to gg$ scattering is given by,
\ba
\dsigma_{q\bar{q},LO}&=&{\cal{N}}_{LO}\ \int{\rm d}\Phi_{2}(p_3,p_{4};p_1,p_2)\ \frac{1}{2!} \sum_{P(i,j)}\nn\\
&&\biggl\{\biggl(B_{4}^{0}(\hat{1}_{q},i_{g},j_{g},\hat{2}_{\bar{q}})-\frac{1}{2N^{2}}\ \wt{B}_{4}^{0}(\hat{1}_{q},{i}_{g},{j}_{g},\hat{2}_{\bar{q}})\biggr)\ J_{2}^{(2)}(p_{3},p_{4})\biggr\},
\ea
where $P(i,j)$ denotes the two permutations of the gluons 3 and 4 in the colour ordering.  The sub-leading colour contribution is formed from the square of a coherent sum of colour-ordered matrix elements,
\ba
\wt{B}_{4}^{0}(\hat{1}_{q},i_{g},j_{g},\hat{2}_{\bar{q}})&=&|{\cal{B}}_{4}^{0}(\hat{1}_{q},i_{g},j_{g},\hat{2}_{\bar{q}})+{\cal{B}}_{4}^{0}(\hat{1}_{q},j_{g},i_{g},\hat{2}_{\bar{q}})|^{2}
\ea
Here the three gluon coupling involving gluons $i$ and $j$ drops out and this contribution is sometimes referred to as QED-like, or as involving Abelian gluons.

\subsection{Virtual cross section, $\bs{\dsigma_{NLO}^{V}}$}

The $\NF$ independent one-loop contribution to the $q\bar{q}\rightarrow gg$ subprocess is given by,
\ba
\dsigma_{q\bar{q},NLO}^{V}&=&{\cal{N}}_{4}^{1}\ \int{\rm d}\Phi_{2}(p_3,p_{4};p_1,p_2)\frac{{\rm{d}}x_{1}}{x_{1}}\frac{{\rm{d}}x_{2}}{x_{2}}\delta(1-x_{1})\delta(1-x_{2})\ \frac{1}{2} \sum_{P(i,j)}\nn\\
&\biggl\{&B_{4}^{1}(\hb{1}_{q},i_{g},j_{g},\hb{2}_{\bar{q}})-\frac{1}{N^{2}}\bigg[\wt{B_{4,a}^{1}}(\hb{1}_{q},i_{g},j_{g},\hb{2}_{\bar{q}})\nn\\
&+&\frac{1}{2}\wt{B_{4,b}^{1}}(\hb{1}_{q},i_{g},j_{g},\hb{2}_{\bar{q}})\bigg]+\frac{1}{2N^{4}}\wt{\wt{B_{4}^{1}}}(\hb{1}_{q},i_{g},j_{g},\hb{2}_{\bar{q}})\biggr\}\ J_{2}^{(2)}(p_{i},p_{j})\label{eq:oneloopdijet}
\ea
where the overall factor is,
\ba
{\cal{N}}_{4}^{1}&=&{\cal{N}}_{LO}\biggl(\frac{\alpha_{s}N}{2\pi}\biggr)\bar{C}(\eps).
\ea
The various contributions are formed from the projection of one-loop partial amplitudes onto tree-level amplitudes.  The first sub-leading colour contribution is split into two terms, $\wt{B_{4,a}^{1}}$ and $\wt{B_{4,b}^{1}}$ to reflect that fact that the explicit poles of these contributions factor onto different tree-level matrix elements.  The poles of these contributions may be expressed in terms of integrated antenna strings,
\ba
\Poles\biggl[B_{4}^{1}(\hb{1}_{q},i_{g},j_{g},\hb{2}_{\bar{q}})\biggr]&=&-\bs{J}_{4}^{(1)}(\hb{1}_{q},i_{g},j_{g},\hb{2}_{\bar{q}})\ B_{4}^{0}(\hb{1}_{q},i_{g},j_{g},\hb{2}_{\bar{q}}),\nn\\
\Poles\biggl[\wt{B}_{4,a}^{1}(\hb{1}_{q},i_{g},j_{g},\hb{2}_{\bar{q}})\biggr]&=&-\bs{J}_{2}^{(1)}(\hb{1}_{q},\hb{2}_{\bar{q}})\ B_{4}^{0}(\hb{1}_{q},i_{g},j_{g},\hb{2}_{\bar{q}}),\nn\\
\Poles\biggl[\wt{B}_{4,b}^{1}(\hb{1}_{q},{i}_{g},{j}_{g},\hb{2}_{\bar{q}})\biggr]&=&-\big(\bs{J}_{3}^{(1)}(\hb{1}_{q},i_{g},\hb{2}_{\bar{q}})+\bs{J}_{3}^{(1)}(\hb{1}_{q},j_{g},\hb{2}_{\bar{q}})\big)\ \wt{B}_{4}^{0}(\hb{1}_{q},{i}_{g},{j}_{g},\hb{2}_{\bar{q}})\nn\\
&&+\bs{J}_{2}^{(1)}(\hb{1}_{q},\hb{2}_{\bar{q}})\ \wt{B}_{4}^{0}(\hb{1}_{q},{i}_{g},{j}_{g},\hb{2}_{\bar{q}}),\nn\\
\Poles\biggl[\wt{\wt{B_{4}^{1}}}(\hb{1}_{q},i_{g},j_{g},\hb{2}_{\bar{q}})\biggr]&=&-\bs{J}_{2}^{(1)}(\hb{1}_{q},\hb{2}_{\bar{q}})\ \wt{B}_{4}^{0}(\hb{1}_{q},i_{g},j_{g},\hb{2}_{\bar{q}}).
\ea
Following the discussion of Sec.~\ref{sec:nloant}, the integrated antenna strings are formed from sums of integrated dipoles,
\ba
\bs{J}_{4}^{(1)}(\hb{1}_{q},i_{g},j_{g},\hb{2}_{\bar{q}})&=&\bs{J}_{2}^{(1)}(\hb{1}_{q},i_{g})+\bs{J}_{2}^{(1)}(i_{g},j_{g})+\bs{J}_{2}^{(1)}(j_{g},\hb{2}_{\b{q}}),\nn\\
\bs{J}_{3}^{(1)}(\hb{1}_{q},i_{g},\hb{2}_{\b{q}})&=&\bs{J}_{2}^{(1)}(\hb{1}_{q},i_{g})+\bs{J}_{2}^{(1)}(j_{g},\hb{2}_{\b{q}}),\nn\\
\bs{J}_{2}^{(1)}(j_{g},\hb{2}_{\b{q}})&=&\bs{J}_{2}^{(1)}(\hb{2}_{q},j_{g}),
\label{eq:j1dipole}
\ea
where the relevant integrated dipoles are given in Tabs.~\ref{tab:ffintsubterms} and~\ref{tab:ifintsubterms}.

Using these relations, the poles of the $\NF$ independent virtual cross section from the channel $q\bar{q}\rightarrow gg$ can be rewritten in the form,
\ba
\Poles\big(\dsigma_{q\bar{q},NLO}^{V}\big)&=&-{\cal{N}}_{4}^{1}\ \int{\rm d}\Phi_{2}(p_3,p_{4};p_1,p_2)\frac{{\rm{d}}x_{1}}{x_{1}}\frac{{\rm{d}}x_{2}}{x_{2}}\ \frac{1}{2} \sum_{P(i,j)}\biggl\{\nn\\
&\bigg[&\bs{J}_{4}^{(1)}(\hb{1}_{q},i_{g},j_{g},\hb{2}_{\bar{q}})\ B_{4}^{0}(\hb{1}_{q},i_{g},j_{g},\hb{2}_{\bar{q}})\bigg]\nn\\
-\frac{1}{N^{2}}&\bigg[&\bs{J}_{2}^{(1)}(\hb{1}_{q},\hb{2}_{\bar{q}})\ B_{4}^{0}(\hb{1}_{q},i_{g},j_{g},\hb{2}_{\bar{q}})+\bs{J}_{3}^{(1)}(\hb{1}_{q},i_{g},\hb{2}_{\bar{q}})\ \wt{B}_{4}^{0}(\hb{1}_{q},i_{g},j_{g},\hb{2}_{\bar{q}})\bigg]\nn\\
+\bigg(\frac{N^{2}+1}{2N^{4}}\bigg)&\bigg[&\bs{J}_{2}^{(1)}(\hb{1}_{q},\hb{2}_{\bar{q}})\ \wt{B}_{4}^{0}(\hb{1}_{q},{3}_{g},{4}_{g},\hb{2}_{\bar{q}})\bigg]\biggr\}\ J_{2}^{(2)}(p_{i},p_{j}).\label{eq:qqggpoles}
\ea

\subsection{Real cross section, $\bs{\dsigma_{NLO}^{R}}$}

The real emisson contribution to the $\NF$ independent $q\b{q}\to gg$ NLO cross section comes from the tree-level five-parton $q\b{q}\to ggg$ process,
\ba
\dsigma_{q\bar{q},NLO}^{R}&=&{\cal{N}}_{5}^{0}\ {\rm d}\Phi_{3}(p_3,p_{4},p_{5};p_1,p_2)\ \frac{1}{3!} \sum_{P(i,j,k)}\biggl\{\ \biggl[B_{5}^{0}(\hat{1}_{q},{i}_{g},j_{g},k_{g},\hat{2}_{\bar{q}})\nn\\
&-&\frac{1}{N^{2}}\ \wt{B}_{5}^{0}(\hat{1}_{q},i_{g},j_{g},k_{g},\hat{2}_{\bar{q}})+\frac{1}{3!}\biggl(\frac{N^{2}+1}{N^{4}}\biggr)\ \wt{\wt{B_{5}^{0}}}(\hat{1}_{q},{i}_{g},{j}_{g},{k}_{g},\hat{2}_{\bar{q}})\biggr]\ J_{2}^{(3)}(p_{i},p_{j},p_{k})\biggr\},\nn\\\label{eq:m50full}
\ea
where $P(i,j,k)$ is the set of six permutations of the gluons in the colour ordering and the overall factor is given by,
\ba
{\cal{N}}_{5}^{0}&=&{\cal{N}}_{LO}\biggl(\frac{\alpha_{s}N}{2\pi}\biggr)\frac{\bar{C}(\eps)}{C(\eps)}.
\ea
The sub-leading colour matrix element $\wt{B}_{5}^{0}(\hat{1}_{q},i_{g},j_{g},k_{g},\hat{2}_{\bar{q}})$ is given by,
\ba
 &&\wt{B_{5}^{0}}(\hat{1}_{q},{i}_{g},j_{g},k_{g},\hat{2}_{\bar{q}})=\nn\\
 &&\hspace{1cm}|{\cal{B}}_{5}^{0}(\hat{1}_{q},i_{g},j_{g},k_{g},\hat{2}_{\bar{q}})+{\cal{B}}_{5}^{0}(\hat{1}_{q},j_{g},i_{g},k_{g},\hat{2}_{\bar{q}})+{\cal{B}}_{5}^{0}(\hat{1}_{q},j_{g},k_{g},i_{g},\hat{2}_{\bar{q}})|^{2},
\ea
such that the gluon $i$ behaves in an Abelian fashion.  The most sub-leading contribution is given by the QED-like matrix element, formed by averaging over all colour-ordered matrix elements,
\ba
\wt{ \wt{B_{5}^{0}}}(\hat{1}_{q},{i}_{g},{j}_{g},{k}_{g},\hat{2}_{\bar{q}})&=&\biggl|\sum_{P(i,j,k)}{\cal{B}}_{5}^{0}(\hat{1}_{q},i_{g},j_{g},k_{g},\hat{2}_{\bar{q}})\biggr|^{2}.
\ea
All gluons in this contribution are QED-like and are only colour connected to the quark-antiquark endpoints.

\subsection{Real radiation subtraction term, $\bs{\dsigma_{NLO}^{S}}$}

The subtraction term for the real cross section presented in Eq.~\eqref{eq:m50full} is given by,
\ba
\dsigma_{q\bar{q},NLO}^{S}&=&{\cal{N}}_{5}^{0}\ {\rm d}\Phi_{3}(p_3,p_{4},p_{5};p_1,p_2)\ \frac{1}{3!} \sum_{P(i,j,k)}\ \biggl\{\nn\\
&\bigg[&d_{3}^{0}(\hat{1},i,j)\ \Mtreeqggqb(\hat{\bar{1}},\wt{(ij)},k,\hat{2})\ J_{2}^{(2)}(p_{(ij)},p_{k})\nn\\
&+&f_{3}^{0}(i,j,k)\ \Mtreeqggqb(\hat{1},\wt{(ij)},\wt{(jk)},\hat{2})\ J_{2}^{(2)}(p_{(ij)},p_{(jk)})\nn\\
&+&d_{3}^{0}(\hat{2},k,j)\ \Mtreeqggqb(\hat{1},i,\wt{(jk)},\hat{\bar{2}})\ J_{2}^{(2)}(p_{i},p_{(jk)})\biggr]\nn\\
-\frac{1}{N^{2}}&\biggl[&A_{3}^{0}(\hat{1},i,\hat{2})\ \Mtreeqggqb(\hat{\bar{1}},\tilde{j},\tilde{k},\hat{\bar{2}})\ J_{2}^{(2)}(p_{j},p_{k})\nn\\
&+&d_{3}^{0}(\hat{1},j,k)\ \Mttreeqggqb(\hat{\bar{1}},i,\wt{(jk)},\hat{2})\ J_{2}^{(2)}(p_{i},p_{(jk)})\nn\\
&+&d_{3}^{0}(\hat{2},k,j)\ \Mttreeqggqb(\hat{1},i,\wt{(jk)},\hat{\bar{2}})\ J_{2}^{(2)}(p_{i},p_{(jk)})\biggr]\nn\\
+\frac{1}{3!}\biggl(\frac{N^{2}+1}{N^{4}}\biggr)&\biggl[&A_{3}^{0}(\hat{1},i,\hat{2})\ \Mttreeqggqb(\hat{\bar{1}},\tilde{j},\tilde{k},\hat{\bar{2}})\ J_{2}^{(2)}(p_{j},p_{k})\nn\\
&+&A_{3}^{0}(\hat{1},j,\hat{2})\ \Mttreeqggqb(\hat{\bar{1}},\tilde{i},\tilde{k},\hat{\bar{2}})\ J_{2}^{(2)}(p_{i},p_{k})\nn\\
&+&A_{3}^{0}(\hat{1},k,\hat{2})\ \Mttreeqggqb(\hat{\bar{1}},\tilde{i},\tilde{j},\hat{\bar{2}})\ J_{2}^{(2)}(p_{i},p_{j})\biggr]\biggr\}.\label{eq:lcqqgg}
\ea

\subsection{Virtual subtraction term,  $\bs{\dsigma_{NLO}^{T}}$}

Integrating the real subtraction term over the single unresolved phase space and combining with the appropriate NLO mass factorisation kernels allows the virtual subtraction term to be constructed from integrated antenna strings,
\ba
\dsigma_{q\bar{q},NLO}^{T}&=&-{\cal{N}}_{4}^{1}\ {\rm d}\Phi_{2}(p_3,p_{4};p_1,p_2)\ \frac{1}{2} \sum_{P(i,j)}\ \int\frac{\text{d}x_{1}}{x_{1}}\frac{\text{d}x_{2}}{x_{2}}\bigg\{\nn\\
&&\bs{J}_{4}^{(1)}(\hat{\bar{1}}_{q},i_{g},j_{g},\hat{\bar{2}}_{\bar{q}})\ \Mtreeqggqb(\hat{\bar{1}},i,j,\hat{\bar{2}})\nn\\
-\frac{1}{N^{2}}&\biggl[&\bs{J}_{2}^{(1)}(\hat{\bar{1}}_{q},\hat{\bar{2}}_{\bar{q}})\ \Mtreeqggqb(\hat{\bar{1}},i,j,\hat{\bar{2}})+\bs{J}_{3}^{(1)}(\hat{\bar{1}}_{q},i_{g},\hat{\bar{2}}_{\bar{q}})\ \Mttreeqggqb(\hat{\bar{1}},i,j,\hat{\bar{2}})\biggr]\nn\\
+\biggl(\frac{N^{2}+1}{2N^{4}}\biggr)&\biggl[&\bs{J}_{2}^{(1)}(\hat{\bar{1}}_{q},\hat{\bar{2}}_{\bar{q}})\ \Mttreeqggqb(\hat{\bar{1}},i,j,\hat{\bar{2}})\biggr]\biggr\}\ J_{2}^{(2)}(p_{i},p_{j}),
\ea
From the definition of these strings the connection to the unintegrated subtraction term in Eq.~\eqref{eq:lcqqgg} is clear.  This expression exactly matches that of Eq.~\eqref{eq:qqggpoles}, i.e., 
\ba
\Poles\big(\dsigma_{q\bar{q},NLO}^{V}\big)-\Poles\big(\dsigma_{q\bar{q},NLO}^{T}\big)&=&0.
\ea


\section{Quark-antiquark scattering to dijets at NNLO at leading colour}
\label{sec:dijetnnlo}

The leading colour NNLO contribution to dijet production has previously been studied using the antenna subtraction formalism for purely gluonic channels at the double real~\cite{Pires:2010jv}, real-virtual~\cite{GehrmannDeRidder:2011aa} and double virtual~\cite{GehrmannDeRidder:2012dg} levels.

This section will focus on the leading colour NNLO corrections to dijet production from quark-antiquark scattering.  These corrections include the double real tree-level contribution $q\bar{q}\rightarrow gggg$, the real-virtual one-loop contribution $q\bar{q}\rightarrow ggg$ and the double virtual two loop contribution $q\bar{q}\rightarrow g g$, all of which will be examined using the methods of Sec.~\ref{sec:nnloant}.

\subsection{Double real cross section, $\bs{\dsigma_{NNLO}^{RR}}$}
\label{sec:sixparton}

The leading colour contribution to the cross section from the $q\b{q}\to gggg$ subprocess can be written in the form,
\begin{eqnarray}
\label{eq:RIIFFFF}
\dsigma_{q\bar{q},NNLO}^{RR}&=&{\cal{N}}_{6}^{0}\ {\rm d}\Phi_{4}(p_{3},\dots,p_{6};p_{1},p_{2})\ \frac{1}{4!}  \sum_{P(i,j,k,l)}\nn\\
&\times&\ B_{6}^{0}(\hat{1}_{q},i_{g},j_{g},k_{g},l_{g},\hat{2}_{\bar{q}})\, J^{(4)}_{2}(p_{i},p_{j},p_{k},p_{l}),
\end{eqnarray}
where $P(i,j,k,l)$ denotes the 24 possible permutations of gluons in the colour ordering and the overall factor is given by,
\ba
{\cal{N}}_{6}^{0}&=&{\cal{N}}_{LO}\biggl(\frac{\alpha_{s}N}{2\pi}\biggr)^{2}\frac{\bar{C}(\eps)^{2}}{C(\eps)^{2}}.
\ea
It is convenient to rearrange the 24 squared amplitudes present in Eq.~\eqref{eq:RIIFFFF} into three terms,
\begin{eqnarray}
\sum_{P(i,j,k,l)}
M^0_{6}(\hat{1}_q,i_g,j_g,k_g,l_g,\hat{2}_{\bar q})
&=& X^0_{6}(\hat{1}_q,3_g,4_g,5_g,6_g,\hat{2}_{\bar q})\nonumber\\ 
&+& X^0_{6}(\hat{1}_q,3_g,5_g,4_g,6_g,\hat{2}_{\bar q})\nonumber\\ 
&+& X^0_{6}(\hat{1}_q,3_g,4_g,6_g,5_g,\hat{2}_{\bar q}),
\end{eqnarray}
where each $X_6^0$ contains eight colour ordered squared amplitudes obtained by the four cyclic permutations of the final state gluons and their line reversals:
\begin{eqnarray}
X_6^0(\hat{1}_q,3_g,4_g,5_g,6_g,\hat{2}_{\bar q})&=&
B_{6}^0(\hat{1}_q,3_g,4_g,5_g,6_g,\hat{2}_{\bar q})+B_{6}^0(\hat{1}_q,6_g,5_g,4_g,3_g,\hat{2}_{\bar q})\nonumber\\
&+&B_{6}^0(\hat{1}_q,4_g,5_g,6_g,3_g,\hat{2}_{\bar q})+B_{6}^0(\hat{1}_q,3_g,6_g,5_g,4_g,\hat{2}_{\bar q})\nonumber\\
&+&B_{6}^0(\hat{1}_q,5_g,6_g,3_g,4_g,\hat{2}_{\bar q})+B_{6}^0(\hat{1}_q,4_g,3_g,6_g,5_g,\hat{2}_{\bar q})\nonumber\\
&+&B_{6}^0(\hat{1}_q,6_g,3_g,4_g,5_g,\hat{2}_{\bar q})+B_{6}^0(\hat{1}_q,5_g,4_g,3_g,6_g,\hat{2}_{\bar q}).
\label{eq:IIFFFF}
\end{eqnarray}
It is sufficient to apply the subtraction technique to one block of orderings, the other two are related by symmetry and contribute numerically the same result after integration over the phase space. 

\subsection{Real-virtual cross section, $\bs{\dsigma_{NNLO}^{RV}}$}

The leading colour one-loop contribution to the real-virtual cross section for the $q\b{q}\to ggg$ subprocess is given by,
\ba
\dsigma_{q\bar{q},NNLO}^{RV}&=&{\cal{N}}_{5}^{1}\int{\rm d}\Phi_{2}(p_3,p_{4},p_{5};p_1,p_2)\ \frac{1}{3!} \frac{{\rm{d}}x_{1}}{x_{1}}\frac{{\rm{d}}x_{2}}{x_{2}}\sum_{P(i,j,k)}\nn\\
&\times&B_{5}^{1}(\hb{1}_{q},i_{g},j_{g},k_{g},\hb{2}_{\bar{q}})\delta(1-x_{1})\delta(1-x_{2})\ J_{2}^{(2)}(p_{i},p_{j},p_{k}).
\ea
where $P(i,j,k)$ denotes the set of six permutations of gluons in the colour ordering and the overall factor can be derived from \eqref{eq:nfac},
\ba
{\cal{N}}_{5}^{1}&=&{\cal{N}}_{LO}\biggl(\frac{\alpha_{s}N}{2\pi}\biggr)^{2}\frac{\bar{C}(\eps)^{2}}{C(\eps)},
\ea
and
\ba
{\cal{N}}_{NNLO}^{RV} = {\cal{N}}_{LO} \left(\frac{\alpha_s N}{2\pi}\right)^2 \frac{\bar{C}(\eps)^2}{C(\eps)} = {\cal{N}}_{NNLO}^{RV} C(\eps).
\ea

\subsection{Double virtual cross section, $\bs{\dsigma_{NNLO}^{VV}}$}

The leading colour contribution to the two-loop $q\b{q}\to gg$ cross section is given by,
\ba
\dsigma_{q\bar{q},NNLO}^{VV}&=&{\cal{N}}_{4}^{2}\int {\rm d}\Phi_{2}(p_3,p_{4};p_1,p_2)\ \frac{1}{2!} \frac{{\rm{d}}z_{1}}{z_{1}}\frac{{\rm{d}}z_{2}}{z_{2}}\sum_{P(i,j)}\nn\\
&\times&B_{4}^{2}(\hb{1}_{q},i_{g},j_{g},\hb{2}_{\bar{q}})\delta(1-x_{1})\delta(1-x_{2})\ J_{2}^{(2)}(p_{i},p_{j}),
\ea
where the overall factor is,
\ba
{\cal{N}}_{4}^{2}&=&{\cal{N}}_{LO}\biggl(\frac{\alpha_{s}N}{2\pi}\biggr)^{2}\bar{C}(\eps)^{2}.
\ea
As in~\cite{GehrmannDeRidder:2011aa,GehrmannDeRidder:2012dg} and Eq.~\eqref{eq:medefs}, the two-loop matrix element contains the projection of the two-loop amplitude onto the tree-level amplitude and the self-interference of the one-loop amplitude.

\subsection{Double real subtraction term, $\bs{\dsigma_{NNLO}^{S}}$}

The leading colour contribution to the squared matrix element is an incoherent sum of squared colour-ordered partial amplitudes.  The sum over colour orderings can be partitioned into three blocks of orderings as described in Sec.~\ref{sec:sixparton}, such that a subtraction term may be constructed for a block of orderings, rather than the entire squared matrix element.  Following the general discussion of Sec.~\ref{sec:nnloant}, the NNLO subtraction term for the block of orderings in Eq.~\eqref{eq:IIFFFF} is given by,
\begin{eqnarray}
&&\dsigma_{NNLO}^{S,X_6}={\cal{N}}_{6}^{0}\, {\rm d}\Phi_{4}(p_{3},\dots,p_{6};p_{1},p_{2})\ \frac{1}{4!}\, \sum_{P_{C}(ijkl)}\ \Biggl\{\nn\\
%
%
%
%
\ph{a2}&&+f_3^0(i,j,k)\,\Mtreeqgggqb(\hat{1},\widetilde{(i,j)},\widetilde{(j k)},l,\hat{2})\,J_2^{(3)}(p_{(ij)},p_{(jk)},p_{l})\nn \\ 
\ph{a3}&&+f_3^0(j,k,l)\,\Mtreeqgggqb(\hat{1},i,\widetilde{(j k)},\widetilde{(k l)},\hat{2})\,J_2^{(3)}(p_{(i)},p_{(j k)},p_{(k l)})\nn \\ 
\ph{a1}&&+d_3^0(\hat{1},i,j)\,\Mtreeqgggqb(\hat{\bar{1}},\widetilde{(i j)},k,l,\hat{2})\,J_2^{(3)}(p_{(i j)},p_{k},p_{l})\nn \\ 
\ph{a4}&&+d_3^0(\hat{2},l,k)\,\Mtreeqgggqb(\hat{1},i,j,\widetilde{(k l)},\hat{\bar{2}})\,J_2^{(3)}(p_{i},p_{j},p_{(k l)})\nn \\ 
\nn\\
\ph{a6}&&+f_3^0(l,k,j)\,\Mtreeqgggqb(\hat{1},\widetilde{(l k)},\widetilde{(k j)},i,\hat{2})\,J_2^{(3)}(p_{(l k)},p_{(k j)},p_{i})\nn \\ 
\ph{a7}&&+f_3^0(k,j,i)\,\Mtreeqgggqb(\hat{1},l,\widetilde{(k j)},\widetilde{(j i)},\hat{2})\,J_2^{(3)}(p_{l},p_{(k j)},p_{(ji)})\nn \\ 
\ph{a5}&&+d_3^0(\hat{1},l,k)\,\Mtreeqgggqb(\hat{\bar{1}},\widetilde{(l k)},j,i,\hat{2})\,J_2^{(3)}(p_{(l k)},p_{j},p_{i})\nn \\ 
\ph{a8}&&+d_3^0(\hat{2},i,j)\,\Mtreeqgggqb(\hat{1},l,k,\widetilde{(j i)},\hat{\bar{2}})\,J_2^{(3)}(p_{l},p_{k},p_{(j i)})\nn \\\nn\\
%
%
%
%
\ph{a9}&&+F_{4,a}^0(i,j,k,l)\,\Mtreeqggqb(\hat{1},\widetilde{(i j k)},\widetilde{(j k l)},\hat{2})\,J_2^{(2)}(p_{(i j k)},p_{(j k l)})\nn \\ 
\ph{a12}&&+F_{4,b}^0(i,j,l,k)\,\Mtreeqggqb(\hat{1},\widetilde{(i j l)},\widetilde{(j l k)},\hat{2})\,J_2^{(2)}(p_{(i j k)},p_{(j l k)})\nn \\ 
\ph{a14}&&+F_{4,a}^0(l,k,j,i)\,\Mtreeqggqb(\hat{1},\widetilde{(l k j)},\widetilde{(k j i)},\hat{2})\,J_2^{(2)}(p_{(l k j)},p_{(k j i)})\nn \\ 
\ph{a17}&&+F_{4,b}^0(l,k,i,j)\,\Mtreeqggqb(\hat{1},\widetilde{(l k i)},\widetilde{(k i j)},\hat{2})\,J_2^{(2)}(p_{(l k i)},p_{(k i j)})\nn \\ 
\ph{a19}&&+D_4^0(\hat{1},i,j,k)\,\Mtreeqggqb(\hat{\bar{1}},\widetilde{(i j k)},l,\hat{2})\,J_2^{(2)}(p_{(i j k)},p_{l})\nn \\ 
\ph{a25}&&+D_4^0(\hat{2},l,k,j)\,\Mtreeqggqb(\hat{1},i,\widetilde{(l k j)},\hat{\bar{2}})\,J_2^{(2)}(p_{i},p_{(l k j)})\nn \\ 
\ph{a31}&&-\tilde{A}_4^0(\hat{1},i,k,\hat{2})\,\Mtreeqggqb(\hat{\bar{1}},\tilde{j},\tilde{l},\hat{\bar{2}})\,J_2^{(2)}(p_{\tilde{j}},p_{\tilde{l}})\nn \\\nn\\
%
%
%
%
\ph{a10}&&-f_3^0(i,j,k)\,f_3^0(\widetilde{(i j)},\widetilde{(j k)},l)\,\Mtreeqggqb(\hat{1},\widetilde{((i j)(j k))},\widetilde{((j k) l)},\hat{2})\,J_2^{(2)}(p_{((i j)(j k))},p_{((j k) l)})\nn \\ 
\ph{a11}&&-f_3^0(j,k,l)\,f_3^0(i,\widetilde{(j k)},\widetilde{(k l)})\,\Mtreeqggqb(\hat{1},\widetilde{(i(j k))},\widetilde{((j k)(k l))},\hat{2})\,J_2^{(2)}(p_{((i j)(j k))},p_{((j k) l)})\nn \\
\ph{a13}&&-\,f_3^0(i,j,k)\,f_3^0(\widetilde{(ij)},l,\widetilde{(j k)})\,\Mtreeqggqb(\hat{1},\widetilde{((i j) l)},\widetilde{((j k) l)},\hat{2})\,J_2^{(2)}(p_{((i j) l)},p_{((j k) l)})\nn \\ \nn\\
\ph{a15}&&-f_3^0(l,k,j)\,f_3^0(\widetilde{(l k)},\widetilde{(kj)},i)\,\Mtreeqggqb(\hat{1},\widetilde{((l k)(k j))},\widetilde{((k j) i)},\hat{2})\,J_2^{(2)}(p_{((l k)(k j))},p_{((k j) i)})\nn \\ 
\ph{a16}&&-f_3^0(k,j,i)\,f_3^0(l,\widetilde{(k j)},\widetilde{(j i)})\,\Mtreeqggqb(\hat{1},\widetilde{(l(k j))},\widetilde{((k j)(j i))},\hat{2})\,J_2^{(2)}(p_{(kj)},p_{((k j)(j i))})\nn \\ 
\ph{a18}&&-\,f_3^0(l,k,j)\,f_3^0(\widetilde{(l k)},i,\widetilde{(k j)})\,\Mtreeqggqb(\hat{1},\widetilde{((l k) i)},\widetilde{((k j) i)},\hat{2})\,J_2^{(2)}(p_{((l k)i)},p_{((k j) i)})\nn \\ \nn\\
\ph{a20}&&-d_3^0(\hat{1},i,j)\,D_3^0(\hat{\bar{1}},\widetilde{(ij)},k)\,\Mtreeqggqb(\hat{\bar{\bar{1}}},\widetilde{((i j) k)},l,\hat{2})\,J_2^{(2)}(p_{((i j), k)},p_{l})\nn \\ 
\ph{a21}&&-f_3^0(i,j,k)\,D_3^0(\hat{1},\widetilde{(i, j)},\widetilde{(j k)})\,\Mtreeqggqb(\hat{\bar{1}},\widetilde{((i j)(j k))},l,\hat{2})\,J_2^{(2)}(p_{((i j)(j k))},p_{l})\nn \\ 
\ph{a22}&&-d_3^0(\hat{1},k,j)\,D_3^0(\hat{\bar{1}},\widetilde{(kj)},i)\,\Mtreeqggqb(\hat{\bar{\bar{1}}},\widetilde{((k j) i)},l,\hat{2})\,J_2^{(2)}(p_{((k j) i)},p_{l})\nn \\ \nn\\
\ph{a26}&&-d_3^0(\hat{2},l,k)\,D_3^0(\hat{\bar{2}},\widetilde{(l k)},j)\,\Mtreeqggqb(\hat{1},i,\widetilde{((l k) j)},\hat{\bar{\bar{2}}})\,J_2^{(2)}(p_{((l k) j)},p_{i})\nn \\ 
\ph{a27}&&-f_3^0(l,k,j)\,D_3^0(\hat{2},\widetilde{(l k)},\widetilde{(k j)})\,\Mtreeqggqb(\hat{1},i,\widetilde{((lk)(k j))},\hat{\bar{2}})\,J_2^{(2)}(p_{i},p_{((lk)(k j))})\nn \\ 
\ph{a28}&&-d_3^0(\hat{2},j,k)\,D_3^0(\hat{\bar{2}},\widetilde{(j k)},l)\,\Mtreeqggqb(\hat{1},i,\widetilde{((j k) l)},\hat{\bar{\bar{2}}})\,J_2^{(2)}(p_{i},p_{((j k) l)})\nn \\ \nn\\
\ph{a32}&&+\,A_3^0(\hat{1},i,\hat{2})\,A_3^0(\hat{\bar{1}},\tilde{k},\hat{\bar{2}})\,\Mtreeqggqb(\hat{\bar{\bar{1}}},\tilde{\tilde{j}},\tilde{\tilde{l}},\hat{\bar{\bar{2}}})\,J_2^{(2)}(p_{\tilde{j}},p_{\tilde{l}})\nn \\ 
\ph{a33}&&+\,A_3^0(\hat{1},k,\hat{2})\,A_3^0(\hat{\bar{1}},\tilde{i},\hat{\bar{2}})\,\Mtreeqggqb(\hat{\bar{\bar{1}}},\tilde{\tilde{j}},\tilde{\tilde{l}},\hat{\bar{\bar{2}}})\,J_2^{(2)}(p_{\tilde{\tilde{j}}},p_{\tilde{\tilde{l}}})\nn \\\nn\\
%
%
%
%
\ph{a13}&&+\frac{1}{2}\,f_3^0(i,l,k)\,f_3^0(\widetilde{(i l)},j,\widetilde{(lk)})\,\Mtreeqggqb(\hat{1},\widetilde{((i l) j)},\widetilde{((lk) j)},\hat{2})\,J_2^{(2)}(p_{((i l) j)},p_{((l k) j)})\nn \\ 
\ph{a42}&&-\frac{1}{2}\,d_3^0(\hat{1},l,i)\,f_3^0(\widetilde{(li)},j,k)\,\Mtreeqggqb(\hat{\bar{1}},\widetilde{((il) j)},\widetilde{(jk)},\hat{2})\,J_2^{(2)}(p_{((il) j)},p_{(jk)})\nn \\ 
\ph{a44}&&-\frac{1}{2}\,d_3^0(\hat{2},l,k)\,f_3^0(i,j,\widetilde{(k l)})\,\Mtreeqggqb(\hat{1},\widetilde{(i j)},\widetilde{(j(k l))},\hat{\bar{2}})\,J_2^{(2)}(p_{(i j)},p_{(j(k l))})\nn \\ 
&&-\frac{1}{2}\biggl[\ \bigl(S_{(il)l(lk)}-S_{((il)j)l(j(lk))}\bigr)
-\bigl(S_{1l(il)}-S_{1l((il)j)}\bigr)
-\bigl(S_{2l(lk)}-S_{2l(j(lk))}\bigr)\ \biggr]\nn\\
&&\times f_{3}^{0}(\widetilde{(il)},j,\widetilde{(lk)})\ 
\Mtreeqggqb(\hat{1},\widetilde{((il)j)},\widetilde{(j(lk))},\hat{2})\ J_{2}^{(2)}(p_{((il)j)},p_{(j(lk))})\nn\\
\nn\\
\ph{a18}&&+\frac{1}{2}\,f_3^0(l,i,j)\,f_3^0(\widetilde{(li)},k,\widetilde{(ij)})\,\Mtreeqggqb(\hat{1},\widetilde{((li) k)},\widetilde{((ij) k)},\hat{2})\,J_2^{(2)}(p_{((li) k)},p_{((ij) k)})\nn \\ 
\ph{a46}&&-\frac{1}{2}\,d_3^0(\hat{1},i,l)\,f_3^0(\widetilde{(il)},k,j)\,\Mtreeqggqb(\hat{\bar{1}},\widetilde{((il) k)},\widetilde{(k j)},\hat{2})\,J_2^{(2)}(p_{((il) k)},p_{(kj)})\nn \\ 
\ph{a48}&&-\frac{1}{2}\,d_3^0(\hat{2},i,j)\,f_3^0(l,k,\widetilde{(i j)})\,\Mtreeqggqb(\hat{1},\widetilde{(l k)},\widetilde{(k (i j))},\hat{\bar{2}})\,J_2^{(2)}(p_{(l k)},p_{(k(i j))})\nn \\ 
&&-\frac{1}{2}\biggl[\ \bigl(S_{(li)i(ij)}-S_{((li)k)i(k(ij))}\bigr)
-\bigl(S_{1i(li)}-S_{1i((li)k)}\bigr)
-\bigl(S_{2i(ij)}-S_{2i(k(ij))}\bigr)\ \biggr]\nn\\
&&\times f_{3}^{0}(\widetilde{(li)},k,\widetilde{(ij)})\ 
\Mtreeqggqb(\hat{1},\widetilde{((li)k)},\widetilde{(k(ij))},\hat{2})\ J_{2}^{(2)}(p_{((li)k)},p_{(k(ij))})\nn\\\nn\\
\ph{a24}&&+\frac{1}{2}\,d_3^0(\hat{1},k,j)\,d_3^0(\hat{\bar{1}},i,\widetilde{(k j)})\,\Mtreeqggqb(\hat{\bar{\bar{1}}},\widetilde{((k j) i)},l,\hat{2})\,J_2^{(2)}(p_{((k j) i)},p_{l})\nn \\ 
\ph{a43}&&-\frac{1}{2}\,f_3^0(j,k,l)\,d_3^0(\hat{1},i,\widetilde{(j k)})\,\Mtreeqggqb(\hat{\bar{1}},\widetilde{(i(j k))},\widetilde{(k l)},\hat{2})\,J_2^{(2)}(p_{(i(j k))},p_{(k l)})\nn \\
\ph{a36}&&-\frac{1}{2}\,A_3^0(\hat{1},k,\hat{2})\,d_3^0(\hat{\bar{1}},\tilde{i},\tilde{j})\,\Mtreeqggqb(\hat{\bar{\bar{1}}},\widetilde{(i j)},\tilde{l},\hat{\bar{2}})\,J_2^{(2)}(p_{(i, j)},p_{\tilde{l}})\nn \\ 
&&-\frac{1}{2}\biggl[\ \bigl(S_{1k(jk)}-S_{\bar{1}k(i(jk))}\bigr)
-\bigl(S_{(jk)k(kl)}-S_{(i(jk))k(kl)}\bigr)
- \underbrace{\bigl(S_{1k2}-S_{\bar{1}k2}\bigr)}_{0}
\ \biggr]\nn\\
&&\times d_{3}^{0}(\hat{1},i,\widetilde{(jk)})\ 
\Mtreeqggqb(\hat{\bar{1}},\widetilde{(i(jk))},\widetilde{(kl)},\hat{2})\ J_{2}^{(2)}(p_{(i(jk))},p_{(kl)})\nn\\\nn\\
\ph{a23}&&+\frac{1}{2}\,d_3^0(\hat{1},i,j)\,d_3^0(\hat{\bar{1}},k,\widetilde{(i j)})\,\Mtreeqggqb(\hat{\bar{\bar{1}}},\widetilde{((i j) k)},l,\hat{2})\,J_2^{(2)}(p_{((i j) k)},p_{l})\nn \\  
\ph{a47}&&-\frac{1}{2}\,f_3^0(j,i,l)\,d_3^0(\hat{1},k,\widetilde{(ji)})\,\Mtreeqggqb(\hat{\bar{1}},\widetilde{(k(ji))},\widetilde{(il)},\hat{2})\,J_2^{(2)}(p_{(k,(j,i))},p_{(i,l)})\nn \\ 
\ph{a34}&&-\frac{1}{2}\,A_3^0(\hat{1},i,\hat{2})\,d_3^0(\hat{\bar{1}},\tilde{k},\tilde{j})\,\Mtreeqggqb(\hat{\bar{\bar{1}}},\widetilde{(kj)},\tilde{l},\hat{\bar{2}})\,J_2^{(2)}(p_{(kj)},p_{\tilde{l}})\nn \\ 
&&-\frac{1}{2}\biggl[\ \bigl(S_{1i(ji)}-S_{\bar{1}i(k(ji))}\bigr)
-\bigl(S_{(k(ji))i(il)}-S_{(ji)i(il)}\bigr)- \underbrace{\bigl(S_{1i2}-S_{\bar{1}i2}\bigr)}_{0}
\ \biggr]\nn\\
&&\times d_{3}^{0}(\hat{1},k,\widetilde{(ji)})\ \Mtreeqggqb(\hat{\bar{1}},\widetilde{(k(ji))},\widetilde{(il)},\hat{2})\ J_{2}^{(2)}(p_{(k(ji))},p_{(il)})\nn\\\nn\\
\ph{a30}&&+\frac{1}{2}\,d_3^0(\hat{2},j,k)\,d_3^0(\hat{\bar{2}},l,\widetilde{(jk)})\,\Mtreeqggqb(\hat{1},i,\widetilde{((j k)l)},\hat{\bar{\bar{2}}})\,J_2^{(2)}(p_{i},p_{((j k) l)})\nn \\ 
\ph{a45}&&-\frac{1}{2}\,f_3^0(i,j,k)\,d_3^0(\hat{2},l,\widetilde{(j k)})\,\Mtreeqggqb(\hat{1},\widetilde{(i j)},\widetilde{(l (j k))},\hat{\bar{2}})\,J_2^{(2)}(p_{(i, j)},p_{(l,(j, k))})\nn \\ 
\ph{a40}&&-\frac{1}{2}\,A_3^0(\hat{1},j,\hat{2})\,d_3^0(\hat{\bar{2}},\tilde{l},\tilde{k})\,\Mtreeqggqb(\hat{\bar{1}},\tilde{i},\widetilde{(lk)},\hat{\bar{\bar{2}}})\,J_2^{(2)}(p_{\tilde{i}},p_{(lk)})\nn \\ 
&&-\frac{1}{2}\biggl[\ \bigl(S_{2j(jk)}-S_{\bar{2}j(l(jk))}\bigr)
-\bigl(S_{(ij)j(jk)}-S_{(ij)j((jk)l)}\bigr)
- \underbrace{\bigl(S_{1j2}-S_{1j\bar{2}}\bigr)}_{0}
\ \biggr]\nn\\
&&\times d_{3}^{0}(\hat{2},l,\widetilde{(jk)})\ 
\Mtreeqggqb(\hat{1},\widetilde{(ij)},\widetilde{((jk)l)},\hat{\bar{2}})\ J_{2}^{(2)}(p_{(ij)},p_{((jk)l)})\nn\\\nn\\
\ph{a29}&&+\frac{1}{2}\,d_3^0(\hat{2},l,k)\,d_3^0(\hat{\bar{2}},j,\widetilde{(l k)})\,\Mtreeqggqb(\hat{1},i,\widetilde{((l k)j)},\hat{\bar{\bar{2}}})\,J_2^{(2)}(p_{((l k) j)},p_{i})\nn \\ 
\ph{a49}&&-\frac{1}{2}\,f_3^0(i,l,k)\,d_3^0(\hat{2},j,\widetilde{(lk)})\,\Mtreeqggqb(\hat{1},\widetilde{(il)},\widetilde{(j(lk))},\hat{\bar{2}})\,J_2^{(2)}(p_{(il)},p_{(j(lk))})\nn \\ 
\ph{a38}&&-\frac{1}{2}\,A_3^0(\hat{1},l,\hat{2})\,d_3^0(\hat{\bar{2}},\tilde{j},\tilde{k})\,\Mtreeqggqb(\hat{\bar{1}},\tilde{i},\widetilde{(jk)},\hat{\bar{\bar{2}}})\,J_2^{(2)}(p_{\tilde{i}},p_{(jk)})\nn \\ 
&&-\frac{1}{2}\biggl[\ \bigl(S_{2l(lk)}-S_{\bar{2}l(j(lk))}\bigr)
-\bigl(S_{(il)l((lk)j)}-S_{(il)l(lk)}\bigr)- \underbrace{\bigl(S_{1l2}-S_{1l\bar{2}}\bigr)}_{0}
\ \biggr]\nn\\
&&\times d_{3}^{0}(\hat{2},j,\widetilde{(lk)})\ 
\Mtreeqggqb(\hat{1},\widetilde{(il)},\widetilde{((lk)j)},\hat{\bar{2}})\ J_{2}^{(2)}(p_{(il)},p_{((lk)j)})\nn\\ \nn\\
\ph{a32}&&-\frac{1}{2}\,A_3^0(\hat{1},i,\hat{2})\,A_3^0(\hat{\bar{1}},\tilde{k},\hat{\bar{2}})\,\Mtreeqggqb(\hat{\bar{\bar{1}}},\tilde{\tilde{j}},\tilde{\tilde{l}},\hat{\bar{\bar{2}}})\,J_2^{(2)}(p_{\tilde{\tilde{j}}},p_{\tilde{\tilde{l}}})\nn \\ 
\ph{a37}&&+\frac{1}{2}\,d_3^0(\hat{1},i,j)\,A_3^0(\hat{\bar{1}},k,\hat{2})\,\Mtreeqggqb(\hat{\bar{\bar{1}}},\widetilde{(i
j)},\tilde{l},\hat{\bar{2}})\,J_2^{(2)}(p_{(\widetilde{i j})},p_{\tilde{l}})\nn \\
\ph{a41}&&+\frac{1}{2}\,d_3^0(\hat{2},i,l)\,A_3^0(\hat{1},k,\hat{\bar{2}})\,\Mtreeqggqb(\hat{\bar{1}},\tilde{j},\widetilde{(i l)},\hat{\bar{\bar{2}}})\,J_2^{(2)}(p_{\tilde{j}},p_{(\widetilde{i l})})\nn \\
&&+\frac{1}{2}\biggl[\ \bigl(S_{1i2}-S_{\bar{1}\tilde{i}\bar{2}}\bigr)
-\bigl(S_{1i(ji)}-S_{\bar{1}\tilde{i}(\widetilde{ji})}\bigr)
-\bigl(S_{2i(il)}-S_{\bar{2}\tilde{i}(\widetilde{il})}\bigr)\ \biggr]\nn\\
&&\times A_{3}^{0}(\hat{1},k,\hat{2})\ 
\Mtreeqggqb(\hat{\bar{1}},\widetilde{(ji)},\widetilde{(il)},\hat{\bar{2}})
\ J_{2}^{(2)}(p_{(\widetilde{ji})},p_{(\widetilde{il})})\nn\\\nn\\
\ph{a33}&&-\frac{1}{2}\,A_3^0(\hat{1},k,\hat{2})\,A_3^0(\hat{\bar{1}},\tilde{i},\hat{\bar{2}})\,\Mtreeqggqb(\hat{\bar{\bar{1}}},\tilde{\tilde{j}},\tilde{\tilde{l}},\hat{\bar{\bar{2}}})\,J_2^{(2)}(p_{\tilde{j}},p_{\tilde{l}})\nn \\
\ph{a35}&&+\frac{1}{2}\,d_3^0(\hat{1},k,j)\,A_3^0(\hat{\bar{1}},i,\hat{2})\,\Mtreeqggqb(\hat{\bar{\bar{1}}},\widetilde{(k j)},\tilde{l},\hat{\bar{2}})\,J_2^{(2)}(p_{(\widetilde{k j})},p_{\tilde{l}})\nn \\ 
\ph{a39}&&+\frac{1}{2}\,d_3^0(\hat{2},k,l)\,A_3^0(\hat{1},i,\hat{\bar{2}})\,\Mtreeqggqb(\hat{\bar{1}},\tilde{j},\widetilde{(k l)},\hat{\bar{\bar{2}}})\,J_2^{(2)}(p_{\tilde{j}},p_{(\widetilde{k l})})\nn \\ 
&&+\frac{1}{2}\biggl[\ \bigl(S_{1k2}-S_{\bar{1}\tilde{k}\bar{2}}\bigr)
-\bigl(S_{1k(jk)}-S_{\bar{1}\tilde{k}(\widetilde{jk})}\bigr)
-\bigl(S_{2k(kl)}-S_{\bar{2}\tilde{k}(\widetilde{kl})}\bigr)\ \biggr]\nn\\
&&\times A_{3}^{0}(\hat{1},i,\hat{2})\ 
\Mtreeqggqb(\hat{\bar{1}},\widetilde{(jk)},\widetilde{(kl)},\hat{\bar{2}})
\ J_{2}^{(2)}(p_{(\widetilde{jk})},p_{(\widetilde{kl})})\nn\\\nn\\
%
%
%
%
\ph{a50}&&-d_3^0(\hat{1},i,j)\,d_3^0(\hat{2},l,k)\,\Mtreeqggqb(\hat{\bar{1}},\widetilde{(i j)},\widetilde{(k l)},\hat{\bar{2}})\,J_2^{(2)}(p_{(i j)},p_{(k l)})\nn \\
\ph{a51}&&-d_3^0(\hat{1},l,k)\,d_3^0(\hat{2},i,j)\,\Mtreeqggqb(\hat{\bar{1}},\widetilde{(l k)},\widetilde{(ij)},\hat{\bar{2}})\,J_2^{(2)}(p_{(l k)},p_{(i j)})\biggr\},\label{eq:RRsub}
\end{eqnarray}
where $P_{C}(i,j,k,l)$ denotes the set of cyclic permutations of the gluons.  

It is interesting to note the appearance of the sub-leading colour quark-antiquark antenna $\tilde{A}_{4}^{0}$ and the accompanying $A_{3}^{0}$ antennae.  The $\tilde{A}_{4}^{0}$ antenna is introduced to remove spurious divergent behaviour of the $D_{4}^{0}$ antenna in the triple collinear limit.  Specifically, the $D_{4}^{0}$ antenna contains the IR divergent limit involving colour-disconnected gluons,
\ba
D_{4}^{0}(i,j,k,l)\stackrel{j||i||l}{\longrightarrow}\tilde{P}_{ijl\rightarrow Q}(x,y,z),\label{eq:D4Ptilde}
\ea
where the antenna tends to the QED-like triple collinear splitting function.  This divergent limit has no analogue in the leading colour physical matrix elements and must be removed.  In QCD the quark resides in the fundamental representation and so always terminates a string of gluons which lie to the right (left) of the quark (antiquark) in the colour ordering, e.g.,
\ba
M_{n}^{0}(\cdots;\hat{1}_{q},i_{g},j_{g},k_{g},\cdots,l_{g},m_{g},n_{g},\hat{2}_{\b{q}};\cdots).\label{eq:mnexample}
\ea
The $D_{4}^{0}$ antenna is derived from the decay of a neutralino to a gluino and gluons, $\tilde{\chi}\rightarrow \tilde{g}ggg$, where the gluino plays the role of the quark.  In contrast to the quarks, the gluino resides in the adjoint representation and so can sit anywhere in the colour ordering, including configurations with gluons either side of the gluino. For example, the $D_{4}^{0}$ is derived from the squared partial amplitude $M_{4}^{0}(i_{\tilde{g}},j_{g},k_{g},l_{g})$ which is symmetric under cyclic permutations of its partons.  The cyclic symmetry of this matrix element gives rise to the spurious divergent limit described in Eq.~\eqref{eq:D4Ptilde}.  

In any QCD calculation involving a quark-antiquark pair, the subtraction term will contain two $D_{4}^{0}$ antennae, one for the quark and the other for the antiquark, each containing a spurious limit of the kind in Eq.~\eqref{eq:D4Ptilde}.  e.g., for the matrix element in Eq.~\eqref{eq:mnexample}, the subtraction term contains the antennae\footnote{Here the quark-antiquark pair is in the initial state (IF $D_{4}^{0}$ antennae) for consistency with the process considered in this section but the argument is generic to FF, IF and II antennae.}, $D_{4}^{0}(\hat{1},i,j,k)$ and $D_{4}^{0}(\hat{2},n,m,l)$. The subtraction term is constructed for a block of colour orderings related by cyclic permutations of gluons.  Using this fact, the subtraction term will always contain the antennae $D_{4}^{0}(\hat{1},i,j,k)$ and $D_{4}^{0}(\hat{2},k,j,i)$, where the arguments of the second antenna are obtained by cyclically permuting the gluons in Eq.~\eqref{eq:mnexample}.  Both of these antennae contain spurious limits of the kind in Eq.~\eqref{eq:D4Ptilde}, involving gluons $i$ and $k$ and the antenna's fermion.  The spurious limits of both $D_{4}^{0}$ antennae can therefore be removed using an $\tilde{A}_{4}^{0}(\hat{1},i,k,\hat{2})$ antenna, as is done in Eq.~\eqref{eq:RRsub}.

Using $D_{4}^{0}$ antennae to isolate divergences associated with the quark and antiquark endpoints of a quark string, forces us to construct the subtraction term for a block of colour orderings related by cyclic permutations of gluons.  Exploiting the cyclic permutations, the spurious limits of the $D_{4}^{0}$ antennae can be systematically removed using an $\tilde{A}_{4}^{0}$ antenna.  It is noted that this argument holds no matter how many gluons are present in the quark string\footnote{The number of gluons should be more than two.  In the case of two gluons an $A_{4}^{0}$ antenna is used instead of the $D_{4}^{0}$ and the issue does not arise.} and the technique is not a special case of the six parton scattering process considered here.

\begin{figure}[t]
\centering
(a)\includegraphics[width=7cm]{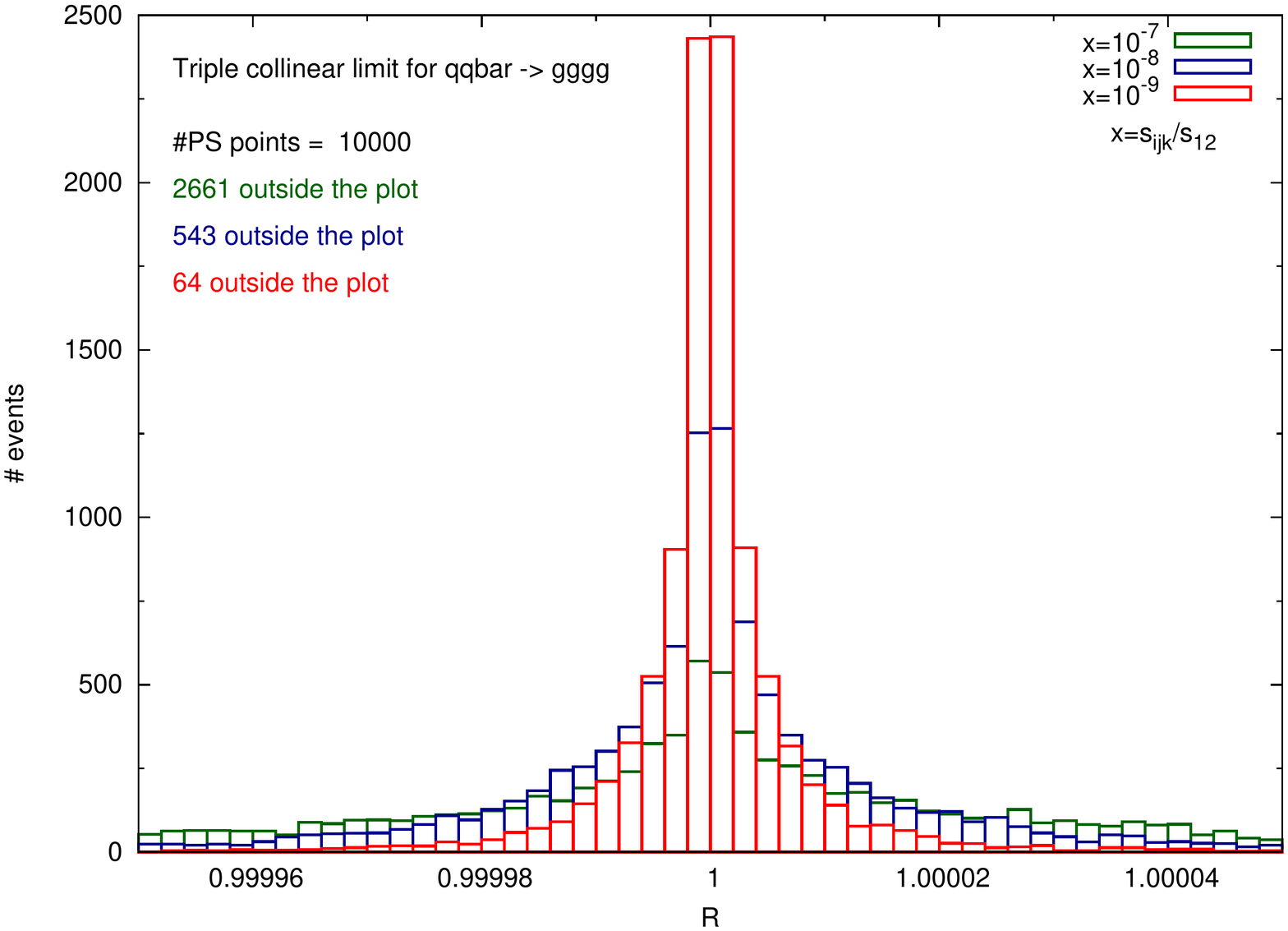}
(b)\includegraphics[width=7cm]{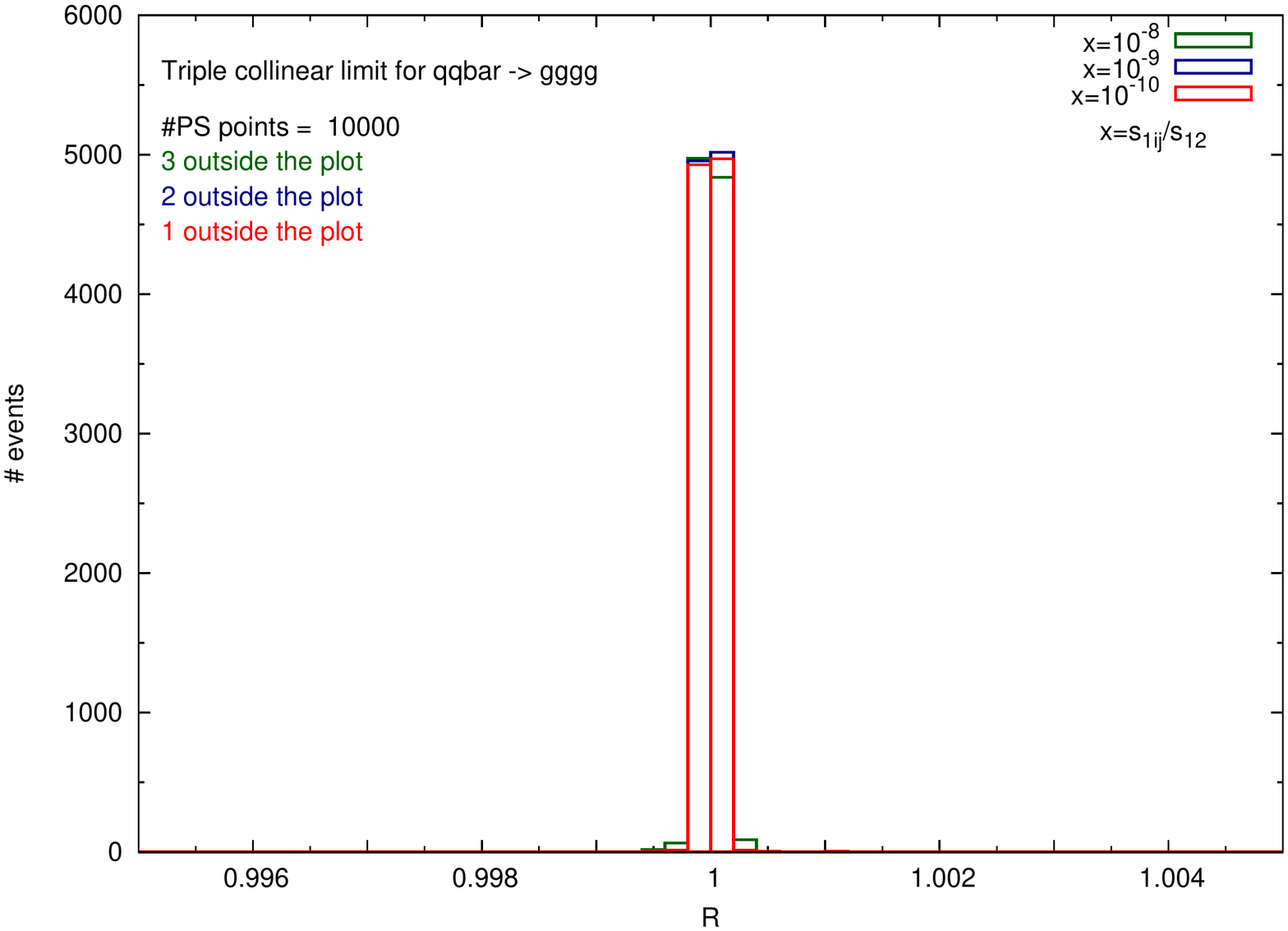}
(c)\includegraphics[width=7cm]{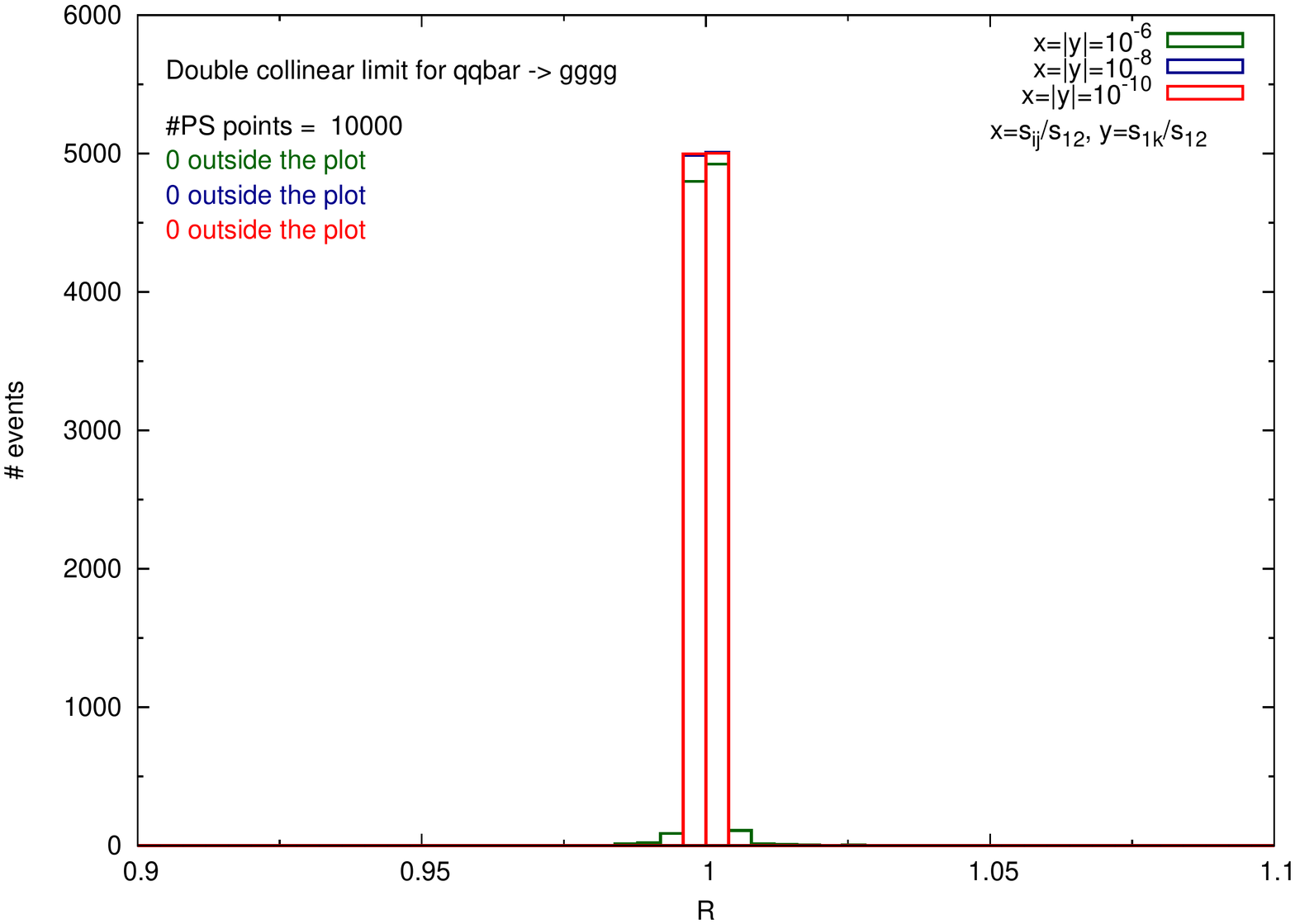}
(d)\includegraphics[width=7cm]{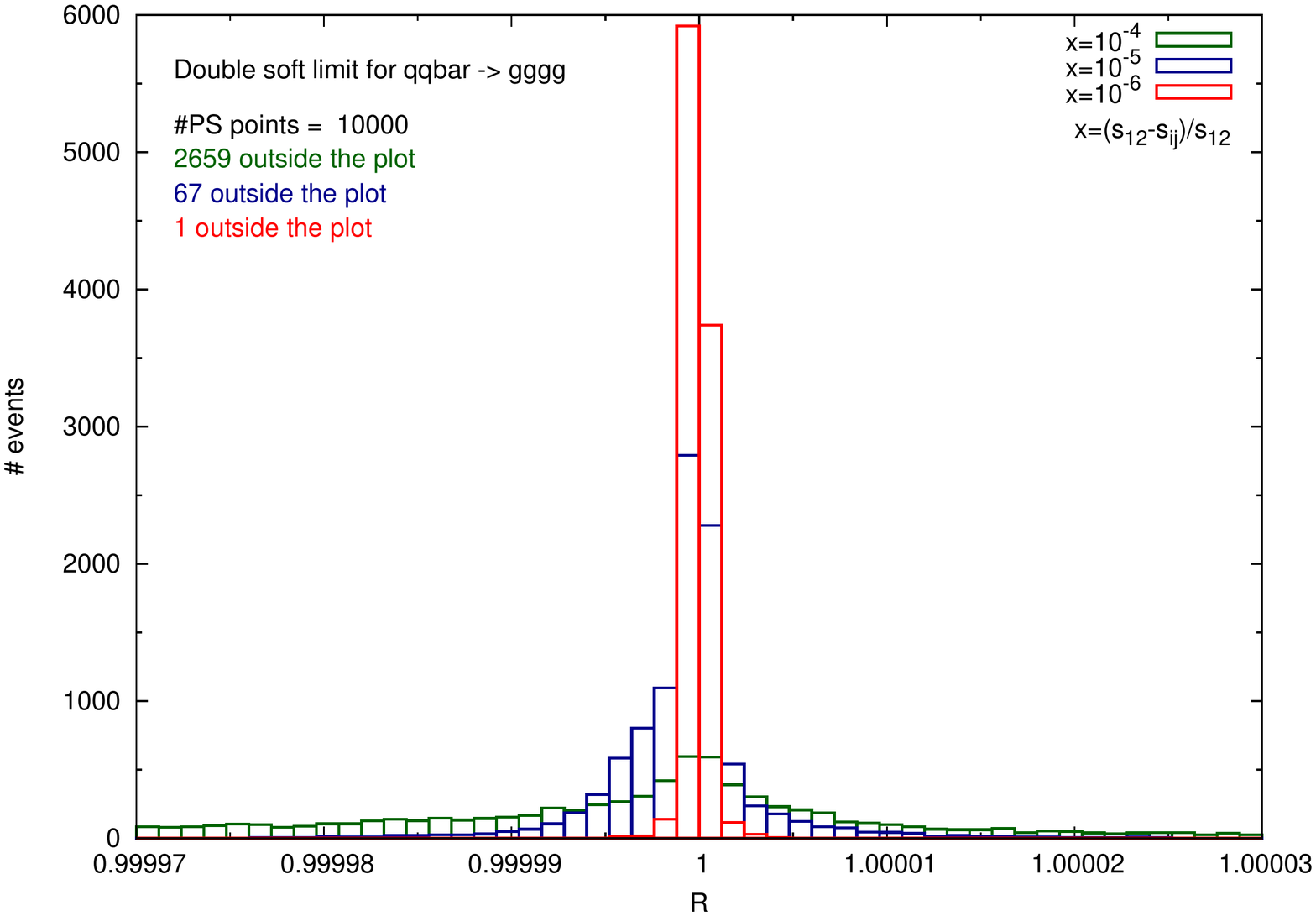}
\caption{Plots displaying the convergence of the subtraction term to the physical matrix element in various unresolved limits.  The green data is furthest from the singular configuration with the blue data closer to the singular region and the red data the closest. (a) Final-state triple collinear limit between partons $i,j,k$ such that $x=s_{ijk}/s_{12}$.  (b) Triple collinear limit between final-state partons $i,j$ and initial-state parton 1, such that $x=s_{1ij}/s_{12}$. (c) Double collinear limit between final-state partons $i,j$ and the initial-final pair $1,k$ such that $x=s_{ij}/s_{12},\ y=s_{1k}/s_{12}$. (d) Double soft limit for soft partons $i,j$ such that $x=(s_{12}-s_{ij})/s_{12}$. }
\label{fig:plots}
\end{figure}

With the modifications due to the spurious limits of the quark-gluon antennae, the double real subtraction term fits the general structure derived in Sec.~\ref{sec:nnloant}. Due to the non-trivial factorisation of the four-parton final-final antennae and the large sum of permutations, it is not straightforward to show analytically that this subtraction term correctly mimics the IR divergent behaviour of the physical cross section.  However, to demonstrate its validity, the subtraction term has been tested numerically.  

For each IR divergent configuration, a set of momenta were generated using \texttt{RAMBO}~\cite{Kleiss:1985gy} such that the momenta fulfil a set of constraints defining the unresolved configuration.  In this configuration, the ratio of the physical matrix element to the subtraction term we compute,
\ba
R&=&\frac{\dsigma_{NNLO}^{RR}}{\dsigma_{NNLO}^{S}}.
\ea
This procedure is then repeated for 10,000 different phase space points in the unresolved configuration defined by the constraints.  The constraints are then tightened to force the phase space points closer to the unresolved singular point and the ratio calculated for another 10,000 points.  The procedure is repeated once more for a set of points even closer to the singular point and the histogram of ratios for the three sets of constraints plotted.

A selection of plots from four different unresolved configurations is shown in Fig.~\ref{fig:plots}.  In each plot we see that the distribution of $R$ becomes more sharply peaked around one as the unresolved singular limit is approached, i.e., as the parameter $x$ gets smaller.  This provides statistical evidence for the convergence of the subtraction term to the physical cross section in the IR divergent limits.
\begin{figure}[t]
\centering
(a)\includegraphics[width=7cm]{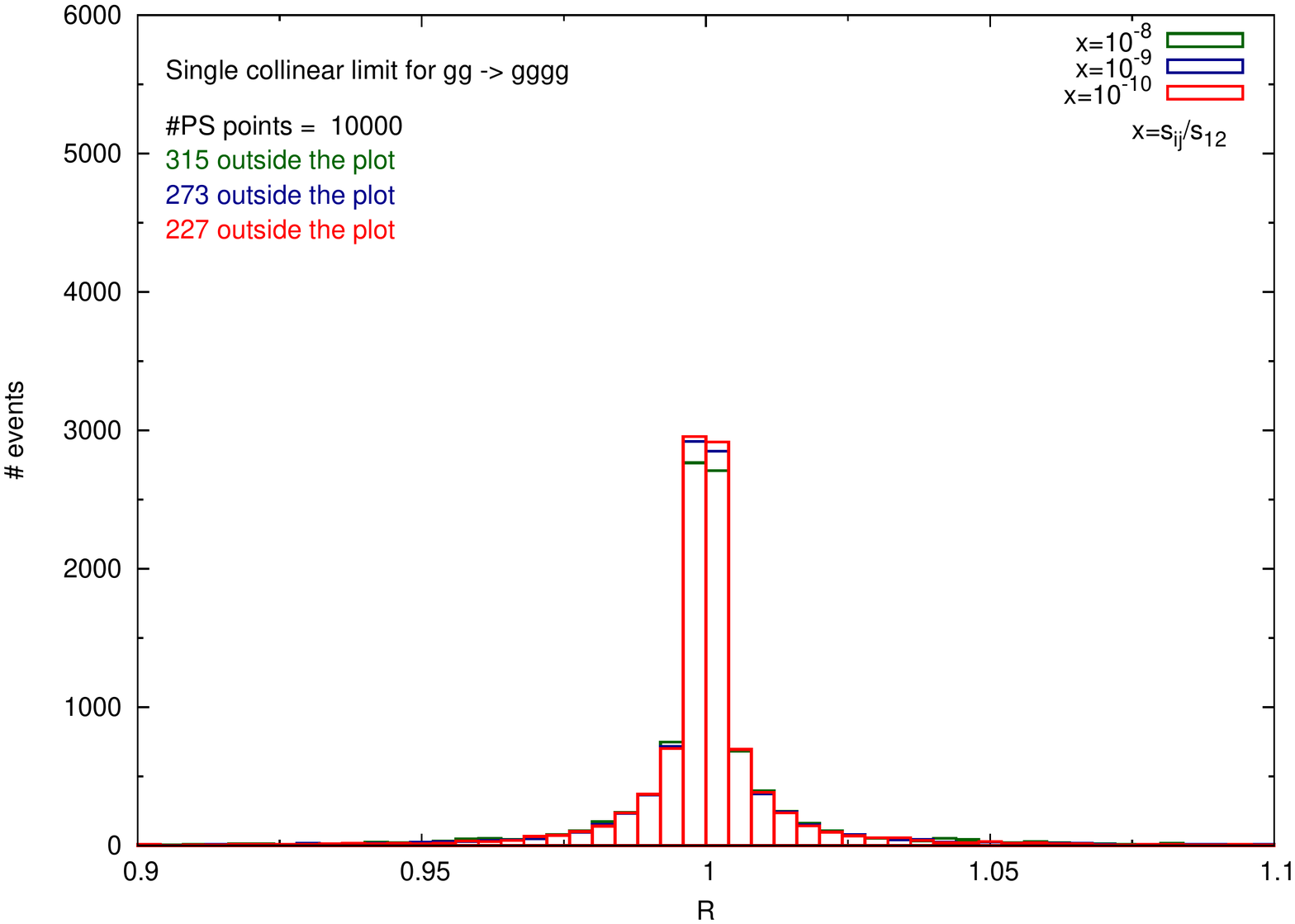}
(b)\includegraphics[width=7cm]{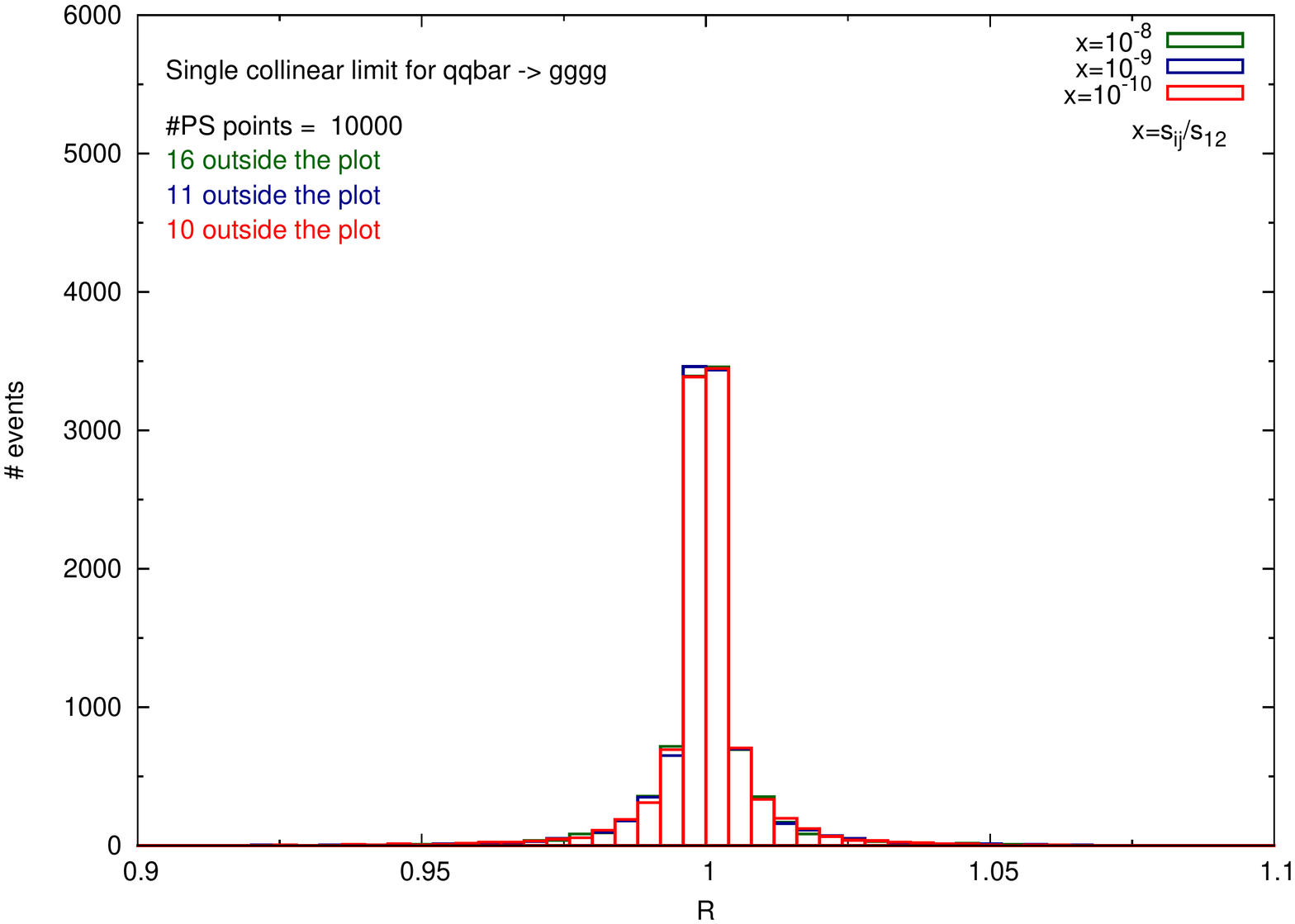}
(c)\includegraphics[width=7cm]{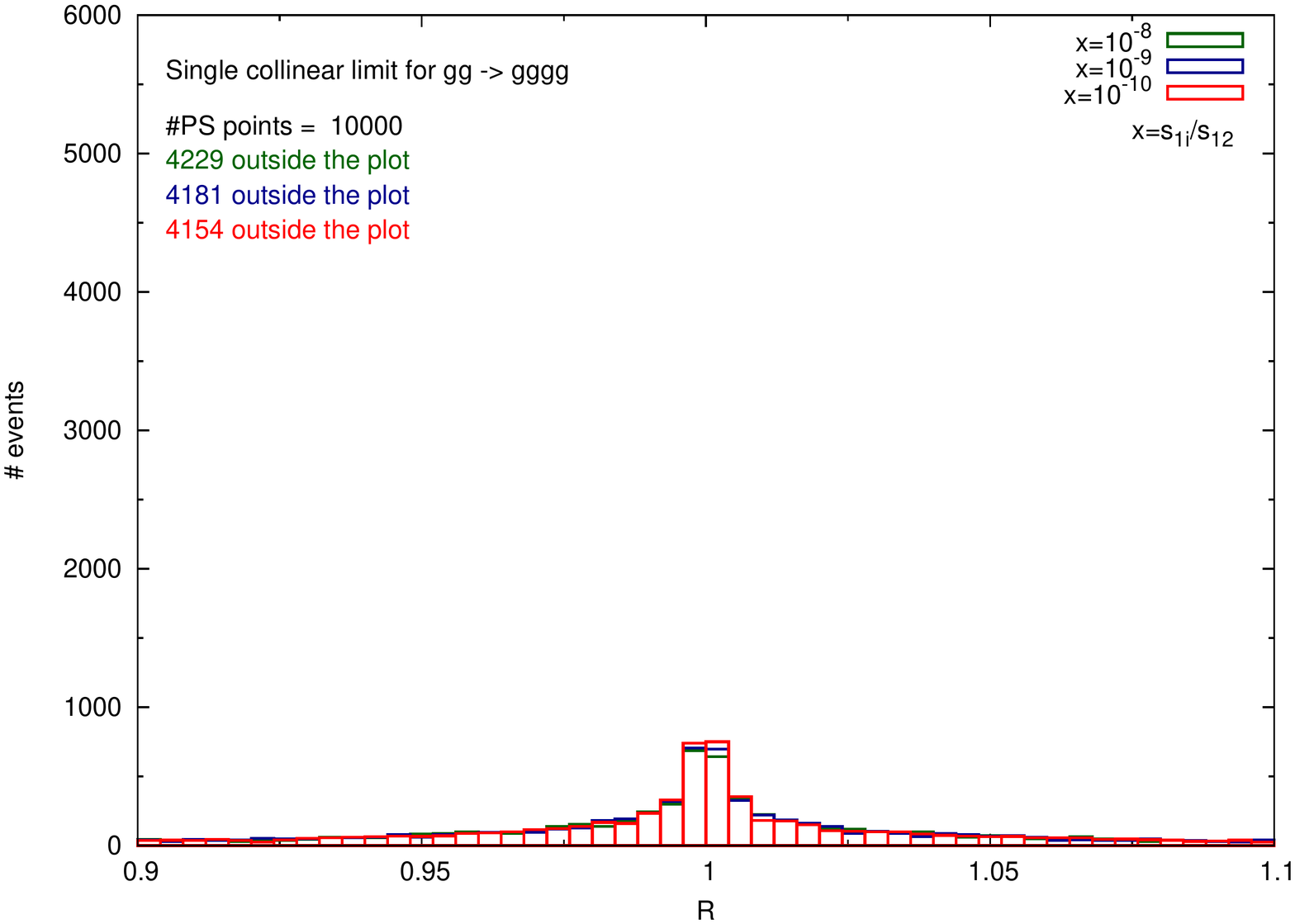}
(d)\includegraphics[width=7cm]{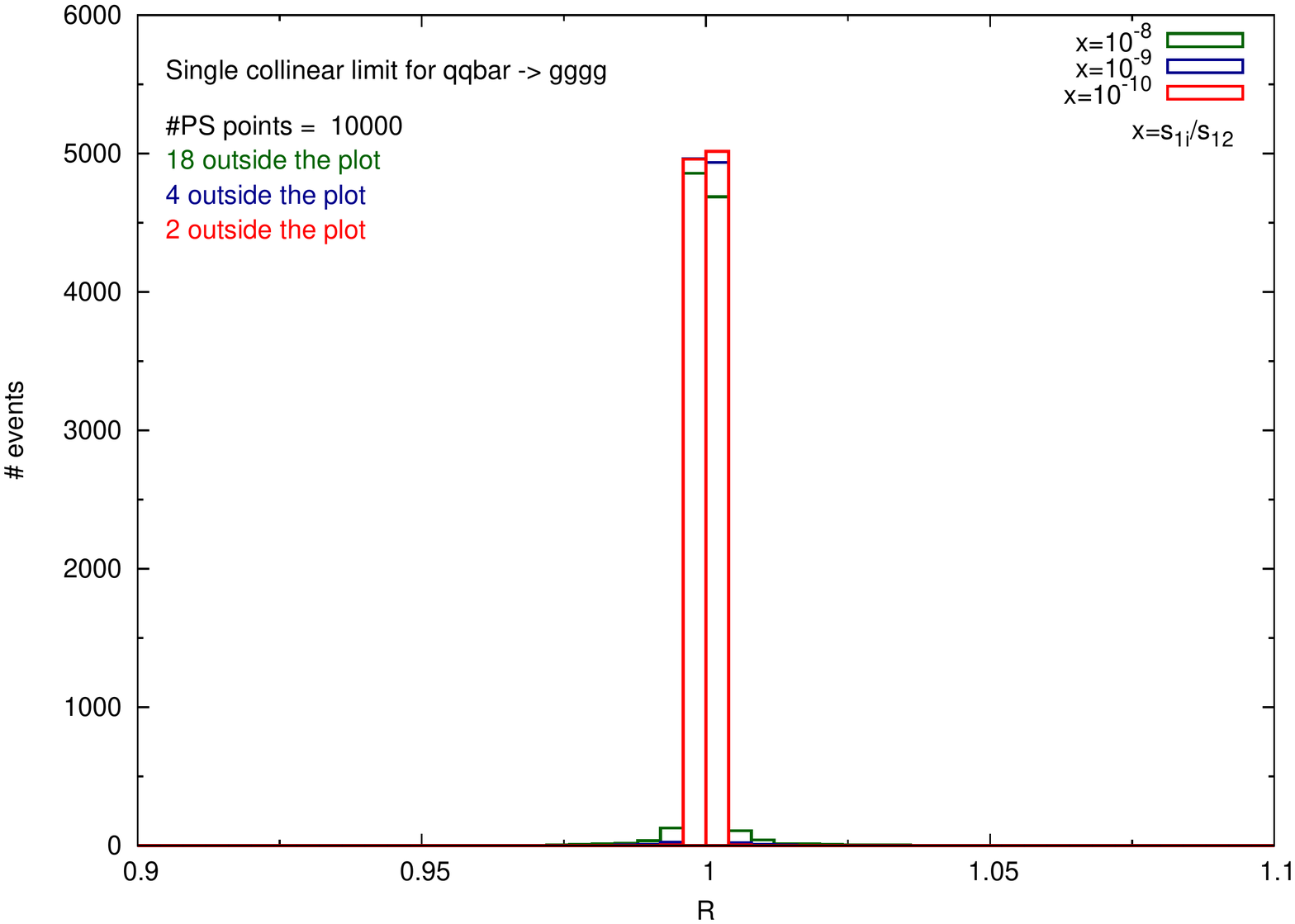}
\caption{Distributions of $R$ without azimuthal angular rotations for single collinear limits: final-final collinear limit for (a) $gg\rightarrow gggg$ and  (b) $q\bar{q}\rightarrow gggg$ and initial-final gluon-gluon collinear limit for (c) $gg\rightarrow gggg$ and (d) $q\bar{q}\rightarrow gggg$. The parameter $x$ controls how closely the singular limit is approached, close (dark green), closer (blue) and closest (red).}
\label{fig:singlecol}
\end{figure}

Testing the subtraction term in this way also gives a clear demonstration of the presence of azimuthal correlations in the single collinear limits. The origin and solution to this problem have been discussed in~\cite{GehrmannDeRidder:2005cm,our3j1,Pires:2010jv}, where it was shown that the azimuthal correlations which spoil the subtraction in the single collinear limit arise from splitting functions where a parent gluon splits into two daughter gluons or a quark-antiquark pair (and the various crossings of this configuration to include initial-state partons).  The $q\bar{q}\rightarrow gggg$ process contains the same final state as the $gg\rightarrow gggg$ process, studied in~\cite{Glover:2010im}.  Therefore by turning off the angular rotations when computing the ratio $R$, both subtraction terms should display similar behaviour when testing the single collinear limit between two final-state partons.  This is seen in Fig.~\ref{fig:singlecol}(a) and (b), where the subtraction terms for both processes display a broad peak, characteristic of the presence of uncompensated angular terms.  An interesting comparison between the two processes is seen in the initial-final collinear limit.  The collinear limit between an initial-state gluon and a final-state gluon does display azimuthal correlations while the collinear limit between an initial-state quark and a final-state gluon  does not. We see that the limit involving the initial-state quark shown in Fig.~\ref{fig:singlecol}(d) is sharply peaked while the corresponding limit in the purely gluonic process  seen in Fig.~\ref{fig:singlecol}(c) displays a broad peak because of the uncompensated azimuthal correlations.

\subsection{Real-virtual subtraction term, $\bs{\dsigma_{NNLO}^{T}}$}

Following the discussion in Sec.~\ref{sec:nnloant}, the real-virtual subtraction term is given by,
\begin{eqnarray}
\dsigma_{q\bar{q},NNLO}^{T}&=&\ {\cal{N}}_{5}^{1}\, \dPSxx{3}{5} \sum_{P(i,j,k)}\ \biggl\{ \nn\\
&-&\bs{J}_{5}^{(1)}(\hat{\bar{1}}_{q},i_{g},j_{g},k_{g},\hat{\bar{2}}_{\bar{q}})\ B_5^0(\hb{1}_q, i_g,j_g,k_g,\hb{2}_{\qb})\ J_2^{(3)}(p_i,p_j,p_k) \nn\\
%
%
&+&d^0_3(\hb{1},i,j)\bigg[B_4^1(\hbb{1}_q,\widetilde{(ij)}_g,{k}_g,\hb{2}_{\qb})\delta(1-x_{1})\delta(1-x_{2})\nn\\
&&+\ \bs{J}_{4}^{(1)}(\hbb{1}_q,\widetilde{(ij)}_g,{k}_g,\hb{2}_{\qb})\ B_4^0(\hbb{1}_q,\wt{(ij)}_g,k_g,\hb{2}_{\qb})\bigg]\ \JET_2^{(2)}(p_{\widetilde{ij}},p_k)\nn\\
&+&f^0_3({i},j,k)\bigg[B_4^1(\hb{1}_q,\widetilde{(ij)}_g,\widetilde{(jk)}_g,\hb{2}_{\qb})\delta(1-x_{1})\delta(1-x_{2})\nn\\
&&+\ \bs{J}_{4}^{(1)}(\hb{1}_q,\wt{(ij)}_g,\wt{(jk)}_g,\hb{2}_{\qb})\ B_4^0(\hb{1}_q,\wt{(ij)}_g,\wt{(jk)}_g,\hb{2}_{\qb})\bigg]\ J_2^{(2)}(p_{\wt{ij}},p_{\widetilde{jk}})\nn\\
&+&d^0_3(\hb{2},k,j)\bigg[B_4^1(\hb{1}_q,{i}_g,\widetilde{(kj)}_g,\hbb{2}_{\qb})\delta(1-x_{1})\delta(1-x_{2})\nn\\
&&+\ \bs{J}_{4}^{(1)}(\hb{1}_q,{i}_g,\wt{(kj)}_g,\hbb{2}_{\qb})\ B_4^0(\hb{1}_q,{i}_g,\wt{(kj)}_g,\hbb{2}_{\qb}) \bigg]\ \JET_2^{(2)}(p_{{i}},p_{\widetilde{jk}}) 
\nn\\
&+&\bigg[d^1_3(\hb{1},i,j)\delta(1-x_{1})\delta(1-x_{2})+\bs{J}_{D}^{(1)}(\hb{1}_g,i_g,j_g)\ d^0_3(\hb{1},i,j)\nn\\
&&-2\ d_{3}^{0}(\hb{1},i,j)\ \bs{J}_{2}^{(1)}(\hbb{1}_{q},\wt{(ij)})\bigg]\ B_4^0(\hbb{1}_q,\widetilde{(ij)}_g,{k}_g,\hb{2}_{\qb})\ \JET_2^{(2)}(p_{\widetilde{ij}},p_k)\nn\\
&+&\bigg[f^1_3(i,j,k)\delta(1-x_{1})\delta(1-x_{2})+\bs{J}_{F}^{(1)}(i_g,j_g,k_g)\ f^0_3(i,j,k)\nn\\
&&-2\ f^0_3(i,j,k)\ \bs{J}_{2}^{(1)}(\wt{(ij)}_{g},\wt{(jk)}_{g})\bigg]\ B_4^0(\hb{1}_q,\widetilde{(ij)}_g,\widetilde{(jk)}_g,\hb{2}_{\qb})\ \JET_2^{(2)}(p_{\widetilde{ij}},p_{\widetilde{jk}})\nn\\
&+&\bigg[d^1_3(\hb{2},k,j)\delta(1-x_{1})\delta(1-x_{2})+\bs{J}_{D}^{(1)}(\hb{2}_{q},k_g,j_{g})\ d^0_3(\hb{2},k,j)\nn\\
&&-2\ d^0_3(\hb{2},k,j)\ \bs{J}_{2}^{(1)}(\hbb{2}_{q},\wt{(kj)}_{g})\bigg]\ B_4^0(\hb{1}_q,{i}_g,\widetilde{(kj)}_g,\hbb{2}_{\qb})\ \JET_2^{(2)}(p_{{i}},p_{\widetilde{jk}})\nn\\
 &-&\biggl[\tilde{A}_{3}^{1}(\hat{\bar{1}},i,\hat{\bar{2}})\delta(1-x_{1})\delta(1-x_{2})+\bs{J}_{\tilde{A}}^{(1)}(\hb{1}_{q},i_{g},\hb{2}_{\bar{q}})\ A_{3}^{0}(\hat{\bar{1}},i,\hat{\bar{2}})\nn\\
 &&- A_{3}^{0}(\hat{\bar{1}},i,\hat{\bar{2}})\ \bs{J}_{2}^{(1)}(\hbb{1}_{q},\hbb{2}_{\b{q}})\biggr]\ B_{4}^{0}(\hat{\bar{\bar{1}}}_{q},\tilde{j}_{g},\tilde{k}_{g},\hat{\bar{\bar{2}}}_{\bar{q}})\ J_{2}^{(2)}(p_{\tilde{j}},p_{\tilde{k}}) \nn\\
&+&b_{0} \log\left( \frac{\mu^2}{|s_{1ij}|}\right)d^0_3(\hb{1},i,j)\delta(1-x_{1})\delta(1-x_{2})\ B_4^0(\hbb{1}_q,\widetilde{(ij)}_g,{k}_g,\hb{2}_{\qb})\JET_2^{(2)}(p_{\widetilde{ij}},p_k)\nn\\
&+&b_{0} \log\left( \frac{\mu^2}{s_{ijk}}\right)f^0_3({i},j,k)\delta(1-x_{1})\delta(1-x_{2})\ B_4^0(\hb{1}_q,\widetilde{(ij)}_g,\widetilde{(jk)}_g,\hb{2}_{\qb})\JET_2^{(2)}(p_{\widetilde{ij}},p_{\widetilde{jk}})\nn\\
&+&b_{0} \log\left( \frac{\mu^2}{|s_{2kj}|}\right)d^0_3(\hb{2},k,j)\delta(1-x_{1})\delta(1-x_{2})\ B_4^0(\hb{1}_q,{i}_g,\widetilde{(jk)}_g,\hbb{2}_{\qb})\JET_2^{(2)}(p_{{i}},p_{\widetilde{jk}})\nn\\
&+&\frac{1}{2}\bigg[ \frac{1}{2}{\cal{D}}_{3,q}^{0}(s_{\dbar{1}\widetilde{(ij)}})
- \frac{1}{2}{\cal{D}}_{3,q}^{0}(s_{\bar{1}j}) 
- \frac{1}{3}{\cal{F}}_{3}^{0}(s_{\wt{(ij)}k}) 
 +  \frac{1}{3}{\cal{F}}_{3}^{0}(s_{jk})+{\cal{A}}_{3,q\bar{q}}^{0}(s_{\bar{1}\bar{2}})-{\cal{A}}_{3,q\bar{q}}^{0}(s_{\bar{\bar{1}}\bar{2}}) \nn\\
&&+ \delta(1-x_{1})\delta(1-x_{2})\Big(
 \sterm{\wt{(ij)}k}{jk} - \stermi{jk}- \sterm{\dbar{1}\wt{(ij)}}{jk}\nn\\
 &&+ \sterm{\bar{1}j}{jk}\Big) \bigg]\ d^0_3(\hb{1},i,j)\ B_4^0(\hbb{1}_q,\wt{(ij)}_g,{k}_g,\hb{2}_{\qb})\ J_2^{(2)}(p_{\widetilde{ij}},p_k)\nn\\
&+&\frac{1}{2}\bigg[ -\frac{1}{2}{\cal{D}}_{3,q}^{0}(s_{\bar{1}\wt{(ij})})+ \frac{1}{2} {\cal{D}}_{3,q}^{0}(s_{\bar{1}i})
- \frac{1}{2}{\cal{D}}_{3,q}^{0}(s_{\bar{2}\wt{(jk)}})+\frac{1}{2}{\cal{D}}_{3,q}^{0}(s_{\bar{2}k}) 
+ \frac{1}{3} \mc{F}{\wt{(ij)}}{\wt{(jk)}}\nn\\
&& - \frac{1}{3}\mc{F}{{i}}{k} + \delta(1-x_{1})\delta(1-x_{2})\Big(- \sterm{\wt{(ij)}\wt{(jk)}}{ik} + \stermi{ik} \nn\\
&&+ \sterm{\bar{1}\wt{(ij)}}{ik}- \sterm{\bar{1}i}{ik} + \sterm{\bar{2}\wt{(jk)}}{ik}\nn\\
&& - \sterm{\bar{2}k}{ik}\Big)\bigg]\ f^0_3({i},j,k)\ B_4^0(\hb{1}_q,\wt{(ij)}_g,\wt{(jk)}_g,\hb{2}_{\qb})\ 
J_2^{(2)}(p_{\widetilde{ij}},p_{\widetilde{jk}})\nn\\
&+&\frac{1}{2}\bigg[ \frac{1}{2}{\cal{D}}_{3,q}^{0}(s_{\dbar{2}\wt{(jk)}})- \frac{1}{2} {\cal{D}}_{3,q}^{0}(s_{\bar{2}j}) 
- \frac{1}{3}{\cal{F}}_{3}^{0}(s_{i\wt{(jk)}})+  \frac{1}{3}{\cal{F}}_{3}^{0}(s_{ij})+{\cal{A}}_{3,q\bar{q}}^{0}(s_{\bar{1}\bar{2}})-{\cal{A}}_{3,q\bar{q}}^{0}(s_{\bar{1}\bar{\bar{2}}})\nn\\
 &&+ \delta(1-x_{1})\delta(1-x_{2})\Big( \sterm{i\wt{(jk)}}{ij} - \stermi{ij}\nn\\
  && - \sterm{\dbar{2}\wt{(kj)}}{ij}+ \sterm{\bar{2}j}{ij}\Big)\bigg]\nn\\
  &&\times\ d^0_3(\hb{2},k,j)\ B_4^0(\hb{1}_q,{i}_g,\wt{(kj)}_g,\hbb{2}_{\qb})\ J_2^{(2)}(p_{{i}},p_{\widetilde{jk}})\nn\\
 &+&\frac{1}{2}\bigg[ 
 - {\cal{A}}_{3,q\bar{q}}^{0}(s_{\dbar{1}\dbar{2}}) + {\cal{A}}_{3,q\bar{q}}^{0}(s_{\bar{1}\bar{2}}) 
 + \frac{1}{2}{\cal{D}}_{3,q}^{0}(s_{\dbar{1}\tilde{j}}) - \frac{1}{2}{\cal{D}}_{3,q}^{0}(s_{\bar{1}j})
 + \frac{1}{2}{\cal{D}}_{3,q}^{0}(s_{\dbar{2}\tilde{k}}) \nn\\
 &&- \frac{1}{2}{\cal{D}}_{3,q}^{0}(s_{\bar{2}k})+\delta(1-x_{1})\delta(1-x_{2})\Big( \sterm{\dbar{1}\dbar{2}}{\tilde{j}\tilde{k}} - \sterm{\bar{1}\bar{2}}{{j}{k}} 
 \nn\\
  &&- \sterm{\dbar{1}\tilde{j}}{\tilde{j}\tilde{k}}+ \sterm{\bar{1}\tilde{j}}{{j}{k}} - \sterm{\dbar{2}\tilde{k}}{\tilde{j}\tilde{k}}\nn\\
 && + \sterm{\bar{2}k}{jk}\Big)\bigg]\ A_3^0(\hb{1}, i, \hb{2})\ B_4^0(\hbb{1}_q, \tilde{j}_g,\tilde{k}_g, \hbb{2}_{\qb})\ J_2^{(2)}(p_{ \tilde{j}},p_{\tilde{k}})\bigg\}.
\end{eqnarray}
The integrated antenna strings used in this subtraction term are given by,
\ba
\bs{J}_{5}^{(1)}(\hb{1}_{q},i_{g},j_{g},k_{g},\hb{2}_{\b{q}})&=&\bs{J}_{2}^{(1)}(\hb{1}_{q},i_{g})+\bs{J}_{2}^{(1)}(i_{g},j_{g})+\bs{J}_{2}^{(1)}(j_{g},k_{g})+\bs{J}_{2}^{(1)}(\hb{2}_{\b{q}},k_{g}),\nn\\
\bs{J}_{D}^{1}(\hb{1}_q,i_g,j_g)&=&\bs{J}_{2}^{(1)}(\hb{1}_{q},i_{g})+\bs{J}_{2}^{(1)}(i_{g},j_{g})+\bs{J}_{2}^{(1)}(\hb{2}_{q},j_{g}),\nn\\
\bs{J}_{F}^{1}(i_g,j_g,k_{g})&=&\bs{J}_{2}^{(1)}(i_{g},j_{g})+\bs{J}_{2}^{(1)}(j_{g},k_{g})+\bs{J}_{2}^{(1)}(i_{g},k_{g}),\nn\\
\bs{J}_{D}^{1}(\hb{2}_q,i_g,j_g)&=&\bs{J}_{D}^{1}(\hb{1}_q,i_g,j_g), (\hb{1}\leftrightarrow \hb{2},\ z_{1}\leftrightarrow z_{2}),\nn\\
\bs{J}_{\tilde{A}}^{1}(\hb{1}_{q},i_{g},\hb{2}_{\b{q}})&=&\bs{J}_{2}^{(1)}(\hb{1}_{q},\hb{2}_{\b{q}}).
\ea
After integration only those terms introduced at the real-virtual level are passed down to the double virtual subtraction term.  

\label{sec:rvsub}

\subsection{Double virtual subtraction term, $\bs{\dsigma_{NNLO}^{U}}$}

Following the discussion in Sec.~\ref{sec:doublevirt}, the double virtual subtraction term is constructed from the remaining terms inherited from the double real and the real-virtual subtraction terms, in addition to the double virtual mass factorisation contribution.  The resulting double virtual subtraction term can be written in terms of single and double unresolved integrated antenna strings,
\ba
\dsigma_{q\bar{q}NNLO}^{U}&=&-{\cal{N}}_{4}^{2}\ {\rm d}\Phi_{2}(p_3,p_{4};p_1,p_2)\ \int\frac{{\rm{d}}z_{1}}{z_{1}}\frac{{\rm{d}}z_{2}}{z_{2}} \frac{1}{2!} \sum_{P(i,j)}\nn\\
&\bigg\{&\bs{J}_{4}^{(1)}(\hat{\bar{1}}_{q},i_{g},j_{g},\hat{\bar{2}}_{\bar{q}})\ \biggl(B_{4}^{1}(\hat{\bar{1}}_{q},i_{g},j_{g},\hat{\bar{2}}_{\bar{q}})-\frac{b_{0}}{\eps}\ B_{4}^{0}(\hat{\bar{1}}_{q},i_{g},j_{g},\hat{\bar{2}}_{\bar{q}})\biggr)\nn\\
&+&\frac{1}{2}\ \bs{J}_{4}^{(1)}(\hat{\bar{1}}_{q},i_{g},j_{g},\hat{\bar{2}}_{\bar{q}})\otimes\bs{J}_{4}^{(1)}(\hat{\bar{1}}_{q},i_{g},j_{g},\hat{\bar{2}}_{\bar{q}})\ B_{4}^{0}(\hat{\bar{1}}_{q},i_{g},j_{g},\hat{\bar{2}}_{\bar{q}})\nn\\
&+&\bs{J}_{4}^{(2)}(\hat{\bar{1}}_{q},i_{g},j_{g},\hat{\bar{2}}_{\bar{q}})\ B_{4}^{0}(\hat{\bar{1}}_{q},i_{g},j_{g},\hat{\bar{2}}_{\bar{q}})\biggr\}\ J_{2}^{(2)}(p_{i},p_{j}).
\ea
The single unresolved integrated antenna string is defined in Eq.~\eqref{eq:j1dipole}. The new ingredient at the double virtual level is the double unresolved integrated antenna string, $\bs{J}_{4}^{(2)}$, which displays the structure discussed in Sec.~\ref{sec:doublevirt}.  In Eq.~\eqref{eq:j1dipole} the single unresolved integrated antenna string was written as a sum over integrated dipoles.  The analogous dipole-like formula for $\bs{J}_{4}^{(2)}$ is given by,
\ba
\bs{J}_{4}^{(2)}(\hat{\bar{1}}_{q},i_{g},j_{g},\hat{\bar{2}}_{\bar{q}})&=&\bs{J}_{2}^{(2)}(\hat{\bar{1}}_{q},i_{g})+\bs{J}_{2}^{(2)}(i_{g},j_{g})+\bs{J}_{2}^{(2)}(j_{g},\hat{\bar{2}}_{q})-\overline{\bs{{J}}}_{2}^{(2)}(\hat{\bar{1}}_{q},\hb{2}_{\b{q}}),
\ea
where the two-parton double unresolved integrated antenna strings are given by:
\ba
\bs{J}_{2}^{(2)}(\hat{\bar{1}}_{q},i_{g})&=&\frac{1}{2}{\cal{D}}_{4,q}^{0}(s_{\b{1}i})+\frac{1}{2}{\cal{D}}_{3,q}^{1}(s_{\b{1}i})+\frac{b_{0}}{2\eps}\bigg(\frac{|s_{\b{1}i}|}{\mu^{2}}\bigg)^{-\eps}{\cal{D}}_{3,q}^{0}(s_{\b{1}i})\nn\\
&-&\frac{1}{4}\big[{\cal{D}}_{3,q}^{0}(s_{\b{1}i})\otimes{\cal{D}}_{3,q}^{0}(s_{\b{1}i})\big](z_{1})-\overline{\Gamma}_{qq}^{(2)}(z_{1})\delta(1-z_{2}),\\
\bs{J}_{2}^{(2)}(i_{g},j_{g})&=&\frac{1}{4}{\cal{F}}_{4}^{0}(s_{ij})+\frac{1}{3}{\cal{F}}_{3}^{1}(s_{ij})+\frac{b_{0}}{3\eps}\bigg(\frac{s_{ij}}{\mu^{2}}\bigg)^{-\eps}{\cal{F}}_{3}^{0}(s_{ij})\nn\\
&-&\frac{1}{9}\big[{\cal{F}}_{3}^{0}(s_{ij})\otimes{\cal{F}}_{3}^{0}(s_{ij})\big], \\
\bs{J}_{2}^{(2)}(i_{g},\hat{\bar{2}}_{\b{q}})&=&\bs{J}_{2}^{(2)}(\hat{\bar{2}}_{q},i_{g}).
\ea

An integrated dipole, $\bs{J}_{2}^{(\ell)}(i,j)$ is associated with a power of $(|s_{ij}|)^{-\ell\e}$, thereby matching specific terms in the Catani representation of the one- and two-loop matrix elements.   However, for the leading colour contribution to $q\b{q} \to gg$~jets, we do not expect a singular contribution that depends on $s_{\b{1}\b{2}}$.

However, the $D_4^0$ antenna contain infrared singular limits that are not present in the matrix elements.   These singularities are compensated by the $\t{A}_{4,q\b{q}}^0$ contribution in 
$\dsigma_{NNLO}^{S}$ and by the $\t{A}_{3,q\b{q}}^1$ and 
and ${\cal A}_{3,q\b{q}}^0\times A_{3}^0$ 
terms in $\dsigma_{NNLO}^{T}$.
Terms of the form $({\cal{D}}_{3,q}^{0}\times A_{3}^{0})$ and $({\cal{A}}_{3,q\bar{q}}^{0}\times d_{3}^{0})$ which were introduced at the real-virtual level, cancel after integration.  
The integrated forms of these spurious contributions reappear in $\dsigma_{NNLO}^U$ and are collected in $\overline{\bs{{J}}}_{2}^{(2)}$,
\ba
\overline{\bs{{J}}}_{2}^{(2)}(\hat{\bar{1}}_{q},\hb{2}_{\b{q}})&=&
\frac{1}{2}
\tilde{\cal{A}}_{4,q\b{q}}^{0}(s_{\b{1}\b{2}})+\tilde{\cal{A}}_{3,q\b{q}}^{1}(s_{\b{1}\b{2}})-\frac{1}{2}\big[{\cal{A}}_{3,q\b{q}}^{0}(s_{\b{1}\b{2}})\otimes{\cal{A}}_{3,q\b{q}}^{0}(s_{\b{1}\b{2}})\big](z_{1},z_{2}),\label{eq:J22bar}
\ea
which is proportional to $(|s_{\b{1}\b{2}}|)^{-2\e}$.
At first sight, the presence of $\overline{\bs{{J}}}_{2}^{(2)}(\hat{\bar{1}}_{q},\hb{2}_{\b{q}})$ appears to be in conflict with our earlier statement about the absence of singularities proportional to $s_{\b{1}\b{2}}$.   
However, the leading singularity in $\overline{\bs{{J}}}_{2}^{(2)}$ is ${\cal O}(1/\e)$, 
\ba
\Poles\left(\e \overline{\bs{{J}}}_{2}^{(2)}(1_{q},2_{\b{q}})\right)&=&0,
\ea
so that, after expansion of the $(|s_{\b{1}\b{2}}|)^{-2\e}$ factor, the singularity does not depend on $s_{\b{1}\b{2}}$ while the remaining finite terms can and do depend on $\ln(s_{\b{1}\b{2}})$.

\subsection{Comparison with the IR structure of the $e^{+}e^{-}\to3$~jet leading colour double virtual subtraction term}

It is interesting to re-examine the form of the double virtual subtraction term for the $e^{+}e^{-}\rightarrow 3j$ process~\cite{our3j1}.  Although the $e^{+}e^{-}$ annihilation calculation involves only final-state partons, the integrated form of the subtraction terms is expected to closely follow the $q\bar{q}\rightarrow gg$ double virtual subtraction term up to mass factorisation contributions and crossed antenna.  The double virtual subtraction term as presented in~\cite{our3j1} can be reformulated in terms of final-state integrated antenna strings to fit the double virtual structure discussed in Sec.~\ref{sec:nnloant}, so that, for example, at leading order in the number of colours,
\ba
\dsigma^{U}_{N^2}&=&{\cal{N}}_3\ N^2 \text{d}\Phi_{3}(p_{1},p_{2},p_{3},q^{2})\   \nn\\
&\bigg\{&\bs{J}_{3}^{(1)}(1_{q},3_{g},2_{\bar{q}})\biggl(M_{3}^{1}(1_{q},3_{g},2_{\bar{q}})-\frac{b_{0}}{\eps}\ M_{3}^{0}(1_{q},3_{g},2_{\bar{q}})\biggr)\nn\\
&+&\frac{1}{2}\ \bs{J}_{3}^{(1)}(1_{q},3_{g},2_{\bar{q}})\otimes\bs{J}_{3}^{(1)}(1_{q},3_{g},2_{\bar{q}})\ M_{3}^{0}(1_{q},3_{g},2_{\bar{q}})\nn\\
&+&\bs{J}_{3}^{(2)}(1_{q},3_{g},2_{\bar{q}})\ M_{3}^{0}(1_{q},3_{g},2_{\bar{q}})\biggr\}\ J_{3}^{(3)}(p_{1},p_{2},p_{3}),
\ea
where $M_{3}^{0}(1_{q},3_{g},2_{\b{q}})$ and $M_{3}^{1}(1_{q},3_{g},2_{\b{q}})$ are the tree-level and one-loop squared partial amplitudes for the process $e^{+}e^{-}\rightarrow\gamma^{*}/Z^{0}\rightarrow qg\b{q}$ and $q^{2}$ is the virtuality of the colourless boson.  The factor ${\cal{N}}_{3}$ carries all overall colour factors, QCD coupling and non-QCD factors,
\ba
{\cal{N}}_{3}&=&4\pi\alpha\sum_{q}e_{q}^{2}g^{2}(N^{2}-1).
\ea
The final-state single unresolved integrated antenna string is given by,
\ba
\bs{J}_{3}^{(1)}(1_{q},3_{g},2_{\bar{q}})&=&\bs{J}_{2}^{(1)}(1_{q},3_{g})+\bs{J}_{2}^{(1)}(3_{g},2_{\b{q}}), \\
\bs{J}_{2}^{(1)}(3_{g},2_{\b{q}})&=&\bs{J}_{2}^{(1)}(2_{{q}},3_{g})
\ea
The double unresolved integrated antenna string for the relevant final-state partons is given by,
\ba
\bs{J}_{3}^{(2)}(1_{q},3_{g},2_{\bar{q}})&=&\bs{J}_{2}^{(2)}(1_{q},3_{g})+\bs{J}_{2}^{(2)}(3_{g},2_{\b{q}})-\overline{\bs{{J}}}_{2}^{(2)}(1_{q},2_{\b{q}}),
\ea
where
\ba
\bs{J}_{2}^{(2)}(1_{q},3_{g})&=&\frac{1}{2}{\cal{D}}_{4}^{0}(s_{13})+\frac{1}{2}{\cal{D}}_{3}^{1}(s_{13})\nn\\
&+&\frac{b_{0}}{2\eps}\bigg(\frac{s_{13}}{\mu^{2}}\bigg)^{-\eps}
{\cal{D}}_{3}^{0}(s_{13})-\frac{1}{4}{\cal{D}}_{3}^{0}(s_{13})\otimes{\cal{D}}_{3}^{0}(s_{13}),\\
\bs{J}_{2}^{(2)}(3_{g},2_{\b{q}})&=&\bs{J}_{2}^{(2)}(2_{q},3_{g}),
\ea
In this example, all integrated antennae are of final-final type and so carry all $z_{1}$ and $z_{2}$ dependence though $\delta$-functions, rendering all convolutions and integration over $z_{1}$ and $z_{2}$ trivial. 
As expected, the relevant integrated antenna strings are composed of two-particle integrated dipoles, $\bs{J}_{2}^{(1)}$ and $\bs{J}_{2}^{(2)}$.

Once again, $\overline{\bs{{J}}}_{2}^{(2)}$ is present to remove the spurious singularities generated by the $D$ antennae,  
\ba
\overline{\bs{{J}}}_{2}^{(2)}(1_{q},2_{\b{q}})&=&\frac{1}{2}\tilde{\cal{A}}_{4}^{0}(s_{12})+\tilde{\cal{A}}_{3}^{1}(s_{12})-\frac{1}{2}{\cal{A}}_{3}^{0}(s_{12})\otimes{\cal{A}}_{3}^{0}(s_{12}).
\label{eq:e+e-}
\ea
It has exactly the same origin, structure and singular behaviour as Eq.~\eqref{eq:J22bar}.

\section{Conclusions and discussion}
\label{sec:conclusions}

In this paper, we made a detailed investigation of the infrared structure of NLO and NNLO pertubative calculations where the
antenna subtraction method is used to isolate the infrared singularities present in the radiative contributions to the cross
section. At NLO, there is a clear and simple relationship between the unintegrated antenna used to subtract the implicit
singularity present in the real radiation contribution and the (singular) integrated antenna that cancels the explicit poles
in the virtual contribution.   At NNLO, the subtraction takes place across the double real, real-virtual and double virtual
contributions and is significantly more complicated than at NLO.   However, the implicit divergence of the double real and
real-virtual contributions is captured by distinct blocks of subtraction terms which follow a predictive structure based on
the colour ordering of the matrix elements under consideration. Fig.~\ref{fig:roadmap} shows explicitly how the integrated and
unintegrated subtraction terms conspire to cancel the explcit divergences in one- and two-loop matrix elements.

Each unintegrated antenna involves two hard radiators that, after integration over the unresolved phase space, appear as a
pair of colour connected final state particles, $I$ and $J$, in the virtual subtraction term.   The singularities associated
with a integrated antenna are proportional to $(|s_{IJ}|)^{-\ell\e}$ and this gives a natural linkage to the one- and two-loop
IR singularity operators $\bs{I}_{ij}^{(1)}(\eps)$ and $\bs{I}_{ij}^{(2)}(\eps)$ proposed by Catani~\cite{Catani:1998bh}.  We
denote the collection of integrated antennae for a particular pair of colour connected particles (plus the relevant mass
factorisation contribution) as an \emph{integrated dipole}.  At one-loop,   $\bs{J}_{2}^{(1)}$ is related to an integrated
three-particle tree-level antenna and describes the unresolved radiation between two colour connected particles. Similarly, at
two-loops, $\bs{J}_{2}^{(2)}$ fulfills the same role for double unresolved radiation and involves integrated four-particle
tree-level antennae, three-particle one-loop antennae and products of three-parton tree-level antennae. Mass factorisation
contributions are absorbed into the integrated dipoles so that the deepest poles contribute only in the soft region. 

The IR structure of the two-loop contribution is determined by \emph{integrated antenna strings} which are simply formed from 
$\bs{J}_{2}^{(1)}$ and $\bs{J}_{2}^{(2)}$. For a given colour ordering of a particular $n$-particle process,
$\bs{J}_{n}^{(1)}$ and $\bs{J}_{n}^{(2)}$ is simply a sum over integrated dipoles that involve colour connected particles. 
Equivalently, the pole structure is identified as a sum of dipole-like terms proportional to
$\left(|s_{ij}|\right)^{-\ell\eps}$  (for $\ell=1,2$) that link the colour connected particles $i$ and $j$. There is one
caveat: When the integrated dipole involves the $D_4^0$ and $D_3^1$ antenna, there is a well defined correction factor
$\overline{\bs{{J}}}_{2}^{(2)}$ that removes the spurious colour connections between the quark and antiquark. 

The integrated antenna strings contribute to the real-virtual and double virtual subtraction terms in a very precise manner,
again as illustrated in Fig.~\ref{fig:roadmap}, leading to the NNLO master formula of Eq.~\eqref{eq:dsu} for the double
virtual subtraction term.  In this form, there is a direct correspondence with Catani's one- and two-loop factorisation
formulae and the explicit pole cancellation against the double virtual contribution is particularly straightforward.  

The form of  $\bs{J}_{n}^{(1)}$ and $\bs{J}_{n}^{(2)}$ in terms of integrated antennae imposes a particular structure on the
unintegrated antennae that make up the real radiation subtraction terms. Understanding the explicit pole structure of virtual
amplitudes in terms of integrated antenna strings gives a direct connection between the block structure of the unintegrated
subtraction terms and the explicit pole structure of virtual contributions, thereby simplifying the construction of the double
real, real-virtual and double virtual subtraction terms.

We anticipate that the structuring of the NNLO subtraction terms $\dsigma_{NNLO}^S$,  $\dsigma_{NNLO}^T$ and 
$\dsigma_{NNLO}^U$, presented here will be of great help towards the evaluation of the NNLO corrections to a number of
important processes involving jets at hadron-hadron and electron-hadron colliders such as $pp \to 2$~jets, $pp\to H+$~jet, $pp
\to V+$~jet and $ep \to 2(+1)$~jets where the experimental interpretation is limited by the theoretical uncertainties.   In
each of these cases, the ultimate goal is a full parton-level Monte Carlo implementation.

\acknowledgments 
We thank Thomas Gehrmann, Aude Gehrmann-De Ridder and Joao Pires for valuable
discussions and a careful reading of the manuscript. This research was supported in part by the UK Science and
Technology Facilities Council and in part by the European Commission through the
``LHCPhenoNet" Initial Training Network PITN-GA-2010-264564. EWNG gratefully
acknowledges the support of the Wolfson Foundation, the Royal Society and the
Pauli Center for Theoretical Studies.


\appendix

\section{Collinear splitting kernels}
\label{sec:split}

\subsection{Leading order splitting kernels}

\label{sec:splitcoltree}

The splitting kernels present in the definition of the mass factorisation terms \eqref{eq:mfnlo}, \eqref{eq:RVMF} and \eqref{eq:VVMF} contain colour factors which can distribute the colour ordered splitting kernels across multiple orders in the colour decomposition.  

The leading order splitting kernels can be decomposed into colour factors $N$ and $\NF$.  Expanding the splitting kernels into this set of colour factors defines a  set of colour stripped splitting kernels:
\begin{eqnarray}
\bs{P}_{qq}^{0}(x)&=&\biggl(\frac{N^{2}-1}{N}\biggr)\ p_{qq}^{0}(x),\nn\\
\bs{P}_{gq}^{0}(x)&=&\biggl(\frac{N^{2}-1}{N}\biggr)\ p_{gq}^{0}(x),\nn\\
\bs{P}_{qg}^{0}(x)&=&p_{qg}^{0}(x),\nn\\
\bs{P}_{gg}^{0}(x)&=&N\ p_{gg}^{0}(x)+\NF\ p_{gg,F}^{0}(x).
\end{eqnarray}
In this decomposition, the colour ordered splitting kernels are given by~\cite{Daleo:2006xa},
\begin{eqnarray}
{p}_{qq}^{0}(x)&=&{\cal{D}}_{0}(x)-\frac{(1+x)}{2}+\frac{3}{4}\delta(1-x)\label{eq:p0},\nn\\
{p}_{gq}^{0}(x)&=&\frac{1}{x}-1+\frac{x}{2},\nn\\
{p}_{qg}^{0}(x)&=&\frac{1}{2}-x+x^{2},\nn\\
{p}_{gg}^{0}(x)&=&2{\cal{D}}_{0}(x)+\frac{2}{x}-4+2x-2x^{2}+b_{0}\delta(1-x),\nn\\
{p}_{gg,F}^{0}(x)&=&b_{0,F}\delta(1-x),\label{eq:LOsplit}
\end{eqnarray}
where the distributions ${\cal{D}}_{n}(x)$ are defined in terms of plus-distributions,
\begin{eqnarray}
{\cal{D}}_{n}(x)&=&\biggl(\frac{\ln^{n}(1-x)}{1-x}\biggr)_{+}.
\end{eqnarray}
The mass factorisation kernels are classified using a similar notation to the splitting kernels,
\begin{eqnarray}
\bs{\Gamma}_{qq}^{(1)}(x)&=&\biggl(\frac{N^{2}-1}{N}\biggr)\ \Gamma_{qq}^{(1)}(x),\nn\\
\bs{\Gamma}_{gq}^{(1)}(x)&=&\biggl(\frac{N^{2}-1}{N}\biggr)\ \Gamma_{gq}^{(1)}(x),\nn\\
\bs{\Gamma}_{qg}^{(1)}(x)&=&\Gamma_{qg}^{(1)}(x),\nn\\
\bs{\Gamma}_{gg}^{(1)}(x)&=&N\ \Gamma_{gg}^{(1)}(x)+\NF\ \Gamma_{gg,F}^{(1)}(x).
\end{eqnarray}
where the mass factorisation kernels in the $\overline{\text{MS}}$ scheme are related to the splitting kernels by~\cite{GehrmannDeRidder:2011aa},
\ba
\bs{\Gamma}_{ij}^{(1)}(x)&=&-\frac{1}{\eps}\bs{P}_{ij}^{0}(x).\label{eq:gammadef}
\ea

\subsection{Next-to-leading order splitting kernels}

\label{sec:splitcolloop}

The one-loop quark-quark and quark-antiquark splitting kernels, $\bs{P}_{q_{i}q_{j}}^{1}$ and $\bs{P}_{q_{i}\bar{q}_{j}}^{1}$, contain a non-trivial flavour structure and are classified according to flavour singlet, $\bs{P}_{ij}^{S,1}$, and flavour non-singlet, $\bs{P}_{ij}^{V,1}$ contributions~\cite{Ellis:1991qj}.  
\begin{eqnarray}
\bs{P}_{q_{i}q_{j}}^{1}(x)&=&\delta_{ij}\bs{P}_{qq}^{V,1}(x)+\bs{P}_{qq}^{S,1}(x),\nn\\
\bs{P}_{q_{i}\bar{q}_{j}}^{1}(x)&=&\delta_{ij}\bs{P}_{q\bar{q}}^{V,1}(x)+\bs{P}_{q\bar{q}}^{S,1}(x),
\end{eqnarray}
where $\bs{P}_{q\bar{q}}^{S,1}(x)=\bs{P}_{qq}^{S,1}(x)$. 

For the purposes of the DGLAP evolution, the parton distributions are decomposed into a flavour basis of non-singlet, $f_{n}(x,\mu_{F}^{2})$, and singlet, $f_{s}(x,\mu_{F}^{2})$, combinations, where $x$ denotes the fraction of the hadron momentum carried by the parton and $\mu_{F}^{2}$ is the factorisation scale.  Suppressing the $x$ and $\mu_{F}^{2}$ dependence of the distributions, the flavour asymmetry distributions are given by,
\ba
f_{n,ij}^{\pm}&=&(f_{i}\pm\bar{f}_{i})-(f_{j}\pm\bar{f}_{j}).
\ea
The total valence distribution is given by,
\ba
f_{n}^{v}&=&\sum_{i=1}^{\NF}\ (f_{i}-\bar{f}_{i}),
\ea 
and the singlet distribution, which mixes with the gluon distribution during DGLAP evolution, is given by,
\ba
f_{s}&=&\sum_{i=1}^{\NF}\ (f_{i}+\bar{f}_{i}).
\ea
Each distribution is evolved between momentum scales using the corresponding DGLAP evolution kernel.  The flavour asymmetry, total valence and sea evolution kernels are denoted  $\bs{P}_{n}^{\pm}(x)$, $\bs{P}_{n}^{v}(x)$ and $\bs{P}_{s}(x)$ respectively.  These kernels are related to the singlet and non-singlet splitting kernels in the following way:
\ba
\bs{P}_{n}^{\pm}(x)&=&\bs{P}_{qq}^{V}(x)\pm\bs{P}_{q\bar{q}}^{V}(x),\nn\\
\bs{P}_{n}^{v}(x)&=&\bs{P}_{n}^{-}(x)+\bs{P}_{n}^{s}(x),\nn\\
\bs{P}_{s}(x)&=&\bs{P}_{n}^{+}(x)+\bs{P}_{s}^{s}(x),\label{eq:PDGLAP}
\ea
where the non-singlet kernel, given by difference between the total valence $\bs{P}_{n}^{v}(x)$ and the ``minus'' $\bs{P}_{n}^{-}(x)$ evolution kernels is,
\ba
\bs{P}_{n}^{s}(x)&=&\NF(\bs{P}_{qq}^{S}(x)-\bs{P}_{q\b{q}}^{S}(x)),
\ea
and the ``pure singlet'' kernel, given by the difference of the sea and ``plus'' evolution kernels is,
\ba
\bs{P}_{s}^{s}(x)&=&\NF(\bs{P}_{qq}^{S}(x)+\bs{P}_{q\b{q}}^{S}(x)).
\ea
The singlet and non-singlet splitting kernels have the perturbative expansions:
\ba
\bs{P}_{qq}^{V}(x)&=&\bs{P}_{qq}^{0}(x)+\biggl(\frac{\alpha_{s}}{2\pi}\biggl) \bs{P}_{qq}^{V,1}(x)+{\cal{O}}(\alpha_{s}^{2}),\nn\\
\bs{P}_{q\bar{q}}^{V}(x)&=&\biggl(\frac{\alpha_{s}}{2\pi}\biggl)\bs{P}_{q\bar{q}}^{V,1}(x)+{\cal{O}}(\alpha_{s}^{2}),\nn\\
\bs{P}_{qq}^{S}(x)&=&\biggl(\frac{\alpha_{s}}{2\pi}\biggl) \bs{P}_{qq}^{S,1}(x)+{\cal{O}}(\alpha_{s}^{2}),\nn\\
\bs{P}_{q\bar{q}}^{S}(x)&=&\biggl(\frac{\alpha_{s}}{2\pi}\biggl) \bs{P}_{q\bar{q}}^{S,1}(x)+{\cal{O}}(\alpha_{s}^{2}).
\ea
At one-loop, the relation $\bs{P}_{qq}^{S,1}(x)=\bs{P}_{q\bar{q}}^{S,1}(x)$ simplifies the perturbative expansion of Eq.~\eqref{eq:PDGLAP}:
\ba
\bs{P}_{n}^{\pm,1}(x)&=&\bs{P}_{qq}^{V,1}(x)\pm\bs{P}_{q\bar{q}}^{V,1}(x),\nn\\
\bs{P}_{n}^{v,1}(x)&=&\bs{P}_{n}^{-,1}(x),\nn\\
\bs{P}_{s}^{1}(x)&=&\bs{P}_{n}^{+,1}(x)+\bs{P}_{s}^{s,1}(x),\label{eq:PDGLAP1loop}
\ea
where the one-loop pure singlet kernel is given by,
\ba
\bs{P}_{s}^{s,1}(x)&=&2\NF\bs{P}_{qq}^{S,1}(x).
\ea
The one-loop non-singlet splitting kernels, $\bs{P}_{qq}^{V,1}(x)$ and $\bs{P}_{q\b{q}}^{V,1}(x)$, are presented explicitly in Refs.~\cite{Ellis:1991qj,Moch:2004pa}, and the singlet evolution kernel, $\bs{P}_{s}^{1}(x)$, in Ref.~\cite{Ellis:1991qj}.  Rearranging for $\bs{P}_{qq}^{S,1}(x)$ in Eq.\eqref{eq:PDGLAP1loop} yields the quark-quark singlet splitting kernel, 
\ba
\bs{P}_{qq}^{S,1}(x)&=&\frac{1}{2\NF}\big(\bs{P}_{s}^{1}(x)-\bs{P}_{n}^{+,1}(x)\big).\label{eq:pss1}
\ea
This function can be derived from the forms of $\bs{P}_{qq}^{V,1}(x)$, $\bs{P}_{q\b{q}}^{V,1}(x)$ and $\bs{P}_{s}^{1}(x)$ in Ref.~\cite{Ellis:1991qj} using Eq.~\eqref{eq:pss1} or found explicitly in Ref.~\cite{Vogt:2004mw}.  Given the explicit form of $\bs{P}_{qq}^{V,1}(x)$, $\bs{P}_{q\b{q}}^{V,1}(x)$ and $\bs{P}_{qq}^{S,1}(x)$, the full one-loop splitting kernels, $\bs{P}_{qq}^{1}(x)$ and $\bs{P}_{q\b{q}}^{1}(x)$ are easily obtained.

The one-loop splitting kernels can be decomposed into the $N,\ \NF$ colour factors:
\ba
\bs{P}_{qq}^{1}(x)&=&\biggl(\frac{N^{2}-1}{N}\biggr)\biggl[N\ p_{qq}^{1}(x)+ \tilde{p}_{qq}^{1}(x)+\frac{1}{N}\ \tilde{\tilde{p}}_{qq}^{1}(x)+\NF\ p_{qq,F}^{1}(x)\biggr],\nn\\
\bs{P}_{q\bar{q}}^{1}(x)&=&\biggl(\frac{N^{2}-1}{N}\biggr)\biggl[ p_{q\bar{q}}^{1}(x)+\frac{1}{N}\ \tilde{p}_{q\bar{q}}^{1}(x)\biggr],\nn\\
\bs{P}_{qQ}^{1}(x)&=&\biggl(\frac{N^{2}-1}{N}\biggr)\ p_{qQ}^{1}(x),\nn\\
\bs{P}_{q\bar{Q}}^{1}(x)&=&\biggl(\frac{N^{2}-1}{N}\biggr)\ p_{q\bar{Q}}^{1}(x),\nn\\
\bs{P}_{gq}^{1}(x)&=&\biggl(\frac{N^{2}-1}{N}\biggr)\biggl[N\ p_{gq}^{1}(x)+ \frac{1}{N}\ \tilde{p}_{gq}^{1}(x)+\NF\ p_{gq,F}^{1}(x)\biggr],\nn\\
\bs{P}_{qg}^{1}(x)&=&N\ p_{qg}^{1}(x)+\frac{1}{N}\ \tilde{p}_{qg}^{1}(x),\nn\\
\bs{P}_{gg}^{1}(x)&=&N^{2}\ p_{gg}^{1}(x)+N\NF\ p_{gg,F}^{1}(x)+\frac{\NF}{N}\ \tilde{p}_{gg,F}^{1}(x).\label{eq:NLOsplit}
\ea
At NNLO, two new ingredients are present in the mass factorisation terms: convolutions of one-loop mass factorisation kernels, either with other mass factorisation kernels or integrated antenna functions:
\ba
\big[\Gamma_{ij}^{(1)}\otimes\Gamma_{jk}^{(1)}\big](z_{1},z_{2})&=&\int{\rm{d}}x_{1}{\rm{d}}y_{1}\ \Gamma_{ij}^{(1)}(x_{1})\Gamma_{jk}^{(1)}(y_{1})\delta(z_{1}-x_{1}y_{1})\delta(1-z_{2}),\nn\\
\big[\Gamma_{ij}^{(1)}\otimes{\cal{X}}_{3}^{0}\big](s;z_{1},z_{2})&=&\int{\rm{d}}x_{1}{\rm{d}}y_{1}\ \Gamma_{ij}^{(1)}(x_{1}){\cal{X}}_{3}^{0}(s;y_{1},z_{2})\delta(z_{1}-x_{1}y_{1}),
\ea
and the reduced two-loop mass factorisation kernel, defined as,
\ba
\bs{\overline{\Gamma}}_{ij}^{(2)}(x)&=&-\frac{1}{2\eps}\bigg(\bs{P}_{ij}^{1}(x)+\frac{\beta_{0}}{\eps}\bs{P}_{ij}^{0}(x)\bigg).\label{eq:reduced2loop}
\ea
Decomposing the reduced two-loop mass factorisation kernels into the $N$ $\NF$ colour factors yields:
\ba
\bs{\overline{\Gamma}}_{qq}^{(2)}(x)&=&\bigg(\frac{N^{2}-1}{N}\bigg)\bigg[N\overline{\Gamma}_{qq}^{(2)}(x)+\tilde{\overline{\Gamma}}_{qq}^{(2)}(x)+\frac{1}{N}\tilde{\tilde{\overline{\Gamma}}}_{qq}^{(2)}(x)+\NF\overline{\Gamma}_{qq,F}^{(2)}(x)\bigg],\nn\\
\bs{\overline{\Gamma}}_{q\b{q}}^{(2)}(x)&=&\bigg(\frac{N^{2}-1}{N}\bigg)\bigg[\overline{\Gamma}_{q\b{q}}^{(2)}(x)+\frac{1}{N}\tilde{\overline{\Gamma}}_{q\b{q}}^{(2)}(x)\bigg],\nn\\
\bs{\overline{\Gamma}}_{qQ}^{(2)}(x)&=&\bigg(\frac{N^{2}-1}{N}\bigg)\ \overline{\Gamma}_{qQ}^{(2)}(x),\nn\\
\bs{\overline{\Gamma}}_{q\b{Q}}^{(2)}(x)&=&\bigg(\frac{N^{2}-1}{N}\bigg)\ \overline{\Gamma}_{q\b{Q}}^{(2)}(x),\nn\\
\bs{\overline{\Gamma}}_{gq}^{(2)}(x)&=&\bigg(\frac{N^{2}-1}{N}\bigg)\bigg[N\overline{\Gamma}_{gq}^{(2)}(x)+\frac{1}{N}\tilde{\overline{\Gamma}}_{gq}^{(2)}(x)+\NF\overline{\Gamma}_{gq,F}^{(2)}(x)\bigg],\nn\\
\bs{\overline{\Gamma}}_{qg}^{(2)}(x)&=&\bigg[N\overline{\Gamma}_{qg}^{(2)}(x)+\frac{1}{N}\tilde{\overline{\Gamma}}_{qg}^{(2)}(x)+\NF\overline{\Gamma}_{qg,F}^{(2)}(x)\bigg],\nn\\
\bs{\overline{\Gamma}}_{gg}^{(2)}(x)&=&\bigg[N^{2}\overline{\Gamma}_{gg}^{(2)}(x)+N\NF\overline{\Gamma}_{gg,F}^{(2)}(x)+\frac{\NF}{N}\tilde{\overline{\Gamma}}_{gg,F}^{(2)}(x)+\NF^{2}\overline{\Gamma}_{gg,F^{2}}^{(2)}(x)\bigg],
\ea
where the explicit form of each colour stripped function can be simply inferred from the colour decomposition of the leading order and one-loop splitting functions in Eqs.~\eqref{eq:LOsplit} and \eqref{eq:NLOsplit}, together with the definition of the reduced two-loop kernel in Eq.~\eqref{eq:reduced2loop}, to be:
\ba
\overline{\Gamma}_{qq}^{(2)}(x)&=&\frac{1}{\eps^{2}}\bigg[-\frac{1}{2}b_{0}p_{qq}^{0}(x)\bigg]\nn\\
&+&\frac{1}{\eps}\bigg[\bigg( \frac{67}{72} - \frac{1}{24}\pi^2 + \frac{13}{48}\ln(x) - \frac{1}{4}\ln(x)\ln(1 - x) + \frac{1}{8}\ln(x)^2 \bigg)p_{qq}(x)\nn\\
&+&\frac{25}{24}(1-x)+\frac{1}{16}\bigg(1-3x-(1+x)\ln(x)\bigg)\ln(x)+\bigg(\frac{43}{192}+\frac{13\pi^2}{144}\bigg)\delta(1 - x)\bigg],\nn\\
\tilde{\overline{\Gamma}}_{qq}^{(2)}(x)&=&\frac{1}{\eps}\bigg[(1-3x)-\frac{1}{9}\bigg(\frac{10}{x}-28x^{2}\bigg)-\frac{1}{2}(1+5x)\ln(x)+\frac{4}{3}x^{2}\ln(x)+(1+x)\ln^{2}(x)\bigg],\nn\\
\tilde{\tilde{\overline{\Gamma}}}_{qq}^{(2)}(x)&=&\frac{1}{\eps}\bigg[-\bigg(\frac{3}{16}\ln(x)+\frac{1}{4}\ln(x)\ln(1-x)\bigg)p_{qq}(x)-\frac{5}{8}(1-x)\nn\\
&-&\frac{1}{16}\bigg(3+7x+(1+x)\ln(x)\bigg)\ln(x)-\frac{1}{4}\bigg(\frac{\pi^{2}}{4}-3\zeta_{3}-\frac{3}{16}\bigg)\delta(1-x)\bigg],\nn\\
\overline{\Gamma}_{qq,F}^{(2)}(x)&=&\frac{1}{\eps^{2}}\bigg[-\frac{1}{2}b_{0,F}p_{qq}^{0}(x)\bigg]\nn\\
&+&\frac{1}{\eps}\bigg[\bigg(\frac{5}{36}+\frac{1}{12}\ln(x)\bigg)p_{qq}(x)+\frac{1}{6}(1-x)+\frac{1}{12}\bigg(\frac{1}{4}+\frac{\pi^{2}}{3}\bigg)\delta(1-x)\bigg],\\
\overline{\Gamma}_{q\b{q}}^{(2)}(x)&=&\tilde{\overline{\Gamma}}_{qq}^{(2)}(x),\nn\\
\tilde{\overline{\Gamma}}_{q\b{q}}^{(2)}(x)&=&\frac{1}{\eps}\bigg[\frac{1}{4}S_{2}(x)p_{qq}(-x)+\frac{1}{2}(1-x)+\frac{1}{4}(1+x)\ln(x)\bigg],\\
\overline{\Gamma}_{qQ}^{(2)}(x)&=&\tilde{\overline{\Gamma}}_{qq}^{(2)}(x),\\
\overline{\Gamma}_{q\b{Q}}^{(2)}(x)&=&\tilde{\overline{\Gamma}}_{qq}^{(2)}(x),\\
\overline{\Gamma}_{gq}^{(2)}(x)&=&\frac{1}{\eps^{2}}\bigg[-\frac{1}{2}b_{0}p_{gq}^{0}(x)\bigg]\nn\\
&+&\frac{1}{\eps}\bigg[-\frac{1}{8}\bigg(1-\frac{\pi^{2}}{3}+\frac{13}{3}\ln(1-x)+\ln^{2}(1-x)-4\ln(x)\ln(1-x)+\ln^{2}(x)\bigg)p_{gq}(x),\nn\\
&-&\frac{1}{4}S_{2}(x)p_{gq}(-x)-\frac{67}{144}(1+x)-\frac{11}{9}x^{2}-\frac{1}{4}x\ln(1-x)+\frac{11}{4}\ln(x)\nn\\
&-&\frac{7}{16}x\ln(x)+\frac{10}{3}\ln(x)x^{3}-\frac{7}{8}\ln^{2}(x)-\frac{5}{16}x\ln^{2}(x)\bigg],\nn\\
\tilde{\overline{\Gamma}}_{gq}^{(2)}(x)&=&\frac{1}{\eps}\bigg[-\frac{1}{8}\bigg(3\ln(1-x)+\ln^{2}(1-x)\bigg)p_{gq}^{0}(x)-\frac{1}{16}\bigg(5+7x(1-\ln(x))\bigg)\nn\\
&-&\frac{1}{4}\bigg(x\ln(1-x)-\ln(x)+\frac{1}{2}\ln^{2}(x)\bigg)\bigg],\nn\\
\overline{\Gamma}_{gq,F}^{(2)}(x)&=&\frac{1}{\eps^{2}}\bigg[-\frac{1}{2}b_{0,F}p_{gq}^{0}(x)\bigg]\nn\\
&+&\frac{1}{\eps}\bigg[\frac{1}{6}\bigg(\frac{5}{3}+\ln(1-x)\bigg)p_{gq}(x)+\frac{1}{6}x\bigg],\\
\overline{\Gamma}_{qg}^{(2)}(x)&=&\frac{1}{\eps^{2}}\bigg[-\frac{1}{2}b_{0}p_{qg}^{0}(x)\bigg]\nn\\
&+&\frac{1}{\eps}\bigg[-\frac{1}{2}S_{2}(x)p_{qg}(-x)+\bigg(\frac{173}{36}+\frac{1}{2}\ln\bigg(\frac{1-x}{x}\bigg)-\frac{1}{4}\ln^{2}\bigg(\frac{1-x}{x}\bigg)-\ln(1-x)\nn\\
&+&\frac{1}{2}\ln^{2}(1-x)-\frac{11}{3}\ln(x)+\frac{1}{4}\ln^{2}(x)\bigg)p_{qg}(x)-\frac{50}{9}-\frac{10}{9x}+\frac{53}{72}x\nn\\
&+&\frac{1}{2}\ln(1-x)+\frac{79}{24}\ln(x)-\frac{71}{6}\ln(x)+\frac{5}{8}\ln^{2}(x)+\frac{7}{4}x\ln^{2}(x)\bigg],\nn\\
\tilde{\overline{\Gamma}}_{qg}^{(2)}(x)&=&\frac{1}{\eps}\bigg[\bigg(\frac{5}{4}-\frac{\pi^{2}}{12}-\frac{1}{2}\ln\bigg(\frac{1-x}{x}\bigg)+\frac{1}{4}\ln^{2}\bigg(\frac{1-x}{x}\bigg)\bigg)p_{qg}(x)+\frac{1}{2}-\frac{9}{8}x\nn\\
&+&\frac{1}{2}\ln(1-x)-\frac{1}{8}\ln(x)-\frac{1}{8}\ln^{2}(x)+\frac{1}{2}x\ln(x)+\frac{1}{4}\ln^{2}(x)\bigg],\\
\overline{\Gamma}_{gg}^{(2)}(x)&=&\frac{1}{\eps^{2}}\bigg[-\frac{1}{2}b_{0}p_{gg}^{0}(x)\bigg]\nn\\
&+&\frac{1}{\eps}\bigg[-S_{2}(x)p_{gg}(-x)-\bigg(\frac{67}{18}+\frac{\pi^{2}}{6}+2\ln(x)\ln(1-x)-\frac{1}{2}\ln^{2}(x)\bigg)p_{gg}(x)\nn\\
&-&\frac{27}{4}(1-x)+\frac{67}{18}\bigg(\frac{1}{x}-x^{2}\bigg)+\frac{25}{6}\ln(x)-\frac{11}{6}(1-4x)\ln(x)\nn\\
&-&2(1+x)\ln^{2}(x)-\frac{1}{2}\bigg(\frac{8}{3}+3\zeta_{3}\bigg)\delta(1-x)\bigg],\nn\\
\overline{\Gamma}_{gg,F}^{(2)}(x)&=&\frac{1}{\eps^{2}}\bigg[-\frac{1}{2}b_{0,F}p_{gg}^{0}(x)-\frac{1}{2}b_{0}p_{gg,F}^{0}(x)\bigg]\nn\\
&&\frac{1}{\eps}\bigg[\frac{5}{9}p_{gg}(x)+\frac{3}{2}-\frac{1}{2}x+\frac{5}{9x}-\frac{14}{9}x^{2}+\frac{13}{12}\ln(x)\nn\\
&+&\frac{19}{12}x\ln(x)+\frac{1}{4}(1+x)\ln^{2}(x)+\frac{11}{24}\delta(1-x)\bigg],\nn\\
\tilde{\overline{\Gamma}}_{gg,F}^{(2)}(x)&=&\frac{1}{\eps}\bigg[-2+\frac{1}{6x}+x+\frac{5}{6}x^{2}-\frac{1}{4}(3+5x)\ln{x}-\frac{1}{4}(1+x)\ln^{2}(x)-\frac{1}{8}\delta(1-x)\bigg],\nn\\
\overline{\Gamma}_{gg,F^{2}}^{(2)}(x)&=&\frac{1}{\eps^{2}}\bigg[\frac{5}{9}b_{0,F}p_{gg}(x)+\bigg(\frac{5}{9x}+\frac{1}{2}(3-x)-\frac{14}{9}x^{2}+\frac{13}{12}\ln(x)\nn\\
&+&\frac{19}{12}x\ln(x)+\frac{1}{4}(1+x)\ln^{2}(x)+\frac{11}{24}\delta(1-x)\bigg)b_{0,F}\bigg],
\ea
where,
\ba
p_{qq}(x)&=&\frac{2}{1-x}-1-x,\nn\\
p_{qg}(x)&=&x^{2}+(1-x)^{2},\nn\\
p_{gq}(x)&=&\frac{1+(1-x)^{2}}{x},\nn\\
p_{gg}(x)&=&\frac{1}{1-x}+\frac{1}{x}-2+x(1-x),
\ea
and
\ba
S_{2}(x)&=&-2{\rm{Li}}_{2}(-x)+\frac{1}{2}\ln^{2}(x)-2\ln(x)\ln(1+x)-\frac{\pi^{2}}{6}.
\ea

\section{Integrated large angle soft term: Initial-final configuration}
\label{sec:soft}

In this appendix, we give an expression for the integrated large angle soft terms when the primary mapping is of the initial-final form. This is an integral of the soft antenna function, $$S_{ajc} = \frac{2 s_{ac}}{s_{aj}s_{jc}},$$
over the unresolved initial-final phase space for the mapping $(\hat{i},j,k)\to(\hat{I},K)$ defined by~\cite{Daleo:2009yj},
\begin{eqnarray}
{\cal S}^{IF}(s_{ac},s_{IK},y_{ac,iK}) &=& \frac{1}{C(\e)}\,\int \d \Phi_{X_{\hat{i}jk}} \frac{Q^2}{2\pi}
\,S_{ajc}.
\end{eqnarray}
After integration over the antenna phase space~\cite{Daleo:2009yj} and expanding in distributions,\footnote{An unexpanded expression is given in \cite{Daleo:2009yj}}
\begin{eqnarray}
{\cal S}^{IF}(s_{ac},s_{IK},y_{ac,iK})
&=& \left(\frac{|s_{IK}|}{\mu^2}\right)^{-\e}
\bigg\{
\left[
\frac{1}{\e^2}-\frac{1}{\e}\ln(y_{ac,iK})+\frac{1}{2}\ln^2(y_{ac,iK}) -\frac{\pi^2}{12}\right] \delta(1-x)\nonumber \\
&&+\bigg[\frac{2}{\e}\left(1-{\cal D}_0(x)\right) 
+ 4 {\cal D}_1(x)
+2\ln\left(\frac{s_{ac}|s_{IK}|}{s_{aK}s_{cK}}\right)
-4\ln(1-x)\nonumber \\
&&\phantom{+\bigg[}
-\frac{4x}{(1-x)}\ln(x)
+\frac{2x}{(1-x)}\ln(y_{ac,iK})\nn\\
&&\phantom{+\bigg[}-\frac{2}{(1-x)}\ln\left(\frac{s_{ac}|s_{IK}|}{s_{aK}s_{cK}}\right)\bigg]+{\cal O}(\e)\bigg\},
\end{eqnarray}
where 
\begin{equation}
y_{ac,iK} = \frac{s_{ac}|s_{IK}|}{\left(s_{aK}+(1-x)s_{ai}\right)\,
\left(s_{cK}+(1-x)s_{ci}\right)}\;.
\label{eq:yang}
\end{equation}


\providecommand{\href}[2]{#2}\begingroup\raggedright

\end{document}